\documentclass[twocolumn,superscriptaddress, floatfix,amsmath,amssymb,nofootinbib,prd]{revtex4-1}
\usepackage[english]{babel}
\usepackage{graphicx}% Include figure files
\usepackage{dcolumn}% Align table columns on decimal point
\usepackage{bm}% bold math
\usepackage{verbatim}
\usepackage{mathrsfs}
\usepackage{natbib}
\usepackage{mathtools}
\usepackage{color}
\usepackage{hyperref}
\usepackage{subcaption}
\usepackage[dvipsnames]{xcolor}
\usepackage[utf8]{inputenc}
\pdfoutput=1

\newcommand{\ket}[1]{\left| {#1} \right\rangle}

\newcommand{\ii}{\mathrm{i}}
\newcommand{\dd}{\mathrm{d}}

\newcommand{\tr}{\operatorname{Tr}}

\newcommand{\tcr}{\textcolor{red}}
\newcommand{\tcb}{\textcolor{blue}}

\newcommand{\sgn}{\text{sgn}}

\newcommand{\be}{\begin{equation}}
\newcommand{\ee}{\end{equation}}

\begin{document}
\title{Unruh-DeWitt detectors and entanglement: the anti-de Sitter space}
\author{Keith K. Ng}
\affiliation{Department of Physics and Astronomy, University of Waterloo, Waterloo, ON, N2L 3G1, Canada}
\author{Robert B. Mann}
\affiliation{Department. Physics and Astronomy, University of Waterloo, Waterloo, ON, N2L 3G1, Canada}
\affiliation{Institute for Quantum Computing, University of Waterloo, Waterloo, Ontario, N2L 3G1, Canada}
\affiliation{Perimeter Institute for Theoretical Physics, Waterloo, ON, N2L 2Y5, Canada}
\author{Eduardo Mart\'{i}n-Mart\'{i}nez}
\affiliation{Department of Applied Mathematics, University of Waterloo, Waterloo, ON, N2L 3G1, Canada}
\affiliation{Institute for Quantum Computing, University of Waterloo, Waterloo, Ontario, N2L 3G1, Canada}
\affiliation{Perimeter Institute for Theoretical Physics, Waterloo, ON, N2L 2Y5, Canada}

\begin{abstract}
%\tcr{We investigate entanglement harvesting for Unruh de Witt detectors in AdS$_4$.  Applying the general results of \cite{Ng:2018ilp}, we consider two scenarios:
%one where both detectors are geodesic, with equal redshift; and one where both are static, at unequal redshift. 
 %As expected, at large AdS length $L$, our results approximate flat space. However at smaller $L$ 
 %we observe non-trivial effects for various field boundary conditions. Furthermore, in the static case we observe a novel feature of the entanglement as a function of switching time delay, which we attribute to different (coordinate) frequencies of the detectors. We also find   an island of separability in parameter space, analogous that
 %that observed in AdS$_3$, to which we compare and contrast our other results. The variety of features observed in
 %both cases suggest further study in other spacetimes.}
  
We investigate entanglement harvesting in AdS$_4$.  Applying the general results of \cite{Ng:2018ilp}, we consider two scenarios:
one where two particle detectors are geodesic, with equal redshift; and one where both are static, at unequal redshift. 
 As expected, at large AdS length $L$, our results approximate flat space. However at smaller $L$ we observe non-trivial effects for various field boundary conditions. Furthermore, in the static case we observe a novel feature of the entanglement as a function of switching time delay, which we attribute to different (coordinate) frequencies of the detectors. We also find  an island of separability in parameter space, analogous that
 that observed in AdS$_3$, to which we compare and contrast our other results. The variety of features observed in
 both cases suggest further study in other spacetimes.
  
\end{abstract}

\maketitle

\section{Introduction}
Quantum mechanics and general relativity are two of the most precise and predictive theories of physics ever created. It is therefore of some concern that they are not compatible. In the frontiers of the black hole, physicists have found a contradiction: either general relativity is correct, and that which falls into a black hole is lost forever; or quantum mechanics is correct, and the information contained within the falling object survives. This disagreement is known as the information paradox.

There are a few schools of thought regarding this paradox \cite{Mann:2015luq}. Some physicists believe that general relativity is correct, and thus that information really is lost. While this would make gravity unique among the fundamental interactions, it is not entirely unprecedented: indeed, the de Sitter solution, not unlike the spacetime we live in, does allow information to lose causal contact with us, over cosmological time scales. Other physicists favour quantum mechanics, and quantum field theory: perhaps that which we call a `black hole' is in truth a more complicated structure at its smallest scales. This structure might indeed be capable of holding information inside it, releasing it as the black hole evaporates. Most recently, a few physicists have even proposed that information never falls into a black hole: instead, it is immediately stripped from infalling matter at the horizon by a `firewall' \cite{Almheiri2012}. Debate on this question continues.

One tool that has been employed to answer this question is known as the AdS/CFT duality
\cite{Ramallo:2013bua}. In its original form, the dynamics of general relativity inside an AdS spacetime may be translated into a quantum field theory on its boundary. Since this QFT cannot lose information, a black hole inside the spacetime's bulk must still conserve information. Physicists have then sought to determine the form that this information takes, and whether this might be generalized to our de Sitter spacetime. However, tracking information through the phase transition/collapse of a black hole has proven to be a formidable challenge indeed. Still, this idea has led to a profusion of new avenues of investigation.

Recently, there has been much interest in the entanglement present in the vacuum state of a quantum field on a curved background. Specifically, the nature of the AdS/CFT duality seems to suggest that questions about gravity in a spacetime might be translated into questions about quantum information on its boundaries, or at least on a relatively gravity-free region. For instance, the length of a wormhole, a theoretical possibility within general relativity, has been found to be linked to the entanglement structure of its two horizons \cite{Brown:2015lvg}. However, calculations of the entanglement entropy of a region of space are very computationally expensive, and questions have arisen as to how much of the entanglement is accessible. (For instance, in models with nonzero spin fields, it is unclear whether some degrees of freedom are accessible at all.) Therefore, the Unruh-DeWitt model, and several other detector-type models, have been deployed to study the entanglement of the quantum vacuum.

Employing particle detectors to study entanglement has been until recently a largely unexplored subject, despits
 Valentini's  demonstration some time ago \cite{VALENTINI1991321} (and revisited by Reznik in 2002 \cite{Reznik:2002fz}) that atoms interacting with the electromagnetic field can become entangled, even when spacelike separated for the duration of their interaction.  This process of extracting entanglement from the quantum field is known as `entanglement harvesting', and  research   in this subject has intensified in the past few years \cite{rangefind,Pozas-Kerstjens:2015gta,PozasEM,Pozas-Kerstjens:2017xjr,PetarNON,PetarJonsson}.

More recently, it has been found that the structure of spacetime affects the entanglement extracted \cite{MenicucciCosmo,Martin-MartinezTopo}. The entanglement of Unruh-DeWitt detectors thus becomes a valuable tool for studying the structure of spacetime.

Much progress has been made in characterizing the behaviour of Unruh-DeWitt detectors.  Unruh-DeWitt detectors are good models for the light-matter interaction when exchange of angular momentum does not play a  role in the interaction \cite{Pozas2016,Martin-Martinez2018}. It is only natural, then, that we bring the UDW detector back to the AdS space, in order (amongst other things) to further understand the AdS/CFT duality. In many ways, AdS is quite `friendly': it may be conformally mapped into a compact space, which implies it has a discrete spectrum, and its high degree of symmetry implies a simple vacuum structure. However, some fundamental questions still remain. In particular, we choose to study the entanglement which UDW detectors may harvest. While this is in some sense comparable to studying entanglement harvesting in a box, the addition of spacetime curvature adds some interesting features, which we will explore.

Understanding the entanglement structure of Anti-de Sitter space is also an important step towards understanding AdS/CFT. Moreover, the methods we develop in this paper may be employed in any static spacetime, and may be of use to studying the entanglement structure of any static spacetime. AdS is thus both a useful spacetime to study, and a testbed towards future investigations.

In this paper we analyze the entanglement harvested by two UDW detectors in AdS$_4$, in two configurations: one where both are in geodesic motion at the same redshift, and another where both are static at different redshifts.   As one would expect, at large curvatures, the geodesic case approximates flat space quite well. However, at smaller curvatures, the effects of the boundary conditions become more apparent.  As well, a number of novel features are visible in the static case, most strikingly an island of separability in parameter space at intermediate values of
the AdS length, whose origin remains to be understood. This feature has also been concurrently observed in AdS$_3$ \cite{HHMSZ}, and we compare and contrast our other results to this case.  The variety of features we
observe warrant further study in other model spacetimes: this is especially important for the static case as it is the only generally applicable case.  Our application of the findings of \cite{Ng:2018ilp} to the AdS$_4$ case serve both as a demonstration of the method's potential, and as a foundation for further investigation.

\section{Fourier transforms and switching functions}

Consider a spacetime manifold with a global timelike Killing symmetry, i.e. a Killing time $t$, with a scalar field. For simplicity, we treat the spacetime as a background: that is, we will not consider the effects of the field's gravitation on the spacetime. Then, the presence of a global time allows us to have a well-defined notion of particles everywhere on the manifold. Specifically, let us consider solutions to the massless Klein-Gordon equation,
\begin{equation}
    (\Box - \zeta R)\Phi=0,
\end{equation}
where $\zeta$ describes the coupling of the scalar field $\Phi$ to the Ricci curvature of the manifold. (In 3+1 dimensions, $\zeta=1/6$ defines the conformal coupling.) Then, because the spacetime is static, we can define a basis of positive-frequency solutions as
\begin{equation}
    \Phi_{n l m}(t,x)=\frac{1}{\sqrt{2\omega_{n l m}}}e^{-\ii\omega_{n l m} t}\varphi_{n l m}(x),
\end{equation}
where $n,l,m$ are indices which enumerate the space of solutions. One can then show that any solution to the field equations can be represented by a linear superposition of the particle solutions and their negative-frequency complex conjugates. We can also use these particle solutions to quantize the scalar field; the quantum field state in which no particles exist is the vacuum. More generally, all states of a free quantum field may be characterized by their one- and two-point correlation functions, $\langle\hat\Phi(t,x)\rangle$ and $W(t,x,t',x')=\langle \Psi | \hat\Phi(t,x) \hat\Phi(t',x') | \Psi \rangle$. Given these functions (more properly distributions), we can calculate any observable quantity.

Next, consider two Unruh-DeWitt detectors, A and B. These detectors are simple two-level systems,  with energy gaps $\Omega_{\textsc{a}}, \Omega_{\textsc{b}}$, coupled to the scalar field via coupling constants $\lambda_{\textsc{a}}, \lambda_{\textsc{b}}$, and have their interaction with the scalar field switched on and off according to switching functions $\chi_{\textsc{a}}(\tau_{\textsc{a}}(t)), \chi_{\textsc{b}}(\tau_{\textsc{b}}(t))$. For now, we will \textit{not} assume any of these are equal. We can then write an interaction Hamiltonian density governing their interactions with the scalar field. With respect to the Killing time $t$, this is
\begin{align}
    \hat{H}_I(t)=&\sum_{I=A,B}\lambda_I
    \chi_I(\tau_I(t))\hat{\mu}_I(t)\nonumber\\
    &\times F_I(\tau_I,\bm{\xi}_I)\hat\Phi(t,x),
\end{align}
where $F_I(\tau_I,\bm{\xi}_I)$ is the smearing function of detector I, and
\begin{equation}
    \hat\mu_I(t)=e^{\ii\Omega_I \tau_I(t)}\hat\sigma^+_I + e^{-\ii\Omega_I \tau_I(t)}\hat\sigma^-_I
\end{equation}
is the monopole operator of detector I in the interaction picture. By calculating how the state of these detectors evolve in time, we hope to understand what information an observer may gain about the structure of the background spacetime.

More specifically, suppose we initialize the detectors in their respective ground states, and the field in the $|\Psi\rangle$ state. Therefore, the initial joint density matrix of the detectors and the field becomes $\hat\rho(t_0)=|g_{\textsc{a}} g_{\textsc{b}} \rangle |\Psi\rangle \langle\Psi| \langle g_{\textsc{a}} g_{\textsc{b}}|$. We then wish to calculate the state of the detectors alone, as a function of $t$: this is accomplished by time-evolving the initial density matrix, then tracing out the field. Given the unitary evolution operator $\hat U(t,t_0)$, our final state is then 
\begin{equation}
    \hat\rho_{\textsc{ab}}(t)=\tr_\phi [\hat U(t,t_0)\hat\rho(t_0)\hat U^\dagger(t,t_0)].
\end{equation}

In practice, we do not have access to the exact unitary evolution operator $\hat U(t).$ Instead, we use the Dyson expansion to express the unitary evolution operator in terms of the interaction Hamiltonian, then truncate the expansion at finite order:
\begin{align}
    \hat U(t,t_0)=&\sum_{n=0}^\infty \hat U_n(t,t_0),\\
    \hat U_n(t,t_0)=&\frac{(-i)^n}{n!}\int_{t_0}^t \dd t_1 \int_{t_0}^t \dd t_2 \cdots \int_{t_0}^t \dd t_n \nonumber\\
    &\mathcal{T}\left(\hat H_I(t_1)\hat H_I(t_2)\cdots \hat H_I(t_n)\right).
\end{align}
We take the convention that $\hat U_0=\openone.$
In order to express the final density matrix of the detectors we then sum up: 
%\tcb{\bf[Not sure what the correct convention is. does rho2 refer to (a) the density matrix up to second order, or (b) the second order terms only?]}
\begin{equation}
    \hat \rho_{\textsc{ab}}(t)=\tr_\phi \left[\sum_{n,m=0}^{n+m\le 2} \hat U_n(t,t_0)\hat \rho(t_0) \hat U_m^\dagger (t,t_0)\right]
    + \cdots
\end{equation}
valid to second order in the interactions.
Note the inclusion of both the second-order unitary operator, as in $U_2\rho$, and the first-order unitary operator, as in $\hat U_1\hat\rho \hat U_1^\dagger$; both of these terms are needed in order to maintain the unit trace   of the density matrix (and its positivity at second order). In general, the Dyson expansion maintains positivity to all orders; however, one must take care that all relevant terms are included in the sum. It is also important to note that for any quantum state where the one-point function is zero, $\hat\rho^{(1)}_{\textsc{ab}}(t)=\hat\rho_0;$ this is the case for vacuum field states, Fock states, and free thermal states. This justifies our use of the second-order density matrix, as it is in fact the leading-order density matrix for such initial field states.

Finally, let us select the following basis for the detector state:
\begin{equation}
    \begin{matrix}
        |g_{\textsc{a}} g_{\textsc{b}}\rangle=(1,0,0,0)^\dagger & |e_{\textsc{a}} g_{\textsc{b}}\rangle=(0,1,0,0)^\dagger\\
        |g_{\textsc{a}} e_{\textsc{b}}\rangle=(0,0,1,0)^\dagger & |e_{\textsc{a}} e_{\textsc{b}}\rangle=(0,0,0,1)^\dagger.
    \end{matrix}
\end{equation}
It can then be shown (e.g. \cite{Sachs:2017exo}) that the terms of the density matrix can be written as an integral transform of the Wightman function.
Specifically, quantizing with respect to the Killing time $t$, we find \cite{Pozas-Kerstjens:2015gta}
\begin{equation}\label{rho2}
\hat{\rho}_{\textsc{ab}}=
\begin{bmatrix}
    1-\mathcal{L}_{\textsc{aa}}-\mathcal{L}_{\textsc{bb}} & 0 & 0 & \mathcal{M}^* \\
    0 & \mathcal{L}_{\textsc{aa}} & \mathcal{L}_{\textsc{ab}} & 0 \\
    0 & \mathcal{L}_{\textsc{ba}} & \mathcal{L}_{\textsc{bb}} & 0 \\
    \mathcal{M} & 0 & 0 & 0
\end{bmatrix}
+O(\lambda_I^4),
\end{equation}
where
\begin{align}
\mathcal{M}=&-\lambda_{\textsc{a}} \lambda_{\textsc{b}} \int_{-\infty}^{\infty} \dd t \int_{-\infty}^{t}\dd t' \int \dd^n\!\bm{x}\int \dd^n\!\bm{x'}\nonumber\\ &\times \sqrt{g(t,\bm x)g(t',\bm x')} \mathcal{M}(t,\bm{x},t',\bm{x'})W(t,\bm{x},t',\bm{x'})
\label{Mintegral}\\
\mathcal{L}_{IJ}=&\lambda_I \lambda_J \int_{-\infty}^{\infty} \dd t \int_{-\infty}^{\infty}\dd t' \int \dd^n\!\bm{x}\int \dd^n\!\bm{x'}\nonumber\\
&\times  \sqrt{g(t,\bm x)g(t',\bm x')} \mathcal{L}_I(t,\bm{x})\mathcal{L}_J^*(t',\bm{x'})W(t,\bm{x},t',\bm{x'})\label{LLintegral}
\end{align}
and 
\begin{align}
&\mathcal{L}_I(t,\bm{x})=\chi_I(\tau_I(t,\bm x))F_I(\tau_I,\bm{\xi}_I)e^{\ii\Omega_I \tau_I(t,\bm x)} 
 \label{Lintegrand}
\\
&\mathcal{M}(t,\bm{x},t',\bm{x'})=\mathcal{L}_{\textsc{a}}(t,\bm{x})\mathcal{L}_{\textsc{b}}(t',\bm{x'})+\mathcal{L}_{\textsc{a}}(t',\bm{x'})\mathcal{L}_{\textsc{b}}(t,\bm{x})\label{Mintegrand}
\end{align}
where $I = \text{A},\text{B}$. Note that $\mathcal L_{\textsc{aa}}$ is simply the leading-order transition rate for detector A.

It is also possible to determine the causal influence of detector $J$ on $I$. Since we plan to use noncompact switching functions, this allows us to quantify the degree to which the detectors are in causal contact. As shown in detail in \cite{Martin-Martinez:2015psa}, the influence of detector $J$ on detector $I$'s density matrix  is quantified by a integral over the field commutator, $\ii \mathcal C^-(t,x,t',x')\coloneqq[\hat\Phi(t,x),\hat\Phi(t',x')]$ (using the notation of \cite{Ng:2018ilp})  rather than the Wightman function. The magnitude of the result depends on the initial state of $I$ and $J$. An upper bound on this causal influence estimator (that we notate $\mathcal{C}_{IJ}$) \cite{Martin-Martinez:2015psa} is given for a detectors initial states with maximal coherences and is
\begin{align}
    &\mathcal{C}_{IJ}=- \lambda_I\lambda_J \int_{-\infty}^{\infty} \dd t \int_{-\infty}^{t}\dd t' \int \dd^n\bm{x} \int \dd^n\bm{x'} \nonumber\\ 
    &\quad\times \sqrt{g(t,\bm{x})g(t',\bm{x'})} \mathcal{L}_I(t,\bm{x})\Re[\mathcal{L}_J(t',\bm{x'})]\ii \mathcal C^-(t,x,t',x')
\label{caus-est}    
\end{align}

We know that the linearly coupled UDW detector does not diverge if smoothly switched, even if it is pointlike, and so we will assume the detector smearing function
$F_I(\tau_I,\bm{\xi}_I)=\delta(\bm{\xi}_I)=\delta(\bm{x_I})/\sqrt{-h_I(t,\bm{x}_I)}$.
where $h_I(t,\bm{x})$ is the determinant of the spatial three-metric \textit{of the detector's proper frame}. Note that for stationary spacetime trajectories, $\frac{\dd \tau}{\dd t} = \sqrt{{g(t,\bm{x})}/{h(t,\bm{x})}}$.

Since we will eventually consider detectors located at different redshifts, we will use the convention that $t$ is the coordinate time, $\Omega_I$ is the proper gap of the detector, $\tau_I$ is the proper time of the detector, and $\tilde{\Omega}_I$ is the `coordinate frequency' of the detector, i.e. the frequency such that $\tilde{\Omega}_I \dd t = \Omega_I \dd\tau_I.$ where $I=$A,B.  We will also use the convention that $\tilde\chi_I(t)=\chi_I(\tau_I(t))\dd\tau_I/\dd t$ is the `coordinate switching function' of detector $I$, guaranteeing that its integral over time will be constant. Notably, in  our specific coordinates for  AdS$_4$, the coordinates are dimensionless, while the proper measurements  (e.g. proper distance, proper time) are not.

Now consider the special case where the Wightman function $W(t,x,t',x')=\langle 0 | \Phi(t,x) \Phi(t',x') | 0 \rangle$ where $\ket0$ describes a static vacuum state. As noted previously, the presence of the Killing time $t$ allows us to quantize the field with respect to particle modes, and implies that a well-defined vacuum state exists. We can then express the field operator and the vacuum Wightman function with respect to the annihilator $a_{n l m}$ of a particle of indices $n,l,m$:
\begin{align}
&\hat\Phi(t,x)=\sum_{nlm}\left[\hat{a}_{n l m}(t,x)\Phi_{n l m}(t,x)\right.\nonumber\\
&\qquad\qquad\qquad\left.+\hat{a}_{n l m}^\dagger(t,x)\bar{\Phi}_{n l m}(t,x)\right]\\
&W(t,x,t',x')=\sum_{nlm}\frac{1}{2\omega_{nlm}}e^{-\ii\omega_{nlm} (t-t')}\varphi_{nlm}(x)\overline{\varphi}_{nlm}(x')
\end{align}
where our expression for the Wightman function depends on time only through $t-t'$, as expected for a stationary state.
Note that we include a factor $1/\sqrt{2\omega}$ which compensates for the difference between the Klein-Gordon inner product and the usual $L^2$ inner product (i.e., for our mode normalization choice $\int \dd x |\varphi(x)|^2=1$), and include indices $n,l,m$ in anticipation of the separation of variables used later. As well, we use a discrete sum over frequencies, anticipating that our boundary conditions will generate a discrete spectrum; other spacetimes will require an integral over frequencies instead.

Consider a pair of detectors in stationary trajectories. Using the mode expansion, it is fairly simple to rewrite the non-entangling terms of the density matrix:
\begin{align}
\mathcal{L}_{IJ}=&\lambda_J\lambda_I \int_{-\infty}^{\infty} \dd t \int_{-\infty}^{\infty}\dd t'\tilde\chi_{J}(t)\tilde\chi_{I}(t')\sum_{\omega lm} \frac{1}{2\omega}\nonumber\\
&\times e^{-\ii(\omega+\tilde{\Omega}_J)t}e^{i(\omega+\tilde{\Omega}_I)t'}\varphi_{n  lm}(x_J(t))\varphi^*_{\omega lm}(x_I(t'))\nonumber\\
=&\lambda_J\lambda_I \sum_{\omega lm}\frac{\pi}{\omega}\nonumber\\
&\times\mathcal{F}_{\tau_J}[\chi_{J}(\tau_J(t))\varphi_{n  lm}(x_J(t))]\left(-\omega\frac{\dd t}{\dd\tau_J}-\Omega_J\right)\nonumber\\
&\times \mathcal{F}_{\tau_I}[\chi_{I}(\tau_I(t'))\varphi^*_{\omega lm}(x_I(t'))]\left(\omega\frac{\dd t}{\dd\tau_I}+\Omega_I\right),
\end{align}
where $\mathcal{F}_{\tau_I}$ is the Fourier transform with respect to the corresponding \textit{proper} time variable. For instance, in the special case where both detectors are  static, this is just
\begin{align}
\mathcal{L}_{IJ}=&\lambda_J\lambda_I \sum_{\omega lm}\frac{\pi}{\omega}
\varphi_{n  lm}(x_J)\varphi^*_{\omega lm}(x_I)\nonumber\\
&\times\hat{\chi}^*_{J}\left(\omega\frac{\dd t}{\dd\tau_J}+\Omega_J\right)\hat{\chi}_{I}\left(\omega\frac{\dd t}{\dd\tau_I}+\Omega_I\right).
\label{Lnm}
\end{align}
Note that if $J=I$ this becomes the expression found in \cite{Ng:2016hzn} for the transition rate, albeit summing over discrete modes rather than continuous modes.

 We note here that for Anti-de Sitter spacetime in particular, an analytical expression for the Wightman function is known to exist, for any coupling \cite{AIS}. For instance, for the conformally coupled scalar, in the special case where the detectors are at equal angular coordinates one can show \cite{Deser:1997ri} that 
 \begin{align}\label{WightDeser}
8\pi^2 L^2 W(x,x')&=X[1+\varepsilon(1+2X)^{-1}],
\end{align}
where $X$ is twice the inverse square of the geodesic distance in the embedding space, and $\varepsilon$ defines the conformal boundary condition:
\begin{equation}
    \varepsilon=
    \begin{cases}
    -1, & \text{Dirichlet} \\
    1, & \text{Neumann} \\
    0, & \text{transparent}
    \end{cases}
\end{equation}
However, we are also interested in studying less symmetric spacetimes, for which no analytical expression exists. AdS is therefore a test of our methods, for use in more general spacetimes.  Our methods are complementary to
a concurrent study of entanglement harvesting in AdS$_3$ \cite{HHMSZ}, in which a sum over images was used
to evaluate the Wightman function.

\section{A simple expression for $\mathcal{M}$}

The nonlocal term $\mathcal{M}$ provides a more formidable challenge. Because of the asymmetry between $t$ and $t'$, it does not convert into a simple double Fourier transform. It may be written as
\begin{align}
\mathcal{M}=&-\lambda_{\textsc{b}}\lambda_{\textsc{a}}\int_{-\infty}^\infty \dd t \int_{-\infty}^t \dd t' \nonumber\\
&\times(e^{\ii(\tilde{\Omega}_{\textsc{b}} t+\tilde{\Omega}_{\textsc{a}} t')}\tilde\chi_{\textsc{b}}(t)\tilde\chi_{\textsc{a}}(t')W(t,x_{\textsc{b}};t',x_{\textsc{a}})\nonumber\\
&+e^{\ii(\tilde{\Omega}_{\textsc{a}} t+\tilde{\Omega}_{\textsc{b}} t')}\tilde\chi_{\textsc{a}}(t)\chi_{\textsc{b}}(t')W(t,x_{\textsc{a}};t',x_{\textsc{b}})).\nonumber\\
\end{align}
This expression has an intriguing amount of symmetry that can be exploited to simplify its calculation drastically as has recently been shown \cite{Ng:2018ilp}. 
In particular, while the inner integration limit $-\infty<t'<t$ frustrates comparison to a Fourier transform, there are special cases where a Fourier transform is still possible.

In order to utilize the full generality of the following theorem, we will re-express the phase term and switching function with respect to the proper time and gap, as in $\Omega_I \tau_I(t).$ This allows for cases where the relation between proper and coordinate time is not stationary: for instance, the accelerating detector. In these terms,
\begin{align}
\mathcal{M}=&-\lambda_{\textsc{b}}\lambda_{\textsc{a}}\int_{-\infty}^\infty \dd t \int_{-\infty}^t \dd t' \nonumber\\
&\times(e^{\ii(\Omega_{\textsc{b}} \tau_{\textsc{b}}(t)+\Omega_{\textsc{a}} \tau_{\textsc{a}}(t'))}\tilde\chi_{\textsc{b}}(t)\tilde\chi_{\textsc{a}}(t')W(t,x_{\textsc{b}};t',x_{\textsc{a}})\nonumber\\
&+e^{\ii(\Omega_{\textsc{a}} \tau_{\textsc{a}} (t)+\Omega_{\textsc{b}} \tau_{\textsc{b}} (t')}\tilde\chi_{\textsc{a}}(t)\tilde\chi_{\textsc{b}}(t')W(t,x_{\textsc{a}};t',x_{\textsc{b}})).\nonumber\\
\label{intM}
\end{align}

Let us use the commutator and anti-commutator of the field,
\begin{align}
    \ii\mathcal C^-(t,x;t',x')&\coloneqq[\hat\Phi(t,x),\hat\Phi(t',x')]\\
    \mathcal C^+(t,x;t',x')&\coloneqq\{\hat\Phi(t,x),\hat\Phi(t',x')\}
\end{align}
to clarify things. The commutator is clearly both conjugate-symmetric and anti-symmetric under exchange of $(t,x)$ with $(t',x')$, while the anti-commutator is symmetric. Now,
\begin{equation}
2W(t,x;t',x')=\mathcal C^+(t,x;t',x')+\ii\mathcal C^-(t,x;t',x').\label{WDelta}
\end{equation}

With respect to these functions, $\mathcal{M}$ becomes \cite{Ng:2018ilp}
\begin{align}
\mathcal{M}=&\mathcal{M}^++\mathcal{M}^- \\
\mathcal{M}^+=&-\frac{1}{2}\lambda_{\textsc{b}}\lambda_{\textsc{a}}\int_{-\infty}^\infty \dd t \int_{-\infty}^\infty \dd t'\nonumber\\
&\times( e^{\ii(\Omega_{\textsc{b}}\tau_{\textsc{b}}(t)+\Omega_{\textsc{a}}\tau_{\textsc{a}}(t'))}\tilde\chi_{\textsc{b}}(t)\tilde\chi_{\textsc{a}}(t')W(t,x_{\textsc{b}};t',x_{\textsc{a}})\nonumber\\
&+e^{\ii(\Omega_{\textsc{a}} \tau_{\textsc{a}} (t)+\Omega_{\textsc{b}} \tau_{\textsc{b}}(t')}\tilde\chi_{\textsc{a}}(t)\tilde\chi_{\textsc{b}}(t')W(t,x_{\textsc{a}};t',x_{\textsc{b}}))\label{intW}\\
\mathcal{M}^-=&-\frac{1}{2}\lambda_{\textsc{b}}\lambda_{\textsc{a}}\int_{-\infty}^\infty \dd t \int_{-\infty}^t \dd t' \nonumber\\
&\times(e^{\ii(\Omega_{\textsc{b}} \tau_{\textsc{b}}(t)+\Omega_{\textsc{a}} \tau_{\textsc{a}}(t'))}\tilde\chi_{\textsc{b}}(t)\tilde\chi_{\textsc{a}}(t')\ii\mathcal C^-(t,x_{\textsc{b}};t',x_{\textsc{a}})\nonumber\\
&+e^{\ii(\Omega_{\textsc{a}} \tau_{\textsc{a}} (t)+\Omega_{\textsc{b}} \tau_{\textsc{b}} (t')}\tilde\chi_{\textsc{a}}(t)\tilde\tau_{\textsc{b}}(t')\ii\mathcal C^-(t,x_{\textsc{a}};t',x_{\textsc{b}}))\label{intDelta}
\end{align}
where (by exploiting the symmetry of the real part of $W$ under exchange  \cite{Ng:2018ilp}) 
 $\mathcal{M}^+$ is unchanged if we replace $W$ by $\mathcal{C}^+$ (hence the notation).
In AdS$_4$, these expressions for the integral are particularly simple. Since in these spacetimes  Huygen's principle holds for the conformally coupled massless field, the commutator has support on the light cone: therefore, $\mathcal C^-(t,x;t',x')$ will reduce to a combination of delta functions. It is also interesting to note that $\mathcal C^-(t,x;t',x')$ is independent of the state of the field: it depends only on the background metric. In particular, the causality of QFT implies that $\mathcal{M}^-$ vanishes if the detectors are out of causal contact, and therefore $\mathcal{M}=\mathcal{M}^+$. This is, of course, true for spacelike separated detectors. Moreover, in any spacetime satisfying the Huygens principle, \textit{timelike} separated detectors will also satisfy this condition.

Let us now consider $\mathcal{M}^+$, which does not vanish when the detectors are causally disconnected. As shown in \cite{Ng:2018ilp}, it can be re-expressed with respect to $\mathcal L_{\textsc{ab}}:$
\begin{equation}
\mathcal{M}^+=-\frac{1}{2}\left(\mathcal{L}_{\textsc{ab}}(\Omega_{\textsc{a}},-\Omega_{\textsc{b}})+\mathcal{L}_{\textsc{ba}}(\Omega_{\textsc{b}},-\Omega_{\textsc{a}})\right)
\label{Lbaab}
\end{equation}
where $\mathcal{L}_{IJ}(\Omega_I,\Omega_J)$ denotes the local correlation term $\mathcal{L}_{IJ}$ where detectors $I$ and $J$ have proper gaps $\Omega_I$ and $\Omega_J$ respectively. In other words, if the detectors are not causally connected during their switching, then the entangling term can be expressed with respect to the term associated with classical correlations. The fact that one of the detectors has a `negative gap' in the $\mathcal{L}_{IJ}$ term also helps explain why the ideal strategy is to have both $\Omega_I$ positive: in that case, the local noise $\mathcal{L}_{II}(\Omega,\Omega)$ is suppressed twice by the gap while $\mathcal{L}_{IJ}(\Omega,-\Omega)$ is only suppressed once.

This means that when the two detectors are out of causal contact, in the case where $\varphi_{n  l m}(x(t))$ and $\dd\tau_I/\dd t$ are independent of time, we can write an explicit compact expression for $\mathcal{M}$ as
\begin{align}
&\mathcal{M} =-\frac{1}{2}\lambda_{\textsc{b}}\lambda_{\textsc{a}} \sum_{\omega l m}\frac{\pi}{\omega}\nonumber\\
&\times \left[\hat{\chi}_{\textsc{b}}\left(\Omega_{\textsc{b}}-\frac{\dd t}{\dd\tau_{\textsc{b}}}\omega\right)\hat{\chi}_{\textsc{a}}\left(\Omega_{\textsc{a}}+\frac{\dd t}{\dd\tau_{\textsc{a}}}\omega\right)\varphi_{n  l m}(x_{\textsc{b}})\varphi^*_{\omega l m}(x_{\textsc{a}})\right.\nonumber\\
&+\left.\hat{\chi}_{\textsc{a}}\left(\Omega_{\textsc{a}}-\frac{\dd t}{\dd\tau_{\textsc{a}}}\omega\right)\hat{\chi}_{\textsc{b}}\left(\Omega_{\textsc{b}}+\frac{\dd t}{\dd\tau_{\textsc{b}}}\omega\right)\varphi_{n  l m}(x_{\textsc{a}})\varphi^*_{\omega l m}(x_{\textsc{b}})\right]\label{symmetricM}
\end{align}

In summary, if the two detectors are out of causal contact, and have a Killing-like orbit, we find that the entangling term can be quickly calculated in a similar fashion to $\mathcal{L}_{IJ}$. This can occur when the detectors are spacelike separated, or even timelike separated if the spacetime is Huygens. Notably, it does not appear to be necessary that the orbits of the detectors are related; as long as the switching remains causally disconnected, we can consider detectors that move along drastically different trajectories, such as orbits on different planes, or even different proper accelerations. The only requirement, from a computational standpoint, is that the time dependence of  $\varphi_{n  l m}(x(t))$ remains simple enough to convolve with $\hat{\chi}(\omega)$.

Amusingly, there is another way to express \eqref{intM} that may be of use. As shown in \cite{Ng:2018ilp},
\begin{align}
\mathcal{M}=&-\frac{1}{2}\lambda_{\textsc{b}}\lambda_{\textsc{a}}\int_{-\infty}^\infty \dd t \int_{-\infty}^\infty \dd t'\nonumber\\
&\times(e^{\ii(\Omega_{\textsc{b}}\tau_{\textsc{b}}(t)+\Omega_{\textsc{a}}\tau_{\textsc{a}}(t'))}\tilde\chi_{\textsc{b}}(t)\tilde\chi_{\textsc{a}}(t')\ii G_F(t,x_{\textsc{b}};t',x_{\textsc{a}})\nonumber\\
&+e^{\ii(\Omega_{\textsc{a}} \tau_{\textsc{a}} (t)+\Omega_{\textsc{b}} \tau_{\textsc{b}}(t')}\tilde\chi_{\textsc{a}}(t)\tilde\chi_{\textsc{b}}(t')\ii G_F(t,x_{\textsc{a}};t',x_{\textsc{b}})),
\end{align}
where $G_F$ is the usual Feynman Green's function. While it suffers from certain singularities on the light cone, there are cases where analytic solutions exist, in which case this expression may become useful. We emphasize that this expression in $\ii G_F$ is valid even if the detectors are timelike connected. As before, this result is completely independent of the motions of the detectors, and even the state of the field.

\section{Global AdS modes}

Since we are interested in detector responses over many AdS periods, we will use global coordinates and modes. We will use the convention found in Avis, Isham, and Storey \cite{AIS}, namely
\begin{equation}
\dd s^2= L^2 \sec^2 \varrho \left(\dd t^2-d\varrho^2-\sin^2 \varrho (\dd\theta^2 + \sin^2 \theta\, \dd\phi^2)\right)
\end{equation}
 where $L=\sqrt{-3/\Lambda}$ is the AdS length.
Note that this is \textit{not} the conformally flat coordinate system. The conformal infinity of AdS exists at $\varrho=\pi/2$, the speed of light in this system is simply $d\varrho/\dd t=1$, and therefore a lightlike signal emitted from the origin reaches   conformal infinity after a coordinate time $\Delta t=\pi/2,$ regardless of $L$. 

We will  consider three boundary conditions on the timelike boundary of AdS, enumerated according to the value $\varepsilon$, as in \cite{Deser:1997ri}: $\varepsilon=-1$ is Dirichlet, $\varepsilon=1$ is Neumann, and $\varepsilon=0$ is ``transparent''  (i.e. considering AdS embedded within an Einstein static universe \cite{AIS});
we note that in \cite{Deser:1997ri} the latter two boundary conditions were erroneously transposed.
We emphasize that Dirichlet best corresponds to the general case of a massive or non-conformally coupled scalar, which would be trapped away from the boundary by a gravitational potential. 

Therefore, the vacuum is described by the following modes:
\begin{align}
\varphi_{n l m}(x)=&\sqrt{2}^{\varepsilon^2}N_{\omega l}\cos \varrho (\sin \varrho)^l C_{\omega-l-1}^{(l+1)}(\cos \varrho)Y_l^m(\theta,\phi),\\
N_{\omega l}=& \frac{2^l l!}{L}\sqrt{\frac{2\omega(\omega-l-1)!}{\pi (\omega+l)!}},
\end{align}
where $C_{a}^b$ are the Gegenbauer polynomials, and
\begin{equation}
    \omega=\begin{cases}
    l+n+1 &\mbox{if } \varepsilon=0\\
    l+2n+1 &\mbox{if } \varepsilon=1\\
    l+2n+2 &\mbox{if } \varepsilon=-1
    \end{cases}.
\end{equation}
For all boundary conditions, any lightlike signal originating from the center will return after a time $\Delta t=2\pi,$ and sooner if $\varepsilon \neq 0.$ Therefore, any observer can interact with the boundary condition within finite time. Note also that while the AdS space is still Cauchy for the transparent case, the `Cauchy surface' is actually two surfaces, at times $t=0,\pi/2;$ this is because signals from $t=0$ `hide' on the other side of the embedding ESU for half the period of AdS. This is why the normalization for that case is different.
%\tcb{\bf[N.B. For various reasons, it is easier to leave the L factor out of Nwl, and simply place it into the final expression for Lij and M+ in numerics.]}

It is fairly easy to show that in AdS, there exist many geodesics at constant radius. Denoting proper time of the detector as $\tau$, one family of geodesics is described by $t=\tau,\phi=\tau.$ Note that the redshift and angular velocity are completely independent of radius: this is a peculiar property of AdS, and does not hold for most spacetimes (e.g. Schwarzschild-AdS). To highlight this, we note that the \textit{proper} length $\Delta x$ between the center and $\varrho$ are related by
\begin{align}
    \Delta x &= L \log (\tan \varrho + \sec \varrho),    \label{rproperfunc}\\
    \varrho &= \arctan \sinh (\Delta x/L).
    \label{rcoordfunc}
\end{align}
%\tcb{\bf[Not sure if this should be in the text: it turns out that cos rho = sech x/L, while sin rho = tanh x/L.]}

Recall that our expressions for the density matrix contained the key term $F[\chi_I(t)\varphi_{n  l m}(x_I(t))].$ In the case of a circular geodesic detector, the observed frequency depends on axial angular momentum, $m$. Specifically, since $\phi(t)=t$, we find that
\begin{align}
\varphi_{n  l m}(x(t))=&\sqrt{2}^{\varepsilon^2}N_{\omega l}\cos \varrho (\sin \varrho)^l C_{\omega-l-1}^{(l+1)}(\cos \varrho)\nonumber\\
&\times\sqrt{\frac{2l+1}{4\pi}\frac{(l-m)!}{(l+m)!}}P_l(\cos \theta)e^{\ii mt}.
\end{align}
This causes a shift in the Fourier transform mentioned earlier.
  
There is another way to express the Wightman function that preserves the symmetry of AdS \cite{AIS}: namely, one can express the Wightman function in terms of geodesic distances in the embedding space. This implies that geodesic observers, e.g. in circular orbit, observe the same $L_{\textsc{aa}}$ as static observers in the center, despite radically different expressions (see \cite{Kent:2014wda} for further discussion).
 Evaluation of  an orbiting geodesic detector's response can therefore largely be done using  symmetry, whereas
the response of static detectors is more computationally intensive.

 \section{Entangled geodesic detectors}

The symmetries of AdS make it simplest to study the case of two geodesic detectors, where one detector is at the center, say detector A. We will then place detector B at a proper separation $\Delta x$, which can be converted to a coordinate separation via \eqref{rcoordfunc}. Since both detectors are the same on the $L_{\textsc{aa}}$ level (i.e. up to the difference in switching function), we will focus instead on $L_{\textsc{ba}}$ and $M$.

The location of detector A at the center has a particular consequence: only $l=0$ modes are non-zero at the center, and so we will only need to sum over one index, $n$. This also means that we do not need to concern ourselves over $m$ and the revolution of detector B. Furthermore, the fact that $\varphi_{n  00}$ is real will simplify things. For brevity, we will simply write $\varphi_{n}$ and $\omega_n$ where appropriate.
\begin{equation}
\mathcal{L}_{II}=\lambda_I^2\sum_n \frac{\pi}{\omega_n}\varphi^2_{n}(x_{\textsc{a}}) |\hat{\chi}_I((\omega_n)/L + \Omega_I)|^2
\end{equation}
As noted, the symmetries of AdS allow us to place detector B at the center for $L_{II}$ (i.e. the change of coordinates, and the corresponding vacua, are compatible).

Since the $l=0$ mode is spherically symmetric, the value of the mode at B happens to be time-independent.
Therefore, we can write
\begin{align}
\mathcal{L}_{\textsc{ab}}=&\lambda_{\textsc{b}} \lambda_{\textsc{a}} \sum_n \frac{\pi}{\omega_n}\varphi_{n }(x_{\textsc{b}})\varphi_{n }(x_{\textsc{a}})\nonumber\\
&\times\hat\chi_{\textsc{a}}(\omega_n/L+\Omega_{\textsc{a}})\hat\chi_{\textsc{b}}^*(\omega_n/L+\Omega_{\textsc{b}}).
\end{align}

As for $\mathcal M$, recall that we can instead use the simplified expression found in \eqref{symmetricM} to conclude that
\begin{align}
\mathcal{M}^+=&-\lambda_{\textsc{b}}\lambda_{\textsc{a}} \sum_{n=0}^\infty\frac{\pi}{2\omega_n}\varphi_{n }(x_{\textsc{b}})\varphi_{n }(x_{\textsc{a}})\nonumber\\
&\times (\hat{\chi}_{\textsc{b}}(\Omega_{\textsc{b}}-\omega_n/L)\hat{\chi}_{\textsc{a}}(\Omega_{\textsc{a}}+\omega_n/L)\nonumber\\
&+\hat{\chi}_{\textsc{a}}(\Omega_{\textsc{a}}-\omega_n/L)\hat{\chi}_{\textsc{b}}(\Omega_{\textsc{b}}+\omega_n/L))
\end{align}
As noted previously, this particular expression happens to be very similar to $\mathcal{L}_{\textsc{ba}}.$  More precisely, expressing all matrix elements as functions of $\Omega_{\textsc{a}},\Omega_{\textsc{b}},$ it is easy to see that in general
\begin{equation}
   \mathcal{M}^+(\Omega_{\textsc{a}},\Omega_{\textsc{b}})=-\frac{1}{2}\mathcal{L}_{\textsc{ab}}(\Omega_{\textsc{a}},-\Omega_{\textsc{b}})-\frac{1}{2}\mathcal{L}_{\textsc{ba}}(\Omega_{\textsc{b}},-\Omega_{\textsc{a}}).
\end{equation}

In the particular case of geodesic detectors in AdS$_4$, an analytic expression for the commutator also exists, provided one detector is located at the center; the result is  \cite{AIS} 
\begin{align}
\ii\mathcal C^-(x,0)=&-\frac{\ii}{4\pi L^2}\varsigma^0(t)[\delta(\sigma^0(x))-(-1)^j\delta(\sigma^0(x)-2)]
\end{align}
where $\sigma^0=1-\cos t\sec\varrho$ is half of the geodesic distance in the embedding space, the Dirichlet boundary condition corresponds to $j=2$, $\varsigma^0(t)=\sgn(\sin(t))$, and the symmetries of AdS$_4$ imply that the commutator depends only on\footnote{Note that our sign convention differs from that of \cite{AIS}.}  $K\sigma^0$ and $t-t'$.  In other words, the Dirichlet commutator has delta-like support on the light cone, changes sign upon reflection, and is $2\pi$-periodic. This simple form allows us to handle the more general case where the detectors are causally connected.  Noting that $\varrho$ is constant, we may write $\dd\sigma^0/\dd t=\sin t\sec\varrho;$ on the light cone, $t=\varrho$, so $\dd\sigma^0/\dd t=\tan\varrho.$ Therefore
\begin{align}
\ii\mathcal C^-(x,t,0,t')=&\sum_{N=-\infty}^\infty-\frac{\ii}{4\pi L^2}(\tan\varrho)^{-1}\nonumber\\
&[\delta(\Delta t-\varrho-2N\pi )\nonumber\\
&+\varepsilon\delta(\Delta t+\varrho-(2N+1)\pi)\nonumber\\
&-\varepsilon\delta(\Delta t-\varrho-(2N+1)\pi)\nonumber\\
&-\delta(\Delta t+\varrho-(2N+2)\pi)]\label{iDeltax}\\
\ii\mathcal C^-(0,t,x',t')=&\sum_{N=-\infty}^\infty-\frac{\ii}{4\pi L^2}(\tan\varrho')^{-1}\nonumber\\
&[\delta(\Delta t-\varrho'-2N\pi )\nonumber\\
&+\varepsilon\delta(\Delta t+\varrho'-(2N+1)\pi)\nonumber\\
&-\varepsilon\delta(\Delta t-\varrho'-(2N+1)\pi)\nonumber\\
&-\delta(\Delta t+\varrho'-(2N+2)\pi)]\label{iDeltaxp}
\end{align}
where $\Delta t=t-t'.$ Note that, as written, $t>t'$ corresponds to $N \geq 0.$

Finally, let us substitute \eqref{iDeltax}, \eqref{iDeltaxp} into the second half of \eqref{intDelta}. Under the assumption that the detector gaps are equal, we find that the commutator part of $\mathcal{M}$ is equal to
\begin{align}
\mathcal{M}^-&=-\frac{1}{2}\lambda_{\textsc{b}}\lambda_{\textsc{a}}\int_{-\infty}^\infty \dd t \int_{-\infty}^t \dd t' e^{\ii\tilde{\Omega}(t+t')}\nonumber\\
&\times(\tilde\chi_{\textsc{b}}(t)\tilde\chi_{\textsc{a}}(t')\ii\mathcal C^-(t,x_{\textsc{b}};t',x_{\textsc{a}})\nonumber\\
&+\tilde\chi_{\textsc{a}}(t)\tilde\chi_{\textsc{b}}(t')\ii\mathcal C^-(t,x_{\textsc{a}};t',x_{\textsc{b}}))\nonumber\\
=&\frac{\ii\lambda_{\textsc{a}}\lambda_{\textsc{b}}}{8\pi\tan\varrho}\int_{-\infty}^\infty \dd t \,e^{2\ii\tilde{\Omega} t}\sum_{N=0}^\infty\nonumber\\
& (e^{-\ii\tilde{\Omega} (\varrho+2N\pi)}\tilde\chi_{\textsc{b}}(t)\tilde\chi_{\textsc{a}}(t-\varrho-2N\pi)\nonumber\\
&+e^{-\ii\tilde{\Omega} (\varrho+2N\pi)}\tilde\chi_{\textsc{a}}(t)\tilde\chi_{\textsc{b}}(t-\varrho-2N\pi)\nonumber\\
&+\varepsilon e^{-\ii\tilde{\Omega} (-\varrho+(2N+1)\pi)}\tilde\chi_{\textsc{b}}(t)\tilde\chi_{\textsc{a}}(t+\varrho-(2N+1)\pi)\nonumber\\
&+\varepsilon e^{-\ii\tilde{\Omega} (-\varrho+(2N+1)\pi)}\tilde\chi_{\textsc{a}}(t)\tilde\chi_{\textsc{b}}(t+\varrho-(2N+1)\pi)\nonumber\\&-\varepsilon e^{-\ii\tilde{\Omega} (\varrho+(2N+1)\pi)}\tilde\chi_{\textsc{b}}(t)\tilde\chi_{\textsc{a}}(t-\varrho-(2N+1)\pi)\nonumber\\
&-\varepsilon e^{-\ii\tilde{\Omega} (\varrho+(2N+1)\pi)}\tilde\chi_{\textsc{a}}(t)\tilde\chi_{\textsc{b}}(t-\varrho-(2N+1)\pi)\nonumber\\
&-e^{-\ii\tilde{\Omega} (-\varrho+(2N+2)\pi)}\tilde\chi_{\textsc{b}}(t)\tilde\chi_{\textsc{a}}(t+\varrho-(2N+2)\pi)\nonumber\\
&-e^{-\ii\tilde{\Omega} (-\varrho+(2N+2)\pi)}\tilde\chi_{\textsc{a}}(t)\tilde\chi_{\textsc{b}}(t+\varrho-(2N+2)\pi))\nonumber\displaybreak[0]\\
=&\frac{\ii\lambda_{\textsc{a}}\lambda_{\textsc{b}}}{8\pi L^2\tan\varrho}\int_{-\infty}^\infty \dd t \,e^{2\ii\tilde{\Omega} t}\sum_{N=0}^\infty\nonumber\\
&\left(\tilde\chi_{\textsc{b}}\left(t+\frac{\varrho+2N\pi}{2}\right)\tilde\chi_{\textsc{a}}\left(t-\frac{\varrho+2N\pi}{2}\right)\right.\nonumber\\
&+\tilde\chi_{\textsc{a}}\left(t+\frac{\varrho+2N\pi}{2}\right)\tilde\chi_{\textsc{b}}\left(t-\frac{\varrho+2N\pi}{2}\right)\nonumber\\
&+\varepsilon\tilde\chi_{\textsc{b}}\left(t-\frac{\varrho-(2N+1)\pi}{2}\right)\tilde\chi_{\textsc{a}}\left(t+\frac{\varrho-(2N+1)\pi}{2}\right)\nonumber\\
&+\varepsilon\tilde\chi_{\textsc{a}}\left(t-\frac{\varrho-(2N+1)\pi}{2}\right)\tilde\chi_{\textsc{b}}\left(t+\frac{\varrho-(2N+1)\pi}{2}\right)\nonumber\\
&-\varepsilon\tilde\chi_{\textsc{b}}\left(t+\frac{\varrho+(2N+1)\pi}{2}\right)\tilde\chi_{\textsc{a}}\left(t-\frac{\varrho+(2N+1)\pi}{2}\right)\nonumber\\
&-\varepsilon\tilde\chi_{\textsc{a}}\left(t+\frac{\varrho+(2N+1)\pi}{2}\right)\tilde\chi_{\textsc{b}}\left(t-\frac{\varrho+(2N+1)\pi}{2}\right)\nonumber\\
&-\tilde\chi_{\textsc{b}}\left(t-\frac{\varrho-(2N+2)\pi}{2}\right)\tilde\chi_{\textsc{a}}\left(t+\frac{\varrho-(2N+2)\pi}{2}\right)\nonumber\\
&\left.-\tilde\chi_{\textsc{a}}\left(t-\frac{\varrho-(2N+2)\pi}{2}\right)\tilde\chi_{\textsc{b}}\left(t+\frac{\varrho-(2N+2)\pi}{2}\right)\right)\nonumber\displaybreak[0]\\
=&\frac{\ii\lambda_{\textsc{a}}\lambda_{\textsc{b}}}{8\pi L^2\tan\varrho}\int_{-\infty}^\infty \dd t \,e^{2\ii\tilde{\Omega} t}\sum_{N=-\infty}^\infty\varsigma(N+1/2)\nonumber\\
&\left(\tilde\chi_{\textsc{b}}\left(t+\frac{\varrho+2N\pi}{2}\right)\tilde\chi_{\textsc{a}}\left(t-\frac{\varrho+2N\pi}{2}\right)\right.\nonumber\\
&+\tilde\chi_{\textsc{a}}\left(t+\frac{\varrho+2N\pi}{2}\right)\tilde\chi_{\textsc{b}}\left(t-\frac{\varrho+2N\pi}{2}\right)\nonumber\\
&-\varepsilon\tilde\chi_{\textsc{b}}\left(t+\frac{\varrho+(2N+1)\pi}{2}\right)\tilde\chi_{\textsc{a}}\left(t-\frac{\varrho+(2N+1)}{2}\right)\nonumber\\
&\left.-\varepsilon\tilde\chi_{\textsc{a}}\left(t+\frac{\varrho+(2N+1)}{2}\right)\tilde\chi_{\textsc{b}}\left(t-\frac{\varrho+(2N+1)}{2}\right)\right)\nonumber\displaybreak[0]\\
=&\frac{\ii\lambda_{\textsc{a}}\lambda_{\textsc{b}}}{8\pi L^2\tan\varrho}\int_{-\infty}^\infty \dd t \,e^{2\ii\tilde{\Omega} t}\sum_{N=-\infty}^\infty\varsigma(N+1/2)(-\varepsilon)^{p(N)}\nonumber\\
&\left(\tilde\chi_{\textsc{b}}\left(t+\frac{\varrho+N\pi}{2}\right)\tilde\chi_{\textsc{a}}\left(t-\frac{\varrho+N\pi}{2}\right)\right.\nonumber\\
&+\left.\tilde\chi_{\textsc{a}}\left(t+\frac{\varrho+N\pi}{2}\right)\tilde\chi_{\textsc{b}}\left(t-\frac{\varrho+N\pi}{2}\right)\right),
\label{Mminusderivation}
\end{align}
where $\varsigma(n)=\sgn(n)$ is the sign function, $\tilde\Omega\,\dd t=\Omega\,\dd\tau$, $p(n)=0$ if $n$ is even and 1 otherwise, and we use the convention that $0^0=1.$
We have thus expressed the light-cone contribution in terms of the Fourier transform of a certain sum of switching functions. This also makes clear what the necessary conditions for convergence of the light-cone contribution are: namely, that the switching function is both smooth enough (because of the Fourier transform) and decays rapidly enough (because of the sum over translations). Note that strictly speaking, the use of the $\varsigma$ function was also unnecessary: we simply used it to reduce the number of terms inside the sum, by extending the sum to negative infinity.

We should note that this summation over infinite $N$ is unique to AdS$_4$:  Huygen's principle does not apply for most AdS$_D$, nor does it apply to most curved spaces (e.g. Schwarzschild); and the infinite echoes produced by the conformal boundary are suppressed if some other feature is present in the interior (e.g. a black hole, as in Schwarzschild-AdS). In other words, this simple expression for the commutator part will not hold for most other spacetimes, and in the few cases where it does (e.g. Minkowski), it will not involve a sum over infinite translations of the switching function.  However, this form does allow us to separate the divergences associated with the light cone from the other features of the quantum state, which may itself be useful.

In fact, we can reason as to what might happen in the more generic case. To begin, in most curved spacetimes (with the notable exception of dS$_4$ \cite{HuygPetar}), and with non-conformal couplings/masses, the commutator is not limited to the lightcone, but also has (non-singular) support inside it \cite{McLenaghan1974,Blanchet1988,Blanchet1992,Faraoni1992,Bombelli1994,Gundlach1994,Faraoni1999}. One says that the strong Huygens principle is violated.  Notably this is remarkably  important for the physics of detectors and allows for the transmission of information along \textit{timelike} paths, \textit{without energy cost} \cite{Jonsson2015,Blasco2015,Blasco:2015foa,huygJonss,HuygJons2,HuygPetar}. 
This is not radically different from our conformal case, especially since the interior contribution decays with time. In fact, something similar also occurs in any dimension greater than (1+1)-D: the commutator generically contains a decaying term inside the lightcone, which will complicate calculations. (1+1)-D space is special because this term no longer decays with time: questions of convergence, among other things, arise (e.g. the (1+1)-D infrared problem). However, in most cases, the singularities within the commutator are limited to the light cone. Therefore, this method may still be a way of preserving the UV-dependent causal characteristics of the scalar field.

%\tcb{\bf[Section on mutual information moved to later, to be with negativity discussion.]}
We include the  expression for the causality estimator \eqref{caus-est}
\begin{align}
&\mathcal{C}_{IJ}=-\lambda_{\textsc{a}} \lambda_{\textsc{b}} \int_{-\infty}^\infty \dd t  \int_{-\infty}^t \dd t' e^{i\tilde\Omega_I t}\nonumber\\
    &\times \Re[e^{i\tilde\Omega_J t'}]\tilde\chi_I(t)\tilde\chi_J(t')\ii\mathcal C^-(t,x_I;t',x_J)\nonumber\\
    =&\frac{\ii\lambda_{\textsc{a}} \lambda_{\textsc{b}}}{4\pi L^2\tan\rho}\int_{-\infty}^{\infty}dt\,e^{i\tilde\Omega_I t}\sum_{N=0}^\infty\nonumber\\
    &\left(\Re[e^{i\tilde\Omega_J(t-\rho-2N\pi)}]\tilde\chi_I(t)\tilde\chi_J(t-\rho-2N\pi)\right.\nonumber\\
    &+\varepsilon\Re[e^{i\tilde\Omega_J(t+\rho-(2N+1)\pi)}]\tilde\chi_I(t)\tilde\chi_J(t+\rho-(2N+1)\pi)\nonumber\\
    &-\varepsilon\Re[e^{i\tilde\Omega_J(t-\rho-(2N+1)\pi)}]\tilde\chi_I(t)\tilde\chi_J(t-\rho-(2N+1)\pi)\nonumber\\
    &\left.\Re[e^{i\tilde\Omega_J(t+\rho-(2N+2)\pi)}]\tilde\chi_I(t)\tilde\chi_J(t+\rho-(2N+2)\pi)\right).
\end{align}
We then can time-translate in order to get a more symmetric expression.
\begin{widetext}
\begin{align}
   \mathcal{C}_{IJ} =&\frac{\ii\lambda_{\textsc{a}} \lambda_{\textsc{b}}}{8\pi L^2\tan\rho}\int_{-\infty}^{\infty}dt\sum_{N=0}^\infty\nonumber\\
    &\left[(e^{2\ii\tilde\Omega t}+e^{\ii\tilde\Omega(\rho+2N\pi)})\tilde\chi_I\left(t+\frac{\rho+2N\pi}{2}\right)\tilde\chi_J\left(t-\frac{\rho+2N\pi}{2}\right)\right. \nonumber\\
    &+\varepsilon(e^{2\ii\tilde\Omega t}+e^{\ii\tilde\Omega(-\rho+(2N+1)\pi)})\tilde\chi_I\left(t-\frac{\rho-(2N+1)\pi}{2}\right)\tilde\chi_J\left(t+\frac{\rho-(2N+1)\pi}{2}\right)\nonumber\\
    &-\varepsilon(e^{2\ii\tilde\Omega t}+e^{\ii\tilde\Omega(\rho+(2N+1)\pi)})\tilde\chi_I\left(t+\frac{\rho+(2N+1)\pi}{2}\right)\tilde\chi_J\left(t-\frac{\rho+(2N+1)\pi}{2}\right)\nonumber\\
    &\left.-(e^{2\ii\tilde\Omega t}+e^{\ii\tilde\Omega(-\rho+(2N+2)\pi)})\tilde\chi_I\left(t-\frac{\rho-(2N+2)\pi}{2}\right)\tilde\chi_J\left(t+\frac{\rho-(2N+2)\pi}{2}\right)\right] \label{c-est2} 
\end{align}
\end{widetext}
where our third equation uses the fact that in the geodesic configuration, $\tilde\Omega_I=\tilde\Omega_J$.

\section{Static detectors}

In order to complete our characterization of AdS$_4,$ we also need to consider detectors that remain static in our global coordinate system. As previous research has shown (e.g. \cite{Jennings:2010vk}),  in the adiabatic switching limit, these detectors do not become spontaneously excited, despite experiencing proper acceleration; on the other hand, this proper acceleration provides reason to believe that  in the finite time regime,  the response of such detectors should change, especially if they are initialized in an excited state. 
Moreover, in the broader picture, AdS$_4$ is especially symmetric, in that all circular geodesics have the same proper time; if we wish to generalize our analysis to other spacetimes, we must consider more generic configurations. The static configuration thus is a good starting point.

To maximize the remaining symmetries, we place detector A at the center as before, and place B on the z-axis (thus forcing $\delta_{m0}$ into our angular momentum sums). While this implies that the detectors remain at constant separation, and thus that the Wightman function is $t$-time-invariant, the detectors no longer agree on proper time. Specifically, the proper times of detectors A and B are now
\begin{align}
    \tau_{\textsc{a}}&=L t,\\
    \tau_{\textsc{b}}&=L t \sec \varrho_{\textsc{b}},
\end{align}
where $\varrho_{\textsc{b}}$ is the spatial coordinate of detector B. We will also investigate dependence on the curvature parameter; we will then use the convention that the proper gaps $\Omega_I$ and proper switching functions $\chi_I(\tau_I)$ remain the same. Consequently the two detectors will no longer be switched on for the same coordinate time. 

Since A is located at the center, $W(t,x_{\textsc{a}}(t);t',x_{\textsc{b}}(t'))$ remains exactly the same. This is because A couples only to the mode $l=0$, so the angular coordinates of B are irrelevant. Still, the different switching functions lead to some loss of symmetry, and $\mathcal{L}_{\textsc{bb}}$ is no longer equal to $\mathcal{L}_{\textsc{aa}}.$ We therefore write the corresponding expression for the static case $\mathcal{L}_{\textsc{bb}}$, in terms of the modes $\varphi_{n,l,m}(x_{\textsc{b}})$.
\begin{align}
    \mathcal{L}_{\textsc{bb}}=&\lambda_{\textsc{b}}^2 \sum_{n,l} \frac{\pi}{\omega_{nl}}\varphi^2_{n ,l,0}(x_{\textsc{b}})\nonumber\\
    &\times|\hat{\chi}_{\textsc{b}}(\omega_{nl}/(L\sec{\varrho_{\textsc{b}}})+\Omega_{\textsc{b}})|^2.
\end{align}
For the two-detector terms, we simply have to remember to apply the Fourier transforms in proper time: that is, we treat the proper switching function $\chi_{\textsc{b}}(\tau)$ as fixed, and apply the Fourier transform as follows:
\begin{equation}
    \hat{\chi}_{\textsc{b}}(\tau)=\mathcal{F}_{\tau_{\textsc{b}}}[\chi_{\textsc{b}}(\tau_{\textsc{b}}(t))].
\end{equation}

There is another difficulty, however. We now obtain
\begin{align}
\mathcal{M}^-=&-\frac{1}{2}\lambda_{\textsc{b}}\lambda_{\textsc{a}}\int_{-\infty}^\infty \dd t \int_{-\infty}^t \dd t'\nonumber\\
&\times(e^{i\tilde{\Omega}_{\textsc{b}} t}\tilde\chi_{\textsc{b}}(t)e^{\ii\tilde{\Omega}_{\textsc{a}} t'}\tilde\chi_{\textsc{a}}(t')\ii\mathcal C^-(t,x_{\textsc{b}};t',x_{\textsc{a}})\nonumber\\
&+(e^{\ii\tilde{\Omega}_{\textsc{a}} t}\tilde\chi_{\textsc{a}}(t)e^{\ii\tilde{\Omega}_{\textsc{b}} t'}\tilde\chi_{\textsc{b}}(t')\ii\mathcal C^-(t,x_{\textsc{a}};t',x_{\textsc{b}}))\nonumber\\
=&\frac{\ii\lambda_{\textsc{b}}\lambda_{\textsc{a}}}{8\pi L^2\tan{\varrho_{\textsc{b}}}}\int_{-\infty}^\infty \dd t \,e^{\ii(\tilde{\Omega}_{\textsc{a}}+\tilde{\Omega}_{\textsc{b}}) t}\sum_{N=0}^\infty\nonumber\\
& (e^{-\ii\tilde{\Omega}_{\textsc{a}} ({\varrho_{\textsc{b}}}+2N\pi)}\tilde\chi_{\textsc{b}}(t)\tilde\chi_{\textsc{a}}(t-{\varrho_{\textsc{b}}}-2N\pi)\nonumber\\
&+e^{-\ii\tilde{\Omega}_{\textsc{b}} ({\varrho_{\textsc{b}}}+2N\pi)}\tilde\chi_{\textsc{a}}(t)\tilde\chi_{\textsc{b}}(t-{\varrho_{\textsc{b}}}-2N\pi)\nonumber\\
&+\varepsilon e^{-\ii\tilde{\Omega}_{\textsc{a}} ({-\varrho_{\textsc{b}}}+(2N+1)\pi)}\tilde\chi_{\textsc{b}}(t)\tilde\chi_{\textsc{a}}(t+{\varrho_{\textsc{b}}}-(2N+1)\pi)\nonumber\\
&+\varepsilon e^{-\ii\tilde{\Omega}_{\textsc{b}} ({-\varrho_{\textsc{b}}}+(2N+1)\pi)}\tilde\chi_{\textsc{a}}(t)\tilde\chi_{\textsc{b}}(t+{\varrho_{\textsc{b}}}-(2N+1)\pi)\nonumber\\
&-\varepsilon e^{-\ii\tilde{\Omega}_{\textsc{a}} ({\varrho_{\textsc{b}}}+(2N+1)\pi)}\tilde\chi_{\textsc{b}}(t)\tilde\chi_{\textsc{a}}(t-{\varrho_{\textsc{b}}}-(2N+1)\pi)\nonumber\\
&-\varepsilon e^{-\ii\tilde{\Omega}_{\textsc{b}} ({\varrho_{\textsc{b}}}+(2N+1)\pi)}\tilde\chi_{\textsc{a}}(t)\tilde\chi_{\textsc{b}}(t-{\varrho_{\textsc{b}}}-(2N+1)\pi)\nonumber\\
&-e^{-\ii\tilde{\Omega}_{\textsc{a}} (-{\varrho_{\textsc{b}}}+(2N+2)\pi)}\tilde\chi_{\textsc{b}}(t)\tilde\chi_{\textsc{a}}(t+{\varrho_{\textsc{b}}}-(2N+2)\pi)\nonumber\\
&-e^{-\ii\tilde{\Omega}_{\textsc{b}} (-{\varrho_{\textsc{b}}}+(2N+2)\pi)}\tilde\chi_{\textsc{a}}(t)\tilde\chi_{\textsc{b}}(t+{\varrho_{\textsc{b}}}-(2N+2)\pi))
\end{align}
for the $\mathcal{M}^-$ integral, and at this point we have a small problem. In general a phase coefficient exists before each term, unlike  the geodesic case, where $\tilde{\Omega}_{\textsc{a}}=\tilde{\Omega}_{\textsc{b}},$ allowing us to $t$-translate each term to make the $N$ phase cancel. 
Fortunately  we can continue as before, translating by half the displacement, in order to make this expression more symmetric.
Then, after translating and relabeling $t'$ as $t$, and renumbering as in the geodesic case \eqref{Mminusderivation}, we find
\begin{widetext}
\begin{align}
\mathcal{M}^-
&=\frac{\ii\lambda_{\textsc{b}}\lambda_{\textsc{a}}}{8\pi L^2\tan{\varrho_{\textsc{b}}}}\int_{-\infty}^\infty \dd t \,e^{i(\tilde{\Omega}_{\textsc{a}}+\tilde{\Omega}_{\textsc{b}}) t}\sum_{N=-\infty}^\infty(-\varepsilon)^{p(N)}\varsigma(N+1/2)  \nonumber\\
&\qquad \qquad \qquad\left(e^{i(\tilde{\Omega}_{\textsc{b}}-\tilde{\Omega}_{\textsc{a}})({\varrho_{\textsc{b}}}+N\pi)/2}\tilde\chi_{\textsc{b}}\left(t+\frac{{\varrho_{\textsc{b}}}+N\pi}{2}\right)\tilde\chi_{\textsc{a}}\left(t-\frac{{\varrho_{\textsc{b}}}+N\pi}{2}\right)\right.\nonumber\\
&\qquad \qquad \qquad +\left.e^{-\ii(\tilde{\Omega}_{\textsc{b}}-\tilde{\Omega}_{\textsc{a}})({\varrho_{\textsc{b}}}+N\pi)/2}\tilde\chi_{\textsc{a}}\left(t+\frac{{\varrho_{\textsc{b}}}+N\pi}{2}\right)\tilde\chi_{\textsc{b}}\left(t-\frac{{\varrho_{\textsc{b}}}+N\pi}{2}\right)\right)
\end{align}
\end{widetext}
 Note that this expression applies equally to the case where both detectors are geodesic, but have different proper gaps. It appears that the symmetry in $\tilde{\Omega}_I$ is more important than any symmetry in $\Omega_I$.

 The causality estimator derivation follows a similar path, except we cannot assume $\tilde\Omega_I=\tilde\Omega_J.$ We will simply write:
 \begin{align}
     \mathcal{C}_{IJ}&=\frac{\ii\lambda_{\textsc{a}} \lambda_{\textsc{b}}}{4\pi L^2\tan{\varrho_{\textsc{b}}}}\int_{-\infty}^{\infty}\dd t\,e^{i\tilde\Omega_I t}\sum_{N=0}^\infty\nonumber\\
     &\left[\Re[e^{\ii\tilde\Omega_J(t-{\varrho_{\textsc{b}}}-2N\pi)}]\tilde\chi_I(t)\tilde\chi_J(t-{\varrho_{\textsc{b}}}-2N\pi)\right.\nonumber\\
     &+\varepsilon\Re[e^{\ii\tilde\Omega_J(t+{\varrho_{\textsc{b}}}-(2N+1)\pi)}]\tilde\chi_I(t)\tilde\chi_J(t+{\varrho_{\textsc{b}}}-(2N+1)\pi)\nonumber\\
     &-\varepsilon\Re[e^{\ii\tilde\Omega_J(t-{\varrho_{\textsc{b}}}-(2N+1)\pi)}]\tilde\chi_I(t)\tilde\chi_J(t-{\varrho_{\textsc{b}}}-(2N+1)\pi)\nonumber\\
     &-\left.\Re[e^{\ii\tilde\Omega_J(t+{\varrho_{\textsc{b}}}-(2N+2)\pi)}]\tilde\chi_I(t)\tilde\chi_J(t+{\varrho_{\textsc{b}}}-(2N+2)\pi)\right]
     \label{Cab_static}
 \end{align}
 While this still can be written as a ``sum of two Fourier transforms", that form is not significantly simpler. In fact, for computational purposes it is simpler to use this expression directly.

\section{Gaussian switching functions}
At this point, let us make a few simplifying assumptions. As is common practice, we will use Gaussian switching functions, with
\begin{equation}
\chi_I(\tau)=e^{-(\tau-\tau_{0I})^2/2\sigma^2}.
\end{equation}
Note that our parameters $\tau_{0I},\sigma$ are expressed in proper time; the conversion between $t_{0I}$ and $\tau_{0I}$ will depend on the detector configuration.
As before, the Fourier transform of the switching function in its proper time is
\begin{equation}
\hat{\chi}_I(k)=\sigma e^{-k^2 \sigma^2/2+\ii k \tau_{0I} }.
\end{equation}
We can therefore express the density matrix elements in terms of $\sigma,t_{0I}$, and try to measure the dependence of the entanglement on the parameters.

%\tcb{\bf [Changed so that $t0>0$ implies A is first.]}
In the geodesic case, let us assume that $\lambda_{\textsc{a}}=\lambda_{\textsc{b}}=\lambda$, $\tilde{\Omega}_{\textsc{a}}=\tilde{\Omega}_{\textsc{b}}=\Omega L, t_{0A}=-t_{0B}=-t_0/2=-\tau_0/2L$, $\tilde{\sigma}_{\textsc{a}}=\tilde{\sigma}_{\textsc{b}}=\sigma/L.$ For $t_0>0$, this implies A switches first.  When considering different curvatures, we will also hold $\Delta x$ constant, calculating ${\varrho_{\textsc{b}}}$ using \eqref{rcoordfunc}. At this point, we emphasize that all circular geodesics in AdS share a common proper time, which implies that these detectors are identical. In a more general spacetime, this corresponds to setting the \textit{coordinate} gap and switching times equal, but the physical interpretation of this choice is less clear. Substituting the switching function into the $\mathcal{L}$ expressions, we find
\begin{align}
\mathcal{L}_{II}=&\lambda^2 \sum_{n} \frac{\pi}{\omega_n}\varphi^2_{n }(x_{\textsc{a}})\sigma^2 e^{-(\omega_n/L+ \Omega )^2 \sigma^2}\\
\mathcal{L}_{\textsc{ab}}=&\lambda^2 \sum_n \frac{\pi}{\omega_n}\varphi_{n }(x_{\textsc{b}})\varphi_{n }(x_{\textsc{a}})\nonumber\\
&\times \sigma^2 e^{-(\omega_n/L+\Omega)^2\sigma^2-\ii(\omega_n/L+\Omega )\tau_0}.
\end{align}
The $\mathcal{M}$ integral, specifically the commutator term $\mathcal{M}^-$, is rather more complicated. Using the symmetries above, we find
\begin{align}
    &\tilde\chi_{\textsc{b}}(t+({\varrho_{\textsc{b}}}+N\pi)/2)\tilde\chi_{\textsc{a}}(t-({\varrho_{\textsc{b}}}+N\pi)/2)\nonumber\\
    &=L^2e^{-(({\varrho_{\textsc{b}}}+N\pi-t_0)/2)^2/\tilde{\sigma}^2}e^{-t^2/\tilde{\sigma}^2},\\
    &\chi_{\textsc{a}}(t+({\varrho_{\textsc{b}}}+N\pi)/2)\chi_{\textsc{b}}(t-({\varrho_{\textsc{b}}}+N\pi)/2)\nonumber\\
    &=L^2e^{-(({\varrho_{\textsc{b}}}+N\pi+t_0)/2)^2/\tilde{\sigma}^2}e^{-t^2/\tilde{\sigma}^2}.
\end{align}
Evaluating the expressions for $\mathcal{M}^+$ and $\mathcal{M}^-$ yields
\begin{align}
    \mathcal{M}^+=&-\lambda^2\sum_{n=0}^\infty(-\varepsilon)^{p(N)}\frac{\pi}{\omega_n}\varphi_{n }(x_{\textsc{b}})\varphi_{n }(x_{\textsc{a}})\nonumber\\
    &\times \sigma^2 \cos[\omega_n\tau_0/L]e^{- (\Omega^2 +\omega_n^2/L^2)\sigma^2}\\
    \mathcal{M}^-=&\lambda^2 \frac{\ii}{8\pi\tan{\varrho_{\textsc{b}}}}\tilde{\sigma}\sqrt{\pi} e^{-\tilde{\sigma}^2\tilde{\Omega}^2}\sum_{N=-\infty}^\infty(-\varepsilon)^{p(N)}\varsigma(N+1/2)\nonumber\\
    &\times(e^{-(({\varrho_{\textsc{b}}}+N\pi-t_0)/2)^2/\tilde{\sigma}^2}+e^{-(({\varrho_{\textsc{b}}}+N\pi+t_0)/2)^2/\tilde{\sigma}^2})\nonumber\\
    =&\lambda^2 \frac{\ii}{4\sqrt{\pi} L\tan{\varrho_{\textsc{b}}}}\sigma e^{-\sigma^2\Omega^2}\sum_{N=-\infty}^\infty(-\varepsilon)^{p(N)}\varsigma(N+1/2)\nonumber\\
    &\times e^{-(({\varrho_{\textsc{b}}}+N\pi)^2L^2+\tau_0^2)/4\sigma^2} \cosh(2({\varrho_{\textsc{b}}}+N\pi)\tau_0 L/4\sigma^2)
\end{align}
Note that $\mathcal{M}^+$ and   $\mathcal{M}^-$ are pure real and pure imaginary respectively. This leads to interesting consequences if we attempt to calculate them separately, which we explore later.

The causality estimator expressions look rather similar to the $\mathcal{M}^-$ expressions, with a twist. In this case, there is a part that is \textit{not} Gaussian in the gap:
\begin{align}
&    \mathcal{C}_{\textsc{ab}}= \lambda^2\frac{\ii \sigma}{8\sqrt{\pi} L\tan{\varrho_{\textsc{b}}}}\sum_{N=0}^\infty\nonumber\\
    &\times\left[(e^{-\sigma^2\Omega^2}+e^{i\Omega({\varrho_{\textsc{b}}}+2N\pi)L})e^{-(({\varrho_{\textsc{b}}}+2N\pi)L+\tau_0)^2/4\sigma^2}\right.\nonumber\\
    &+\varepsilon(e^{-\sigma^2\Omega^2}+e^{i\Omega(-{\varrho_{\textsc{b}}}+(2N+1)\pi)L})e^{-((-{\varrho_{\textsc{b}}}+(2N+1)\pi)L+\tau_0)^2/4\sigma^2}\nonumber\\
    &-\varepsilon(e^{-\sigma^2\Omega^2}+e^{i\Omega({\varrho_{\textsc{b}}}+(2N+1)\pi)L})e^{-(({\varrho_{\textsc{b}}}+(2N+1)\pi)L+\tau_0)^2/4\sigma^2}\nonumber\\
    &-\left.(e^{-\sigma^2\Omega^2}+e^{i\Omega(-{\varrho_{\textsc{b}}}+(2N+2)\pi)L})e^{-((-{\varrho_{\textsc{b}}}+(2N+2)\pi)L+\tau_0)^2/4\sigma^2}\right]\displaybreak[0]\\
&    \mathcal{C}_{\textsc{ba}}= \lambda^2\frac{\ii \sigma}{8\sqrt{\pi} L\tan{\varrho_{\textsc{b}}}}\sum_{N=0}^\infty\nonumber\\
    &\times\left[(e^{-\sigma^2\Omega^2}+e^{i\Omega({\varrho_{\textsc{b}}}+2N\pi)L})e^{-(({\varrho_{\textsc{b}}}+2N\pi)L-\tau_0)^2/4\sigma^2}\right.\nonumber\\
    &+\varepsilon(e^{-\sigma^2\Omega^2}+e^{i\Omega(-{\varrho_{\textsc{b}}}+(2N+1)\pi)L})e^{-((-{\varrho_{\textsc{b}}}+(2N+1)\pi)L-\tau_0)^2/4\sigma^2}\nonumber\\
    &-\varepsilon(e^{-\sigma^2\Omega^2}+e^{i\Omega({\varrho_{\textsc{b}}}+(2N+1)\pi)L})e^{-(({\varrho_{\textsc{b}}}+(2N+1)\pi)L-\tau_0)^2/4\sigma^2}\nonumber\\
    &-\left.(e^{-\sigma^2\Omega^2}+e^{i\Omega(-{\varrho_{\textsc{b}}}+(2N+2)\pi)L})e^{-((-{\varrho_{\textsc{b}}}+(2N+2)\pi)L-\tau_0)^2/4\sigma^2}\right]
\end{align}

For the static case, we set the detectors to have equal proper gap and switching time. This implies that $\tilde{\Omega}_{\textsc{a}}=\Omega L$, $\tilde{\Omega}_{\textsc{b}}=\Omega (L \sec {\varrho_{\textsc{b}}})$, $\tilde{\sigma}_{\textsc{a}}=\sigma/L$, $\tilde{\sigma}_{\textsc{b}}=\sigma/(L \sec {\varrho_{\textsc{b}}})$.  For the time displacement we again choose $t_{0A}=-t_{0B}=-t_0/2$.
While this still produces the total coordinate-time displacement of $t_0$, we must now be more mindful of the effect on the phases. We also choose to scale the coordinate-time displacement as $t_0=\tau_0/L=- {2}\tau_{0A}/L$, holding the proper time constant; that is, we scale the coordinate-time displacement as though it were measured at the center.

Under this choice of time displacements, we can calculate the {non-entangling} detector responses as follows:
\begin{align}
\mathcal{L}_{II}=&\lambda^2 \sigma^2 \sum_{nl} \frac{\pi}{\omega_{nl}}\varphi^2_{nl}(x_I)  e^{-\sigma^2(\Omega+\omega_{nl}/(L\sec{\varrho_{\textsc{b}}}_I))^2 }\\
\mathcal{L}_{IJ}=&\lambda^2 \sigma^2 \sum_n \frac{\pi}{\omega_n}\varphi_{n }(x_J)\varphi_{n }(x_I) e^{-\sigma^2(\Omega+\omega_n/(L\sec{\varrho_{\textsc{b}}}_I))^2/2} \nonumber\\
&\quad \times  e^{-\sigma^2(\Omega+\omega_n/(L\sec{\varrho_{\textsc{b}}}_J))^2/2}\;  e^{-i(\omega_n+\bar{\Omega})\tau_{0}/L}
\end{align}
where $\bar{\Omega}=(\tilde{\Omega}_{\textsc{a}}+ \tilde{\Omega}_{\textsc{b}})/2=\Omega L (1+\sec{\varrho_{\textsc{b}}})/2$.

Next, for $\mathcal{M}^+$, we find
\begin{align}
    \mathcal{M}^+&=-\frac{\lambda^2}{2}\sum_{n}\frac{\pi}{\omega_n}\varphi_{n }(x_{\textsc{b}})\varphi_{n }(x_{\textsc{a}})\sigma^2\nonumber\\
    &\left(e^{- (\sigma^2(\frac{\omega_n}{L\sec{\varrho_{\textsc{b}}}}-\Omega)^2+\sigma^2(\frac{\omega_n}{L}+\Omega)^2)/2-\ii(2n+2+(\tilde{\Omega}_{\textsc{a}}-\tilde{\Omega}_{\textsc{b}}))t_0}\right.\nonumber\\
    &+\left.e^{- (\sigma^2(\frac{\omega_n}{L}-\Omega)^2+\sigma^2(\frac{\omega_n}{L\sec{\varrho_{\textsc{b}}}}+\Omega)^2)/2+\ii(\omega_n-(\tilde{\Omega}_{\textsc{a}}-\tilde{\Omega}_{\textsc{b}}))t_0}\right)\nonumber\\
    &=-\lambda^2\sigma^2 e^{-\sigma^2\Omega^2+i\Omega \tau_0(\sec{\varrho_{\textsc{b}}}-1)} \sum_{n}\frac{\pi}{\omega_n}\varphi_{n }(x_{\textsc{b}})\varphi_{n }(x_{\textsc{a}})\nonumber\\
    &\times e^{-\sigma^2(\omega_n)^2(1+\cos^2{\varrho_{\textsc{b}}})/2L^2}\nonumber\\
    &\times\cosh(\sigma^2\Omega(\omega_n)(1-\cos{\varrho_{\textsc{b}}})/L-\ii(\omega_n)\tau_0/L)
\end{align}
employing the identity \eqref{Lbaab}, 
where in the final step we use the fact that $\tilde{\sigma}_I \tilde{\Omega}_I = \sigma \Omega.$ It seems that $M_W$ is no longer real.

The switching functions appearing in $\mathcal{M}^-$ become 
\begin{align}
   \tilde\chi_{\textsc{b}}&\left(t+\frac{{\varrho_{\textsc{b}}}+N\pi}{2}\right)\tilde\chi_{\textsc{a}}\left(t-\frac{{\varrho_{\textsc{b}}}+N\pi}{2}\right)\nonumber\\
    =&L^2(\sec{\varrho_{\textsc{b}}}) e^{-(t+\frac{{\varrho_{\textsc{b}}}+N\pi-t_0}{2})^2/2\tilde{\sigma}^2_{\textsc{b}}}e^{-(t-\frac{{\varrho_{\textsc{b}}}+N\pi-t_0}{2})^2/2\tilde{\sigma}^2_{\textsc{a}}}\nonumber\\
    =&L^2(\sec{\varrho_{\textsc{b}}})e^{-\frac{({\varrho_{\textsc{b}}}+N\pi-t_0)^2}{2(\tilde{\sigma}_{\textsc{a}}^2+\tilde{\sigma}_{\textsc{b}}^2)}}
   e^{-\left(t+({\varrho_{\textsc{b}}}+N\pi-t_0)\frac{\tilde{\sigma}_{\textsc{a}}^2-\tilde{\sigma}_{\textsc{b}}^2}{\tilde{\sigma}_{\textsc{a}}^2+\tilde{\sigma}_{\textsc{b}}^2}\right)^2/2\tilde{\sigma}_{\textsc{ab}}^2}\\
    \tilde\chi_{\textsc{a}}&\left(t+\frac{{\varrho_{\textsc{b}}}+N\pi}{2}\right)\tilde\chi_{\textsc{b}}\left(t-\frac{{\varrho_{\textsc{b}}}+N\pi}{2}\right)\nonumber\\
    =&L^2(\sec{\varrho_{\textsc{b}}}) e^{-(t+\frac{{\varrho_{\textsc{b}}}+N\pi+t_0}{2})^2/2\tilde{\sigma}^2_{\textsc{a}}}e^{-(t-\frac{{\varrho_{\textsc{b}}}+N\pi+t_0}{2})^2/2\tilde{\sigma}^2_{\textsc{b}}}\nonumber\\
    =&L^2(\sec{\varrho_{\textsc{b}}})e^{-\frac{({\varrho_{\textsc{b}}}+N\pi+t_0)^2}{2(\tilde{\sigma}_{\textsc{a}}^2+\tilde{\sigma}_{\textsc{b}}^2)}}
   e^{-\left(t-({\varrho_{\textsc{b}}}+N\pi+t_0)\frac{\tilde{\sigma}_{\textsc{a}}^2-\tilde{\sigma}_{\textsc{b}}^2}{\tilde{\sigma}_{\textsc{a}}^2+\tilde{\sigma}_{\textsc{b}}^2}\right)^2/2\tilde{\sigma}_{\textsc{ab}}^2}
   \label{chiachib}
\end{align}
where $\tilde{\sigma}_{\textsc{ab}}^2=\tilde{\sigma}_{\textsc{a}}^2\tilde{\sigma}_{\textsc{b}}^2/(\tilde{\sigma}_{\textsc{a}}^2+\tilde{\sigma}_{\textsc{b}}^2).$
We can then apply the $\dd t$ integral term-by-term. The final result is
\begin{align}
    \mathcal M^-=&\frac{\ii\lambda^2}{8\pi\sin{\varrho_{\textsc{b}}}}\sum_{N=-\infty}^\infty (-\varepsilon)^{p(N)}\varsigma(N+1/2)\nonumber\\
    &\times \tilde{\sigma}_{\textsc{ab}}\sqrt{2\pi}e^{-2\sigma_{\textsc{ab}}^2\bar{\Omega}^2+2\ii t_0\bar{\Omega} \frac{\tilde{\sigma}_{\textsc{a}}^2-\tilde{\sigma}_{\textsc{b}}^2}{\tilde{\sigma}_{\textsc{a}}^2+\tilde{\sigma}_{\textsc{b}}^2}}\nonumber\\
    &\times \left(e^{-\frac{({\varrho_{\textsc{b}}}+N\pi-t_0)^2}{2(\tilde{\sigma}_{\textsc{a}}^2+\tilde{\sigma}_{\textsc{b}}^2)}-\ii({\varrho_{\textsc{b}}}+N\pi) \frac{\tilde{\sigma}_{\textsc{a}}^2\tilde{\Omega}_{\textsc{a}}-\tilde{\sigma}_{\textsc{b}}^2\tilde{\Omega}_{\textsc{b}}}{\tilde{\sigma}_{\textsc{a}}^2+\tilde{\sigma}_{\textsc{b}}^2}}\right.\nonumber\\
    &+\left.e^{-\frac{({\varrho_{\textsc{b}}}+N\pi+t_0)^2}{2(\tilde{\sigma}_{\textsc{a}}^2+\tilde{\sigma}_{\textsc{b}}^2)}+\ii({\varrho_{\textsc{b}}}+N\pi) \frac{\tilde{\sigma}_{\textsc{a}}^2\tilde{\Omega}_{\textsc{a}}-\tilde{\sigma}_{\textsc{b}}^2\tilde{\Omega}_{\textsc{b}}}{\tilde{\sigma}_{\textsc{a}}^2+\tilde{\sigma}_{\textsc{b}}^2}}\right)\nonumber\displaybreak[0]\\
   =&\frac{i\lambda^2}{8\pi\sin{\varrho_{\textsc{b}}}} \tilde{\sigma}_{\textsc{ab}}\sqrt{2\pi}e^{-2\sigma_{\textsc{ab}}^2\bar{\Omega}^2+2\ii t_0\bar{\Omega} \frac{\tilde{\sigma}_{\textsc{a}}^2-\tilde{\sigma}_{\textsc{b}}^2}{\tilde{\sigma}_{\textsc{a}}^2+\tilde{\sigma}_{\textsc{b}}^2}}\nonumber\\
    &\times \sum_{N=-\infty}^\infty(-\varepsilon)^{p(N)} \varsigma(N+1/2)e^{-\frac{({\varrho_{\textsc{b}}}+N\pi)^2+t_0^2}{2(\tilde{\sigma}_{\textsc{a}}^2+\tilde{\sigma}_{\textsc{b}}^2)}}\nonumber\\
    &\times 2\cosh\left[\frac{{\varrho_{\textsc{b}}}+N\pi}{\tilde{\sigma}_{\textsc{a}}^2+\tilde{\sigma}_{\textsc{b}}^2}\left(t_0-\ii (\tilde{\sigma}_{\textsc{a}}^2\tilde{\Omega}_{\textsc{a}}-\tilde{\sigma}_{\textsc{b}}^2\tilde{\Omega}_{\textsc{b}})\right)\right]\displaybreak[0] \nonumber\\
    =&\frac{i\lambda^2}{8\pi L\sin{\varrho_{\textsc{b}}}}\sigma\sqrt{\frac{2\pi}{1+\sec^2{\varrho_{\textsc{b}}}}}\nonumber\\
    &\times e^{-\frac{(1+\sec{\varrho_{\textsc{b}}})^2}{1+\sec^2{\varrho_{\textsc{b}}}}\sigma^2\Omega^2/2+\ii\tau_0\Omega \frac{(1-\cos^2{\varrho_{\textsc{b}}})(1+\sec{\varrho_{\textsc{b}}})}{1+\cos^2{\varrho_{\textsc{b}}}}}\nonumber\\
    &\times \sum_{N=-\infty}^\infty(-\varepsilon)^{p(N)} \varsigma(N+1/2)e^{-\frac{({\varrho_{\textsc{b}}}+N\pi)^2L^2+\tau_0^2}{2\sigma^2(1+\cos^2{\varrho_{\textsc{b}}})}}\nonumber\\
    &\times 2\cosh\left[\frac{({\varrho_{\textsc{b}}}+N\pi)L}{1+\cos^2{\varrho_{\textsc{b}}}}\left(\tau_0/\sigma^2-\ii\Omega(1-\cos{\varrho_{\textsc{b}}}) \right)\right]
\end{align}
Notably, it appears that the asymmetry between A and B has caused $\mathcal M^-$ to no longer be purely imaginary, except in the special case ${\varrho_{\textsc{b}}}=0.$ Unlike the geodesic case, it is not clear whether $\mathcal M^+$ and $\mathcal M^-$ have any particular relative phase.

Finally,   the causality estimator largely follows the $\mathcal{M}^-$ derivation. The full expression may be found in \eqref{Cab_static_full}.
While the result is rather complicated, we can see two terms: one Gaussian on the sum of detector frequencies, and one Gaussian on the difference. In most cases the Gaussian of differences (or the $e^x$ part of $\cosh x$) will dominate.

Next, the logarithmic negativity can be expressed as $\mathcal{N}=\textrm{max}(\mathcal{N}^{(2)},0)$ where
\begin{equation}
\mathcal{N}^{(2)}=-\frac{1}{2}\left(\mathcal{L}_{\textsc{aa}}+\mathcal{L}_{\textsc{bb}}-\sqrt{(\mathcal{L}_{\textsc{aa}}-\mathcal{L}_{\textsc{bb}})^2+4|\mathcal{M}|^2}\right).
\end{equation}
However, in the geodesic case, because $\mathcal{L}_{\textsc{aa}}=\mathcal{L}_{\textsc{bb}},$ this simplifies to
\begin{equation}\label{Neg2}
\mathcal{N}^{(2)}=|\mathcal{M}|-\mathcal{L}_{II}.
\end{equation}
This form demonstrates the competition between nonlocal correlations and local noise. It also suggests that it may be possible to quantify the degree to which the detectors are timelike-connected, by comparing the commutator part of $\mathcal{M}$ to local noise. We will explore this in the next section.

Finally, we will also consider the mutual information, given by
\begin{equation}
    I(\rho_{\textsc{ab}})=S(\rho_{\textsc{a}})+S(\rho_{\textsc{b}})-S(\rho_{\textsc{ab}})
\end{equation}
where $S=-Tr(\rho \log \rho)$ is the von Neumann entropy and $\rho_I=Tr_J(\rho_{IJ})$ is the partial trace of the total state of two detectors. Evaluating in terms of $\mathcal{L}_{JI},$ we find \cite{Pozas-Kerstjens:2015gta}
\begin{align}
    I(\rho_{\textsc{ab}})=&\mathcal{L}_+\log\mathcal{L}_+ + \mathcal{L_-}\log\mathcal{L_-}\nonumber\\
    &-\mathcal{L}_{\textsc{aa}}\log \mathcal{L}_{\textsc{aa}} - \mathcal{L}_{\textsc{bb}} \log \mathcal{L}_{\textsc{bb}}
\end{align}
where 
\begin{equation}
    \mathcal{L}_\pm=\frac{1}{2}\left(\mathcal{L}_{\textsc{aa}}+\mathcal{L}_{\textsc{bb}}\pm\sqrt{(\mathcal{L}_{\textsc{aa}}-\mathcal{L}_{\textsc{bb}})^2+4|\mathcal{L}_{\textsc{ab}}|^2}\right).
\end{equation}
Of course, in the geodesic case, the symmetries of the problem allow us to reduce this to
\begin{equation}
\mathcal{L}_\pm=\mathcal{L}_{\textsc{aa}}\pm |\mathcal{L}_{\textsc{ab}}|.
\end{equation}

\section{Results}
We will begin by analyzing $L_{II}.$ While the transition probability of a uniformly accelerating single detector in AdS$_4$ was previously explored by Jennings \cite{Jennings:2010vk}, our use of finite time switching implies the possibility of qualitatively different behaviour. Furthermore, we will also (briefly) consider negative gaps, corresponding to a situation in which the detector is initially excited. For brevity, we will mostly plot data from the Dirichlet case, $\varepsilon=-1,$ and discuss differences where they occur.

\subsection{Transition Rates}

\begin{figure}[hbtp]
    \centering
    \begin{subfigure}[hbtp]{0.95\columnwidth}
    \caption{$r=0$}
    \includegraphics[width=\textwidth]{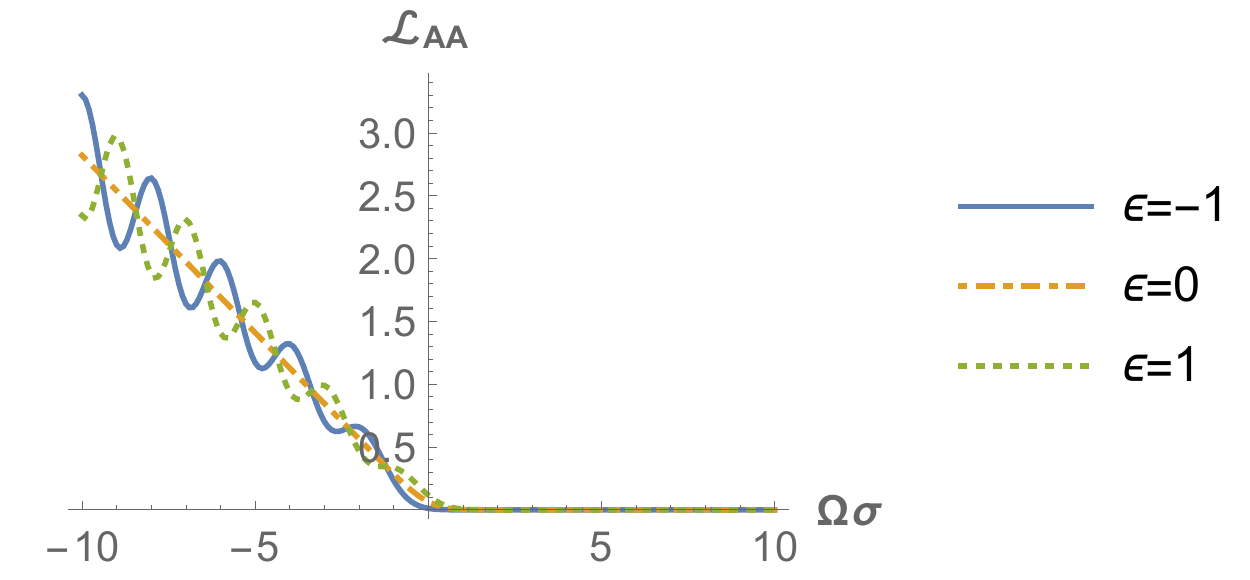}
    \label{Laa_gapeps_graph}
    \end{subfigure}\\
    \begin{subfigure}[hbtp]{0.95\columnwidth}
    \caption{$\Omega=2/\sigma$}
    \includegraphics[width=\textwidth]{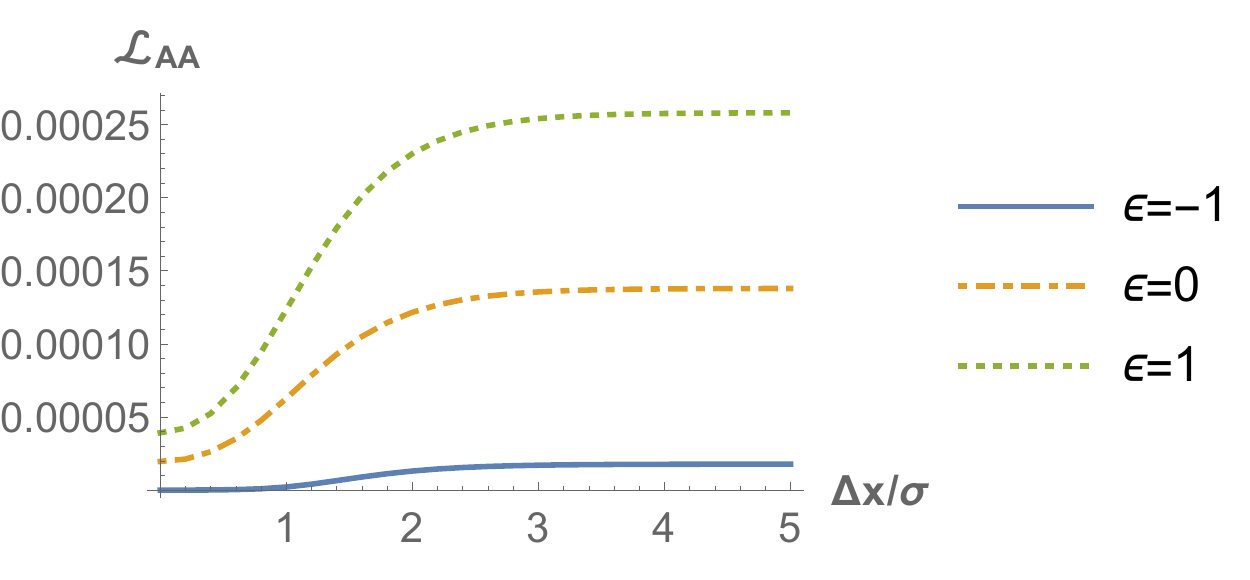}
    \label{Lbb_reps_graph}
    \end{subfigure}
    \caption{The one-detector transition rate for the static detector, $L=\sigma$, varying $\Omega$ and $r$ separately. Note the resonances visible in Fig. \ref{
    Laa_gapeps_graph}}
\end{figure}

Using the expression found previously, we compute $L_{II}$ for a static detector as a function of gap and position, where we use $\sigma$ as our reference  length scale. Notably, we find essentially no variation with respect to the position of the detector.  As one might expect, the effect of a finite switching time is to excite the detector, even if it starts in its ground state. A negative-gap detector, on the other hand, spontaneously de-excites; the greater the negative gap (in absolute value), the faster this occurs.  For small values of $L$, such as  $L=\sigma$, we can also see hints of a resonant effect in Fig. \ref{Laa_gapeps_graph}, as the detector couples to different modes within AdS, much as it would couple to different modes in a cavity. The main difference between boundary conditions appears to be the location of the resonances: the von Neumann case $\varepsilon=1$ has one set of modes, the Dirichlet case $\varepsilon=-1$ has another, and the transparent case $\varepsilon=0$ has both. Notably, the lowest energy mode is higher in the Dirichlet case: as shown in Fig. \ref{Laa_gapeps_graph}, this also causes the transition rate to decay faster for positive $\Omega$. As well, the resonances become less distinct at increasing distance from the centre, due to the presence of higher spherical harmonics.  However, since we wish to minimize $L_{II},$ we will only consider positive gap when calculating entanglement. (Of course, as discussed earlier, the geodesic detector has the same behaviour regardless of its position.)
We also note that comparison with the analytic expression for the Wightman function yields identical results.

There is one feature of the transition rate that is characteristic of the dimensionality, and \textit{not} of AdS$_n$ generally. For large negative gap, we see that the transition rate is roughly proportional to $|\Omega|,$ modulo the oscillations. This is consistent with previous results in 3+1 dimensional spacetime, e.g. Minkowski, as calculated in \cite{Ng:2016hzn}. However, it is quite different from the picture found by Henderson et.al. in AdS$_3$ \cite{HHMSZ}; in that case, the response of the detector approaches a finite limit for large negative $\Omega$. This difference is not surprising --
it is well-known that the response of a detector exhibits different dependence on $\Omega$ as a function of spacetime dimension due to the short distance behaviour of the Wightman function~\cite{Hodgkinson:2011pc}. An equivalent way to understand this is from the energy scaling of the density of states of  field quanta probed by the detector, which goes as $\Omega^{d-3}$~\cite{densityOfStates}. In specific terms, both AdS$_3$ and AdS$_4$ have an azimuthal mode number $m$, but AdS$_4$ also has an angular momentum mode $l$.

\begin{figure}[hbtp]
    \centering
    \includegraphics[width=0.95\columnwidth]{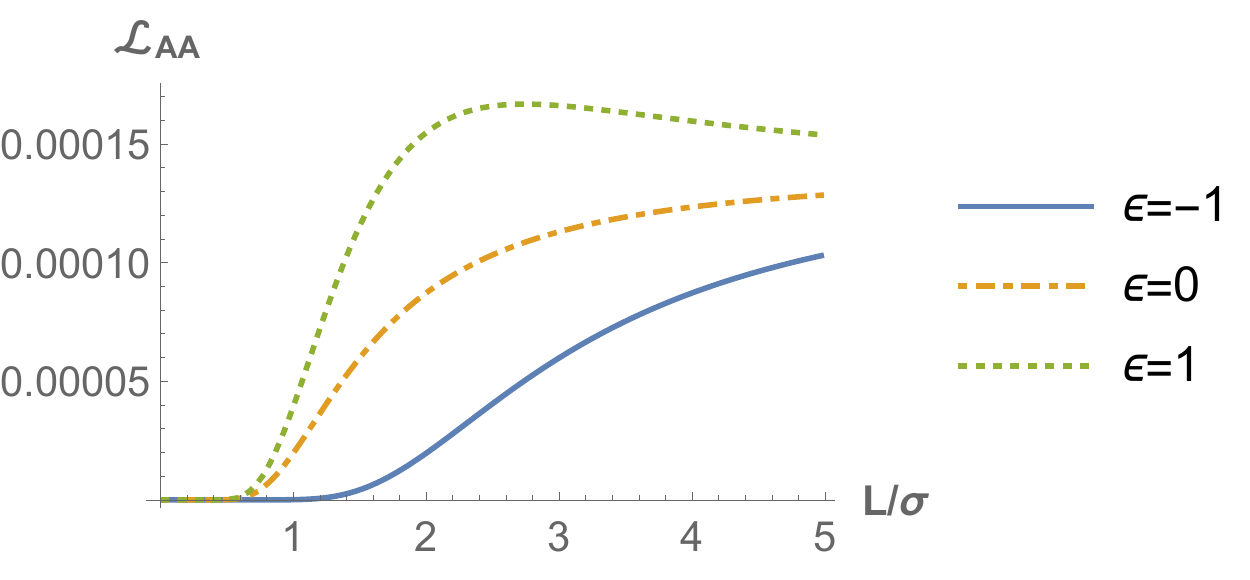}
    \caption{The one-detector transition rate at the centre, with fixed gap $\Omega=2/\sigma,$ varying curvature length $L$.}
    \label{Laa_Leps_graph}
\end{figure}

We also compare different values of $L$ to determine whether the detector can determine the curvature of AdS in a `faster than light' sense discussed in \cite{Ng:2016hzn}.  Specifically, can a detector identify the curvature of spacetime, if $L > \sigma$? This question is not as trivial as it appears: although AdS is usually described as uniformly curved, there is a coordinate system in which it is conformally flat, over a patch covering half an AdS period. We therefore might {na\"{i}vely} expect that the curvature cannot be detected except in timescales larger than that patch. As shown in Fig. \ref{Laa_Leps_graph}, this is not true: while $L_{AA}$ converges as $L/\sigma$ is increased to infinity, there is an observable difference for $L/\sigma$ as large as 5. This confirms our expectations: namely, that AdS approximates flat space in the lower curvature limit, but has some detectable differences.

 Once again, we also compared different boundary conditions over varying values of $L$; while the resonances were once again shifted for different $\varepsilon,$ the more noticeable effect at positive $\Omega$ is the small-$L$ behaviour. Specifically, the Dirichlet condition ($\varepsilon=-1$) decays the fastest. We also observed that when $\Omega<0,$ $\varepsilon$-dependent resonances with varying $L$ were observed, as one would expect.

\subsection{Geodesic Entanglement Harvesting}

\begin{figure}[hbtp]
    \centering
    \begin{subfigure}[hbtp]{0.9\columnwidth}
        \caption{$\mathcal N^{(2)}$}
        \includegraphics[width=\textwidth]{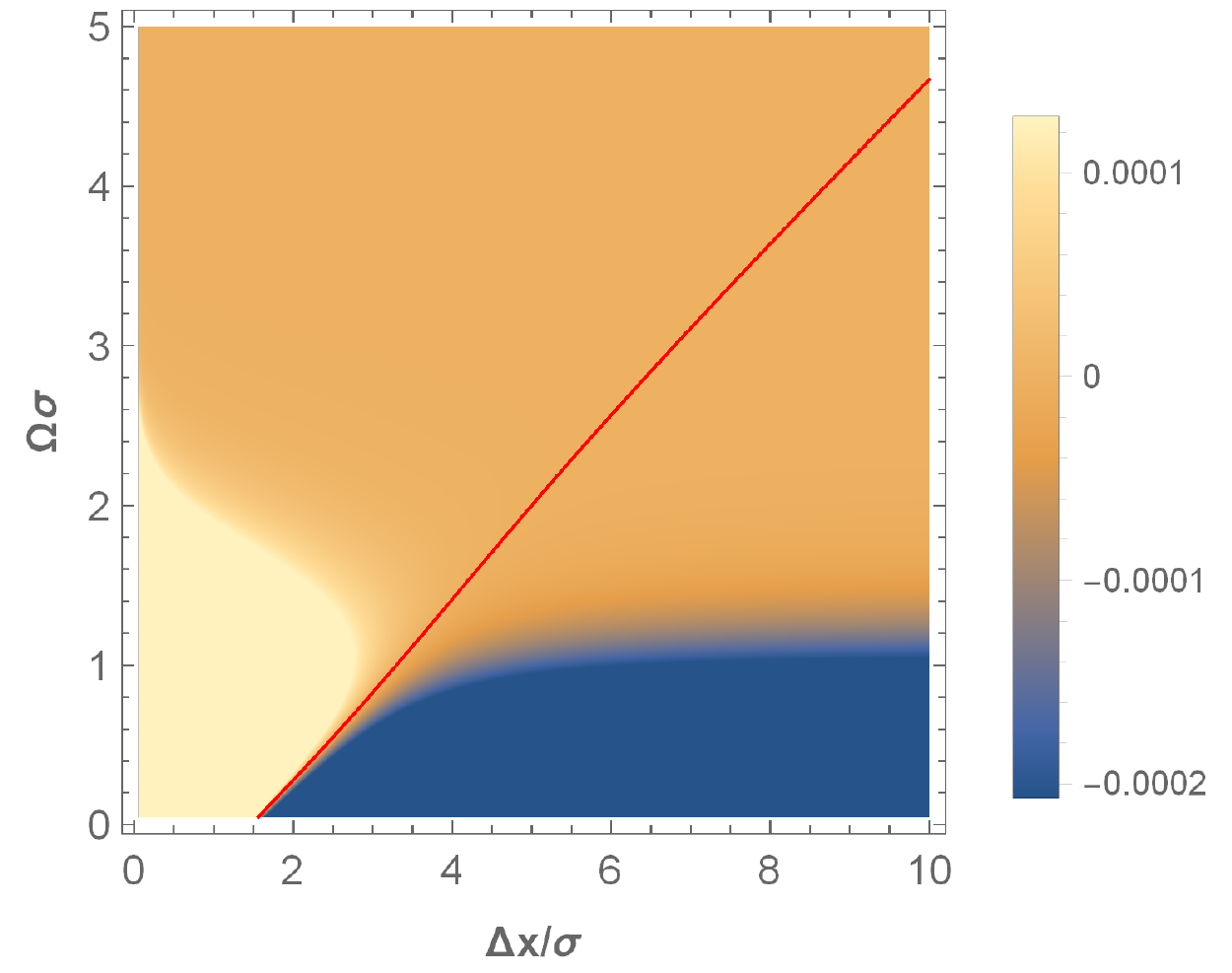}
        \label{n2_rgap_L5_graph}
    \end{subfigure}\\
    \begin{subfigure}[hbtp]{0.9\columnwidth}
        \caption{$\mathcal N^{(2)}-\mathcal C_{\textsc{ab}}$}
        \includegraphics[width=\textwidth]{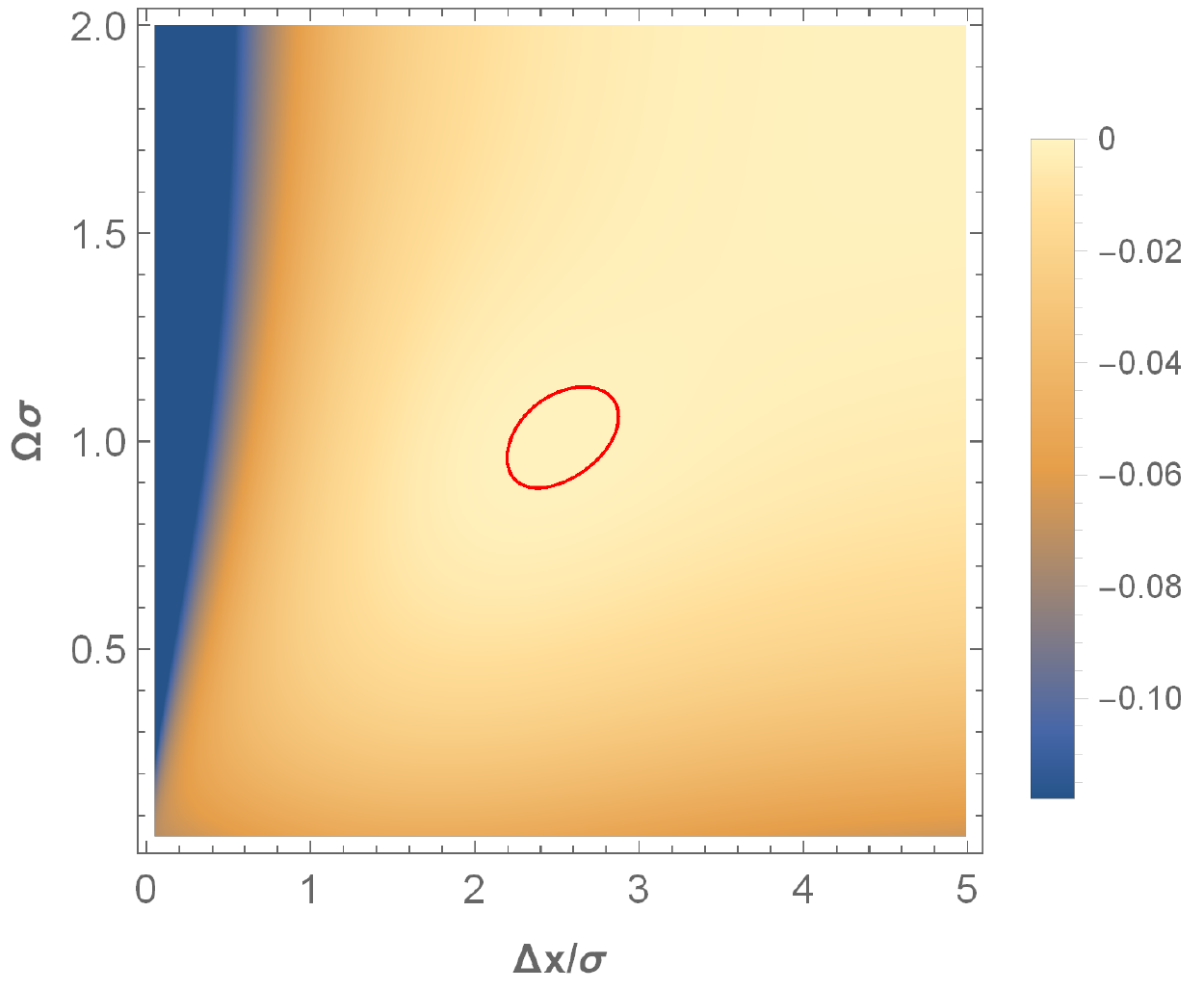}
        \label{n2_minus_Cab}
    \end{subfigure}
    \caption{Negativity values for varying separation and gap. $\varepsilon=-1.$ Region where $\mathcal N^{(2)}>\mathcal C_{\textsc{ab}}$ appears to be limited in size, even as $L \rightarrow \infty$. Zero contour marked in red.}
    \label{n2_rgap_graphs}
\end{figure}

We now analyze the negativity $\mathcal{N}^{(2)}.$ While this scenario was previously analyzed in flat space in \cite{Pozas-Kerstjens:2015gta}, some unique considerations appear in $AdS$. At the simplest and most obvious level, there is one additional parameter, the curvature scale $L$. The flat space limit is $L \rightarrow \infty$, while at the opposite limit, curvature approaches infinity; so one might expect that entanglement behaves very differently if $L$ is as small as $\sigma$ or smaller. At that point, the detector can probe the discrete spectrum of $AdS$, so we might expect some qualitatively different behaviour. However, the question of whether entanglement exists \textit{at all} appears to have a similar answer: namely, for any separation, some entanglement exists for sufficiently high detector gap. We show a gap/separation plot for $L=5\sigma$ in Figure \ref{n2_rgap_L5_graph},  where we have plotted $\mathcal{N}^{(2)}$ in \eqref{Neg2} and not the negativity $\mathcal{N}$.
 Notably, there still appears to be a linear relation between minimum gap and proper separation, independent of the value of $\varepsilon.$ Plotting for $L=\sigma$ shows an almost identical graph, complete with a line of zero negativity in roughly the same place; the implications of this invariance recommend further study.

The fact that the boundary of entanglement is described by a line deserves further examination. For instance, in \cite{Pozas-Kerstjens:2015gta}, the boundary becomes a straight line for both (3+1)-D and (1+1)-D Minkowski space, when the detector is switched via Gaussian, suggesting the dimension is not a factor here. This is further supported by similar findings \cite{HHMSZ} in AdS$_3$, in which this linear relation breaks down at small gap and separation. However, it appears that AdS has a nearly perfect line here; in particular, it appears that even degenerate detectors can harvest entanglement, while this is not possible in Minkowski space. While this is not strictly a violation of \cite{Pozas-Kerstjens:2017xjr}, since the finite switching implies the detectors are in causal contact, it is still surprising that AdS is distinguished from Minkowski space in this way. Further study may be needed here.
 
The question of whether \textit{spacelike} entanglement is possible is more complicated. Recall that we use Gaussian switching functions, rather than compact switching functions. 
%\tcb{\bf[Now I'm wondering what would happen if we used a compact-but-not-sudden switching function...]} 
Since our switching functions do not have compact support, we must therefore be wary of how much the tails of the Gaussians contribute to the entanglement. There are two ways we might attempt to quantify this: we could either simply assert that the detectors are spacelike separated if their switching supports are suitably separated in space, or we could compare the negativity to the ``causal" contribution to the density matrix. Unfortunately, these two methods do \textit{not} yield similar answers: since our density matrix term $\mathcal M$ is Gaussian in the gaps, but our causality estimator $\mathcal C_{IJ}$ is not, we cannot simply increase the gap until the negativity exceeds the causal contribution. Therefore, Fig. \ref{n2_rgap_L5_graph} is a little misleading: if the separation is too large, then even if the gap may be increased until $\mathcal N^{(2)}>0,$ the negativity will still not exceed the causality operator. Note that the parameter space in which $\mathcal N^{(2)}>|\mathcal C_{IJ}|$ is quite restrictive, even as we approach the flat space limit; we plot the difference $\mathcal N^{(2)}-|\mathcal C_{IJ}|$ in Fig. \ref{n2_minus_Cab} for $L=5$. In fact, the size of this ``island of spacelike entanglement" does not increase in size significantly, even for $L=200$ or larger,  in defiance of what we would expect of the flat space limit. We leave the correct characterization of spacelike entanglement to future work. %\tcb{\bf[$L=5$ fixed again. Turns out we were right the first time.]}

Another unique feature of AdS appears if we try to decrease $L$ too far. 
Because we use a conformally-coupled scalar field, signals can reach spatial infinity in finite time; therefore, if two detectors are switched at the same time, there is a switching time $\sigma$ for which the detectors will become timelike connected, no matter how far apart they are in space. (While this is also true for a non-conformally coupled field, strictly speaking, the corresponding finite energy modes are trapped away from the boundary; the `signal' that reaches the boundary can only be faithfully represented in the UV limit. However, two detectors will still become connected for sufficiently small $L$.) This occurs when the characteristic length $L$ is approximately equal to $\sigma$. This unique situation raises some further challenges when determining whether detectors are causally connected: for instance, since the boundary is located at $\rho = \pi/2,$ and the speed of light in these coordinates is $d\rho/dt=1,$ then the proper time for the central detector to send a signal, have it reflect off the boundary, then return, is $L\pi;$ no greater separation is possible. Nevertheless, we may still consider what happens to the negativity for such high curvatures, despite their more classical origin. At such curvatures, the boundary condition $\varepsilon$ becomes relevant, as we will see.

\begin{figure*}[hbtp]
    \centering
    \begin{subfigure}[hbtp]{0.3\textwidth}
    \caption{$\Omega=3/\sigma$,$\varepsilon=-1$}
    \includegraphics[width=\textwidth]{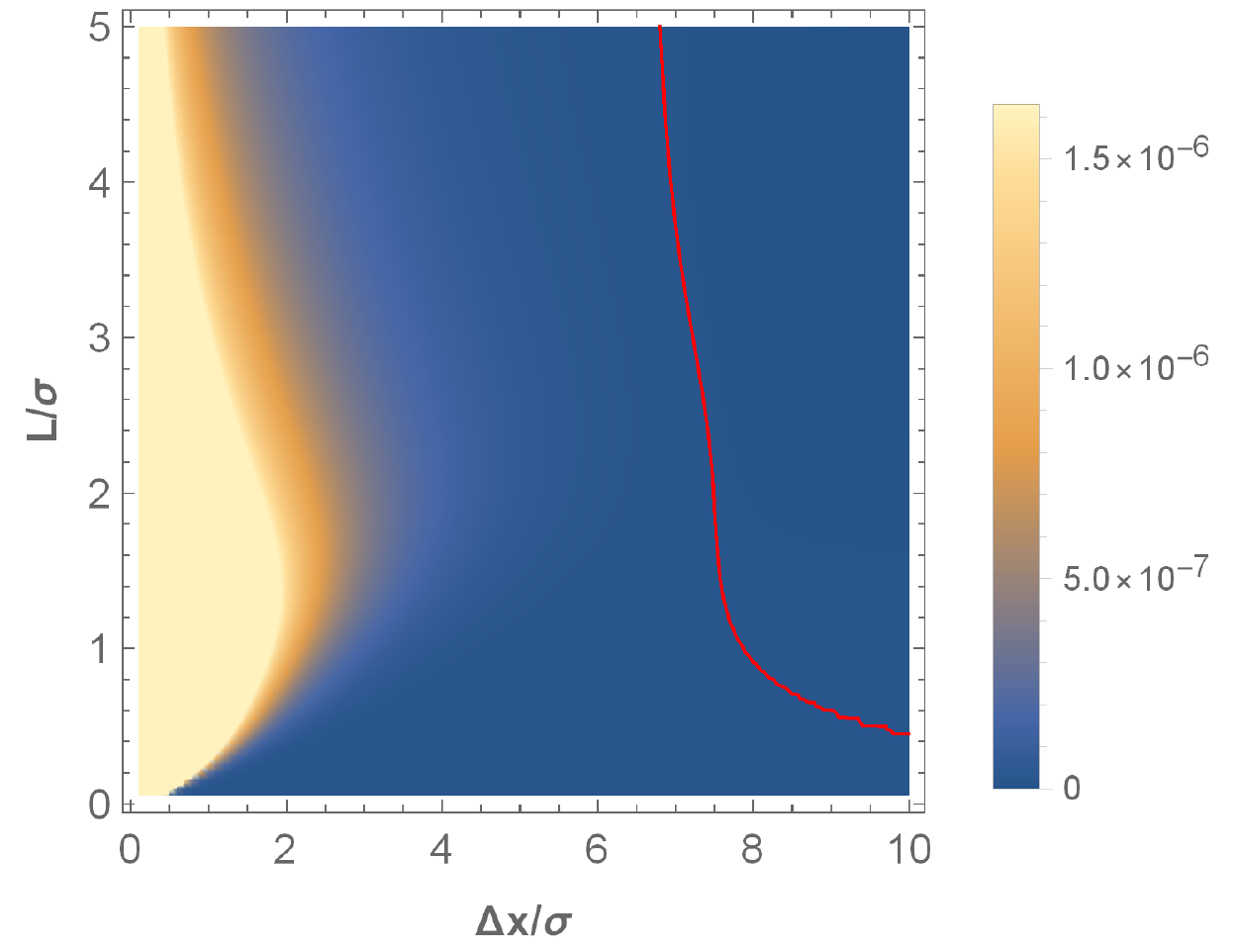}
    \label{n2_rL_E3_graph}
    \end{subfigure}
    \qquad
    \begin{subfigure}[hbtp]{0.3\textwidth}
    \caption{$\Omega=3/\sigma$,$\Delta x=6\sigma$}
    \includegraphics[width=\textwidth]{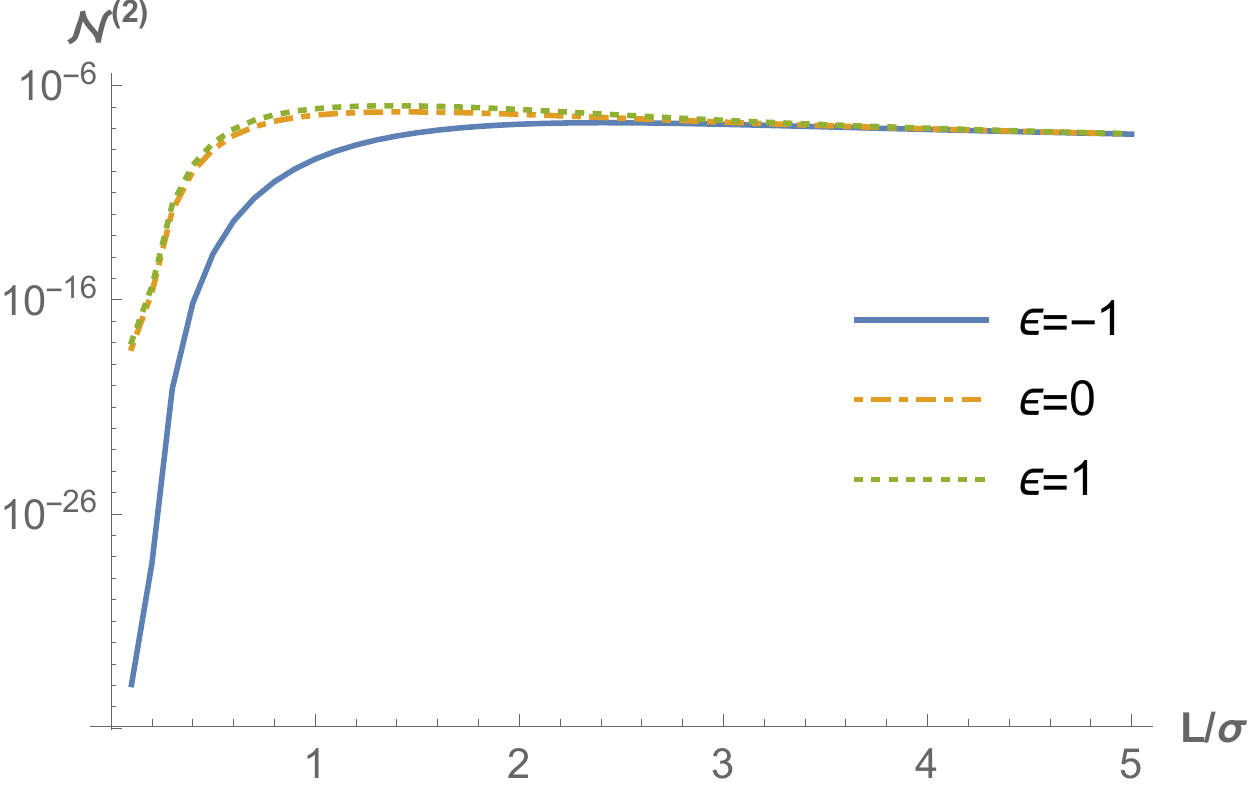}
    \label{n2_Leps_graph}
    \end{subfigure}
    \qquad
    \begin{subfigure}[hbtp]{0.3\textwidth}
    \caption{$\Omega=3/\sigma$,$L=\sigma$}
    \includegraphics[width=\textwidth]{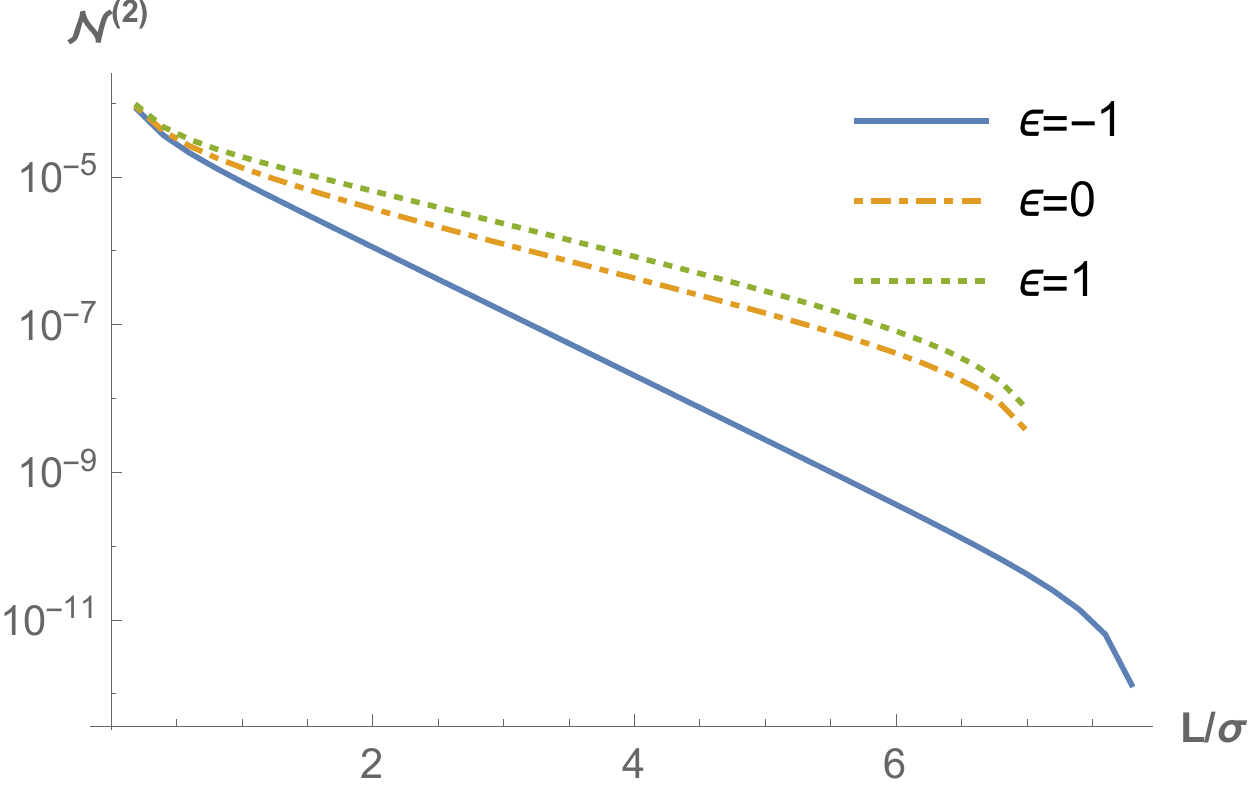}
    \label{n2_reps_graph}
    \end{subfigure}
    \caption{The negativity of two geodesic detectors for varying boundary conditions, one at the origin, one at proper separation $\Delta x.$}
    \label{n2_eps_graph}
\end{figure*}

We also plot the negativity as a function of curvature scale and proper separation, keeping the proper gap constant at $\Omega=3\sigma,$ shown in Fig. \ref{n2_rL_E3_graph}, with selected sections in Figs. \ref{n2_Leps_graph} and \ref{n2_reps_graph}. For sufficiently large curvature scale, the negativity appears to approach an asymptotic limit, as we would expect. It also appears that for very small curvature scales, the negativity trends towards zero. However, contrary to what one might expect, it seems that even as $L\rightarrow 0$, the maximum range of entanglement increases. Of course, as mentioned previously, this is most likely due to the timelike connection phenomenon: in that limit, detectors become causally related, no matter what their proper separation is.

At this point, we compare the negativity obtained for different boundary conditions $\varepsilon$ as a function of curvature scale. While these values converge for large $L$, as we might expect, their behaviour at small $L$ is drastically different. We attribute these differences to the different shapes of the ground states of AdS: namely, while the lowest energy excitation of the Dirichlet case is centered about the origin, the lowest energy excitation of the von Neumann and transparent cases is evenly spread out over AdS, explaining why the Dirichlet entanglement decays so quickly with distance.

\begin{figure*}[hbtp]
    \centering
    \begin{subfigure}[hbtp]{0.3\textwidth}
        \caption{Dirichlet $\mathcal{N}^{(2)}$}
        \includegraphics[width=\textwidth]{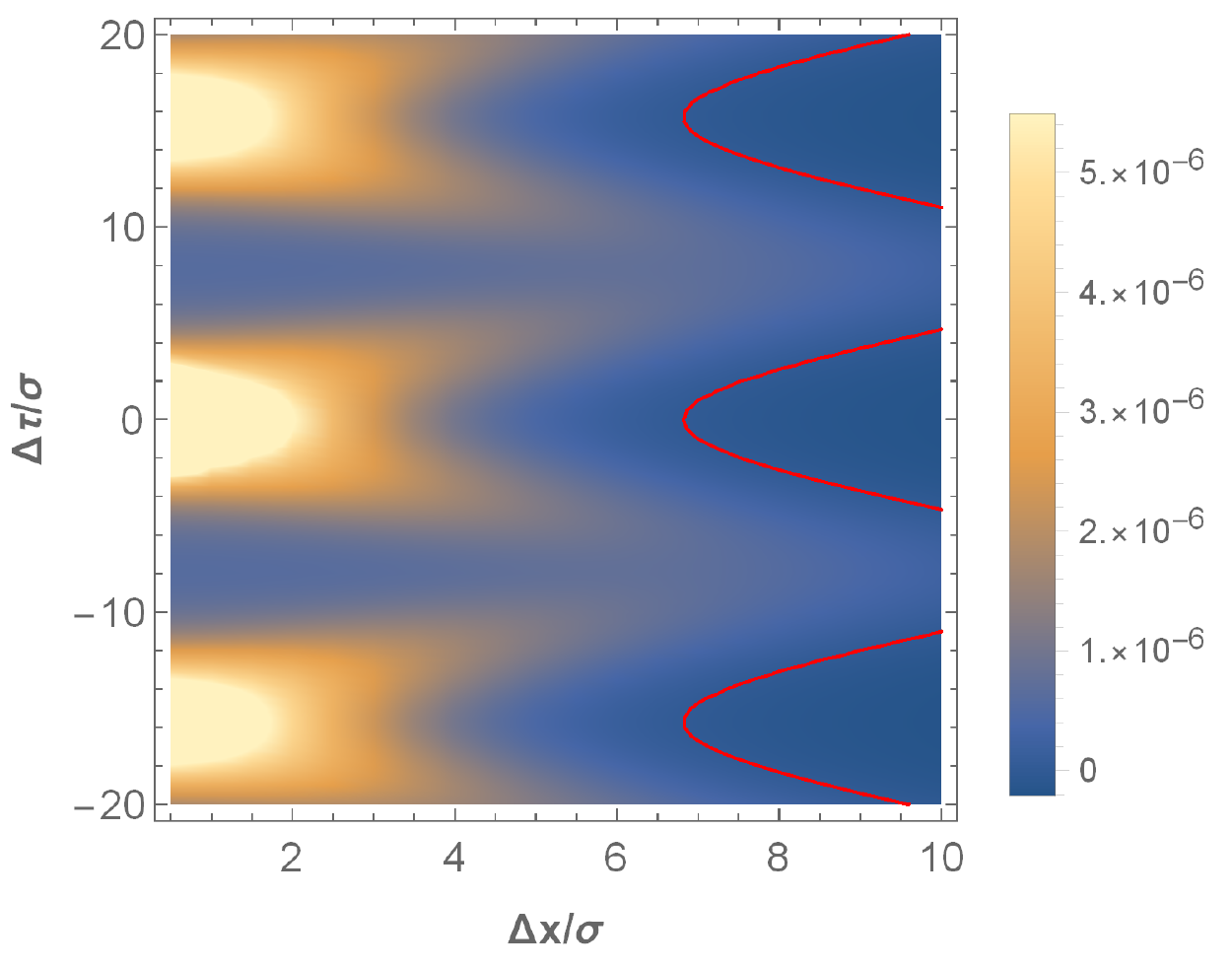}
        \label{n2_rt_L5E3_dirichlet_graph}
    \end{subfigure}
    \qquad
    \begin{subfigure}[hbtp]{0.3\textwidth}
        \caption{Transparent $\mathcal{N}^{(2)}$}
        \includegraphics[width=\textwidth]{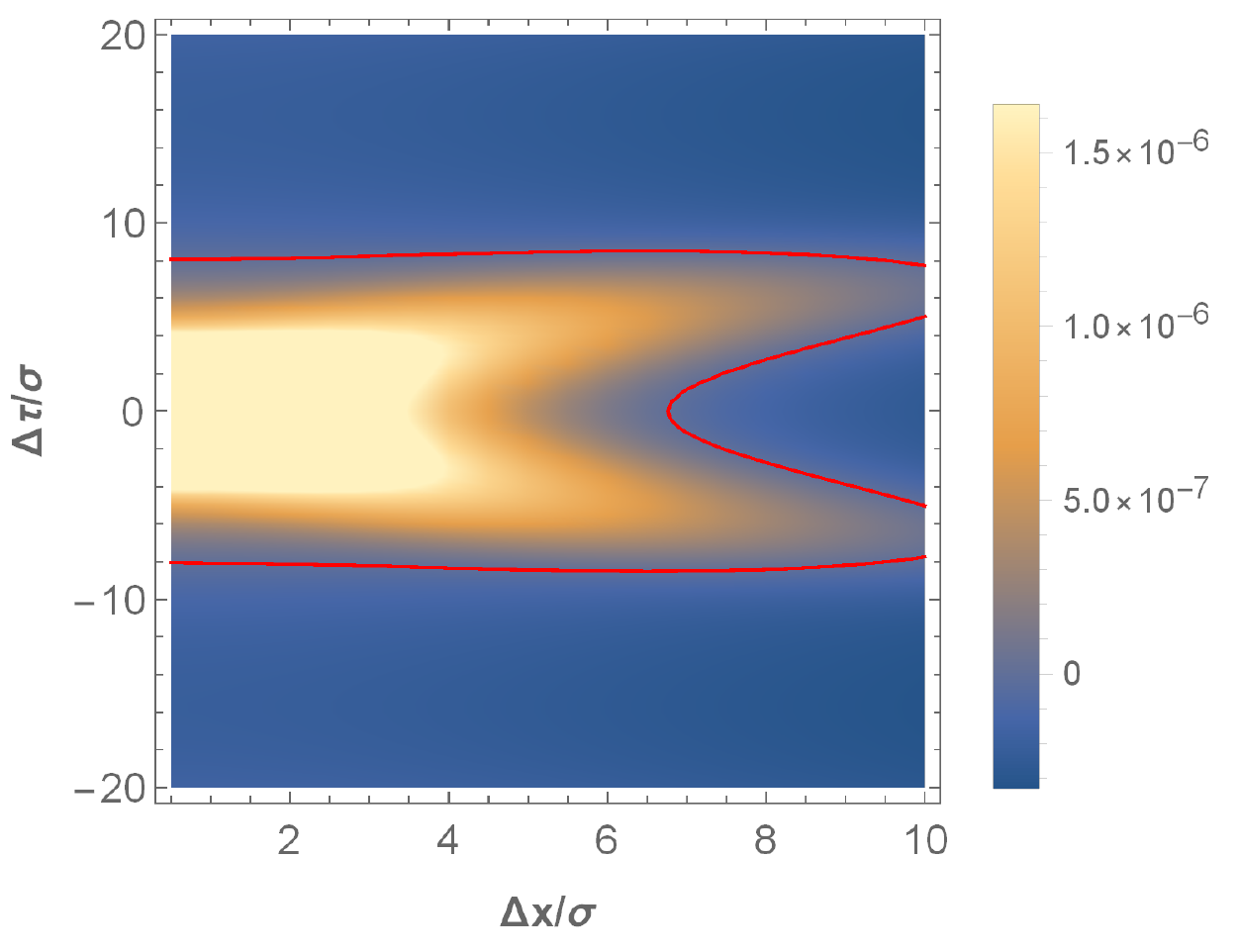}
        \label{n2_rt_L5E3_transparent_graph}
    \end{subfigure}
    \qquad
    \begin{subfigure}[hbtp]{0.3\textwidth}
        \caption{Neumann $\mathcal{N}^{(2)}$}
        \includegraphics[width=\textwidth]{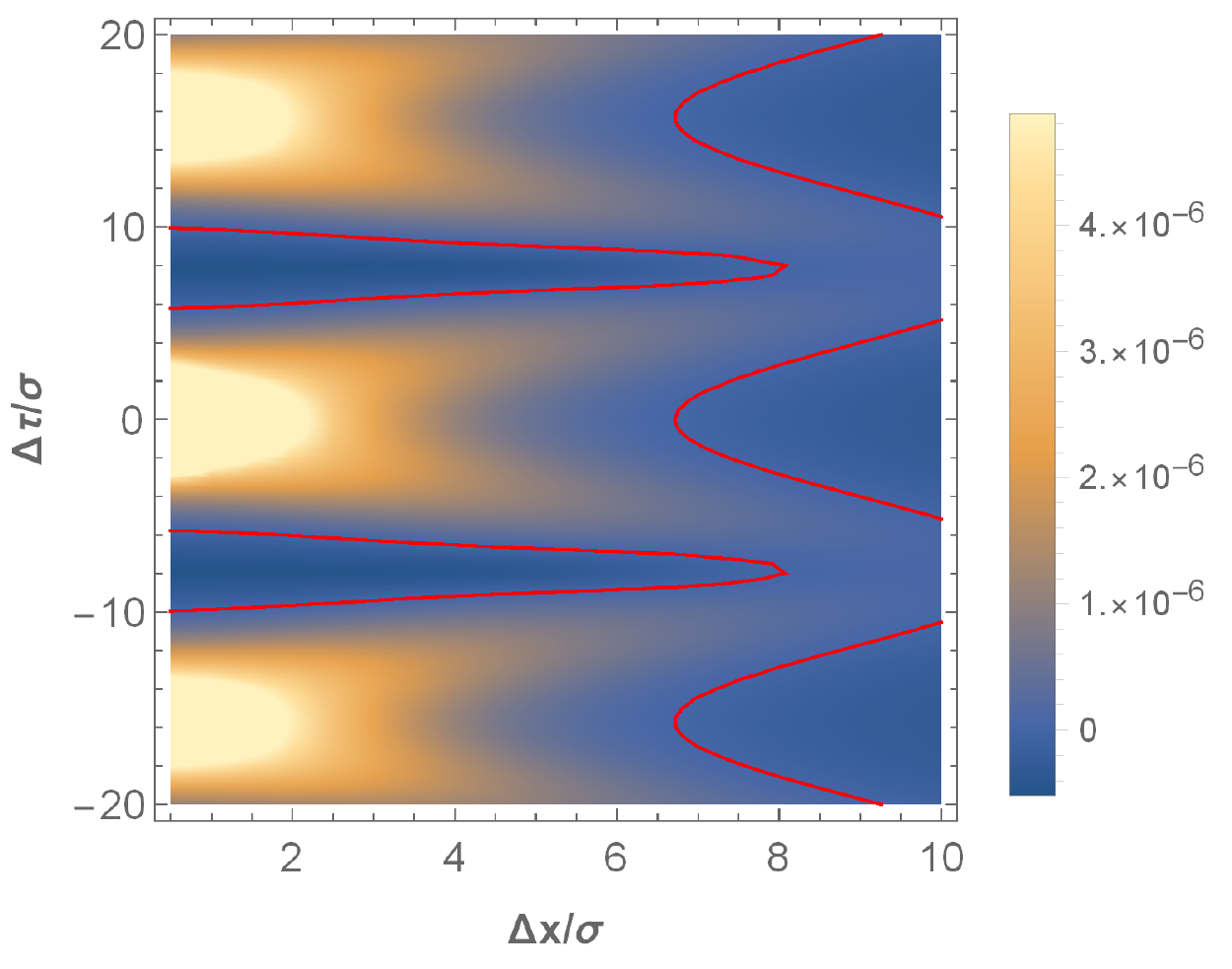}
        \label{n2_rt_L5E3_vonneumann_graph}
    \end{subfigure}
    \qquad
    \begin{subfigure}[hbtp]{0.3\textwidth}
        \caption{Dirichlet $I(\rho_{\textsc{ab}})$}
        \includegraphics[width=\textwidth]{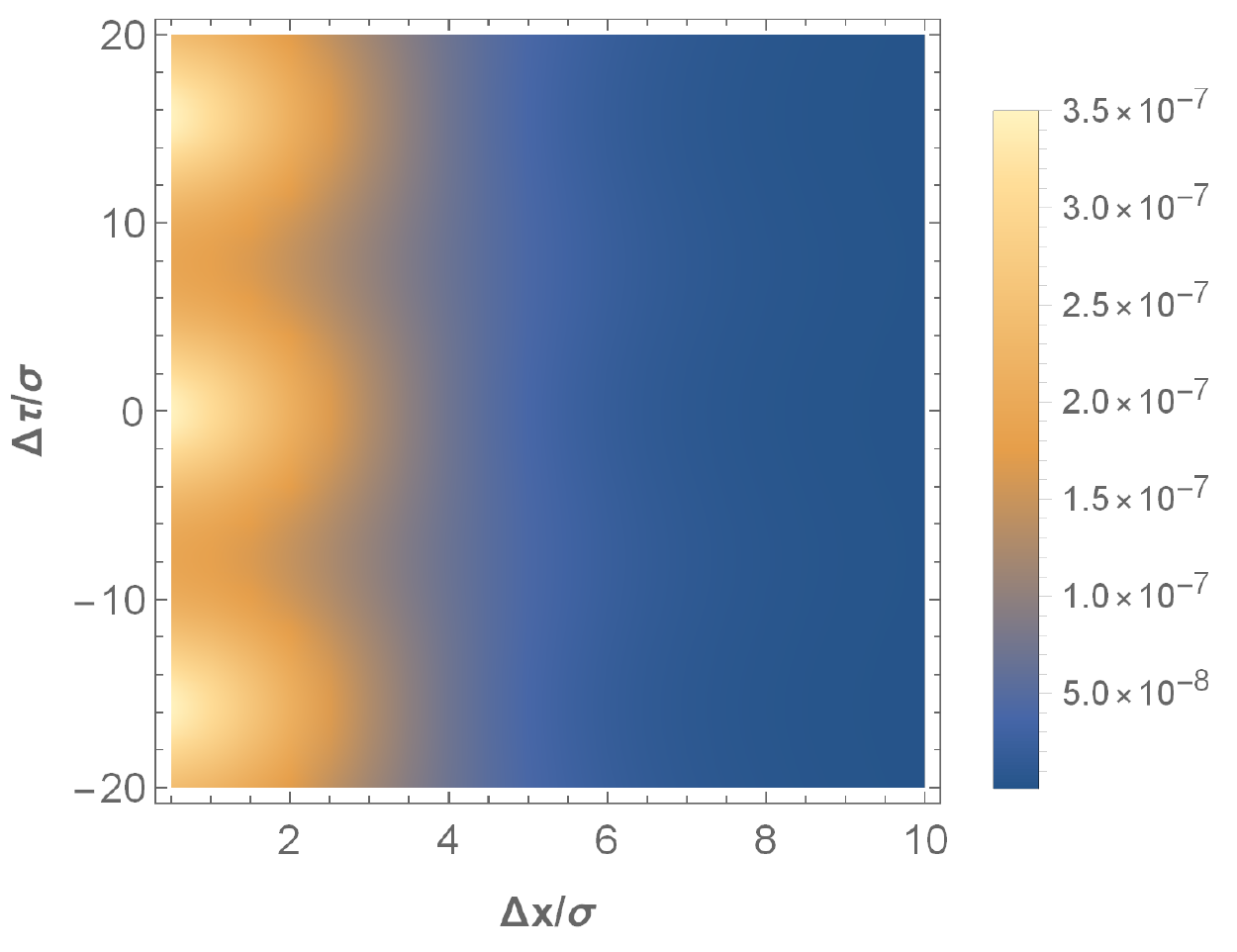}
        \label{Iab_rt_L5E3_dirichlet_graph}
    \end{subfigure}
    \qquad
    \begin{subfigure}[hbtp]{0.3\textwidth}
        \caption{Transparent $I(\rho_{\textsc{ab}})$}
        \includegraphics[width=\textwidth]{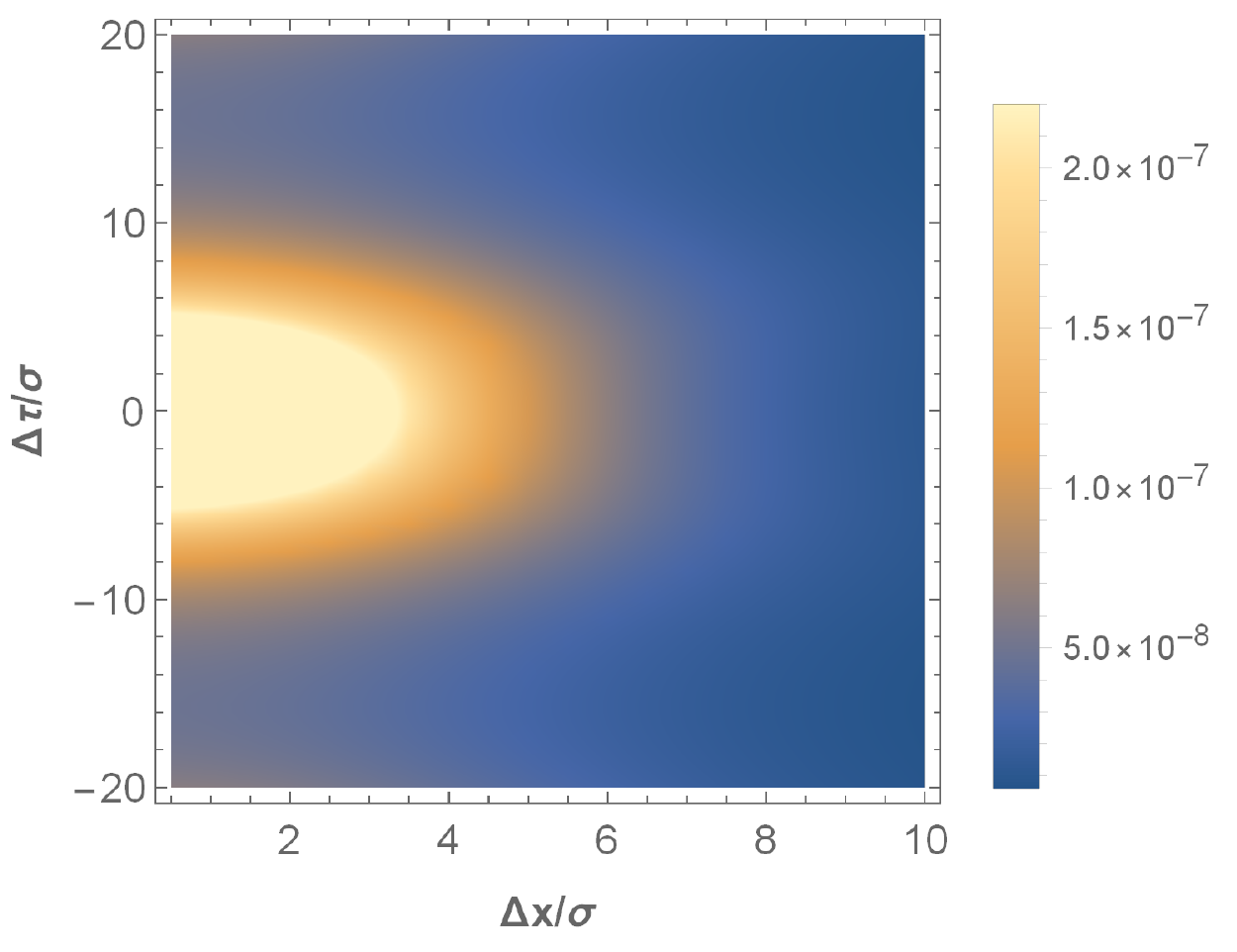}
        \label{Iab_rt_L5E3_transparent_graph}
    \end{subfigure}
    \qquad
    \begin{subfigure}[hbtp]{0.3\textwidth}
        \caption{Neumann $I(\rho_{\textsc{ab}})$}
        \includegraphics[width=\textwidth]{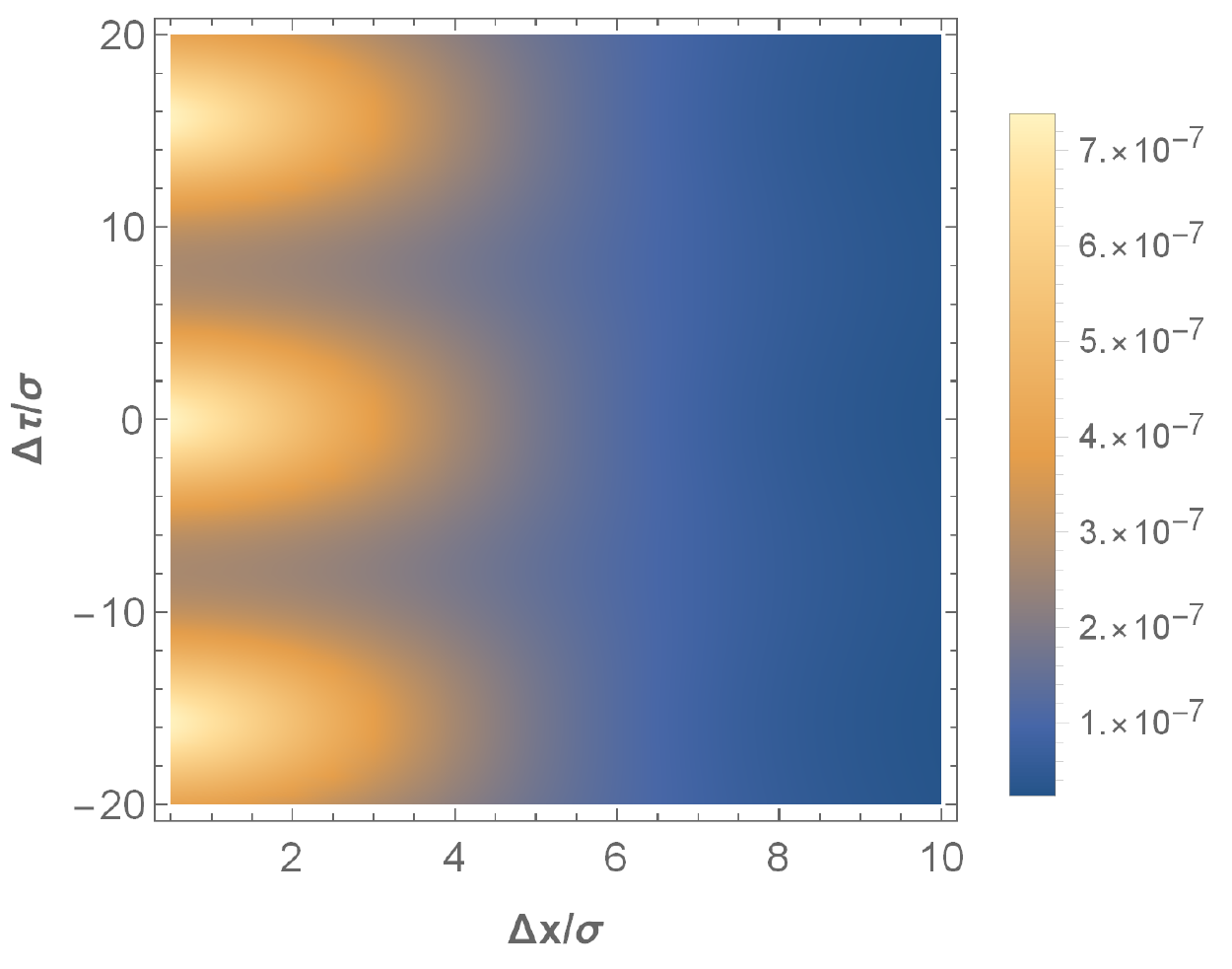}
        \label{Iab_rt_L5E3_vonneumann_graph}
    \end{subfigure}
    \caption{Negativity $\mathcal{N}^{(2)}$
    and mutual information $I(\rho_{AB})$ for geodesic detectors, one at the origin, as a function of separation in space and time. $L=5\sigma, \Omega=3/\sigma.$ Zero contour in red.}
    \label{n2_rt_L5E3_graph}
\end{figure*}

Figure \ref{n2_rt_L5E3_graph} reveals some interesting features of the negativity and mutual information, as a function of separation in time and space. One can see the light cones emanating from the central detector, reflecting off the conformal boundary;  this is highlighted in Fig. \ref{n2_rt_L5E3_dirichlet_graph}. It is also interesting to note the effects of changing the boundary conditions: notably, the von Neumann condition $\varepsilon=1$ in Fig. \ref{n2_rt_L5E3_vonneumann_graph} causes entanglement to vanish between the maxima on the $\Delta x=0$ line, while that does not happen for the transparent and Dirichlet conditions.
As one might expect, because our boundary conditions are reflective, the negativity is almost periodic. (Of course, the transparent case in Fig. \ref{n2_rt_L5E3_transparent_graph} does not have reflection off the boundary \textit{per se}, but it is still almost periodic, albeit with a period twice as long.)  While this neatly explains the behaviour of the light-cone part of the entanglement, this still leaves a small mystery: namely, why is the spacelike part $\mathcal M^+$ also affected? This is highlighted in Figs. \ref{Iab_rt_L5E3_dirichlet_graph}-\ref{Iab_rt_L5E3_vonneumann_graph}; while no trace of the light cones remains, the mutual information still appears sensitive to the boundary condition. While the commutator has an obvious dependence on $\varepsilon$, the dependence of $\mathcal{M}^+$ is a matter for further study.

\begin{figure*}[hbtp]
    \centering
    \begin{subfigure}[hbtp]{0.3\textwidth}
        \caption{Dirichlet boundary condition}
        \includegraphics[width=\textwidth]{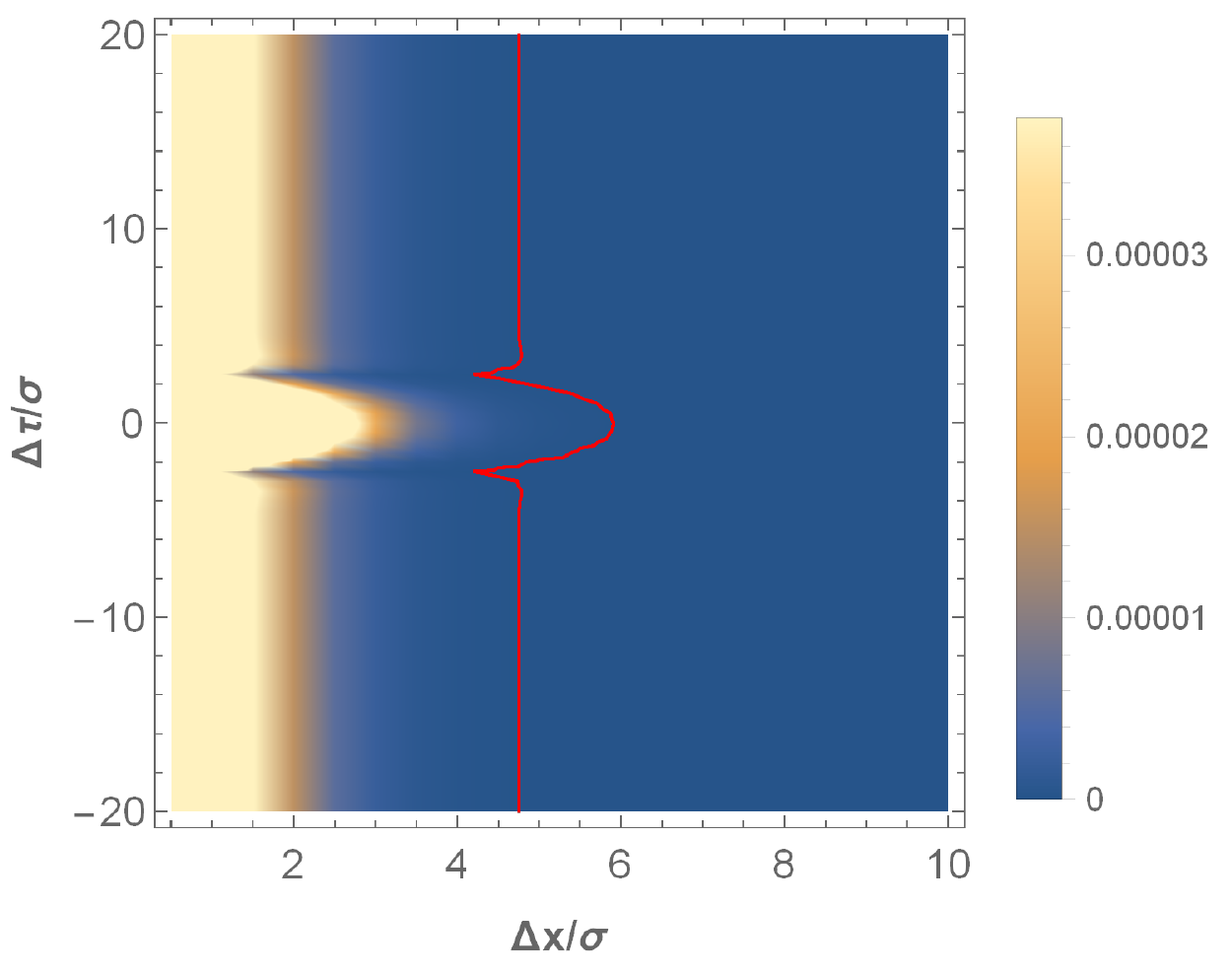}
        \label{n2_rt_L1E2_dirichlet_graph}
    \end{subfigure}
    \qquad
    \begin{subfigure}[hbtp]{0.3\textwidth}
        \caption{Transparent boundary condition}
        \includegraphics[width=\textwidth]{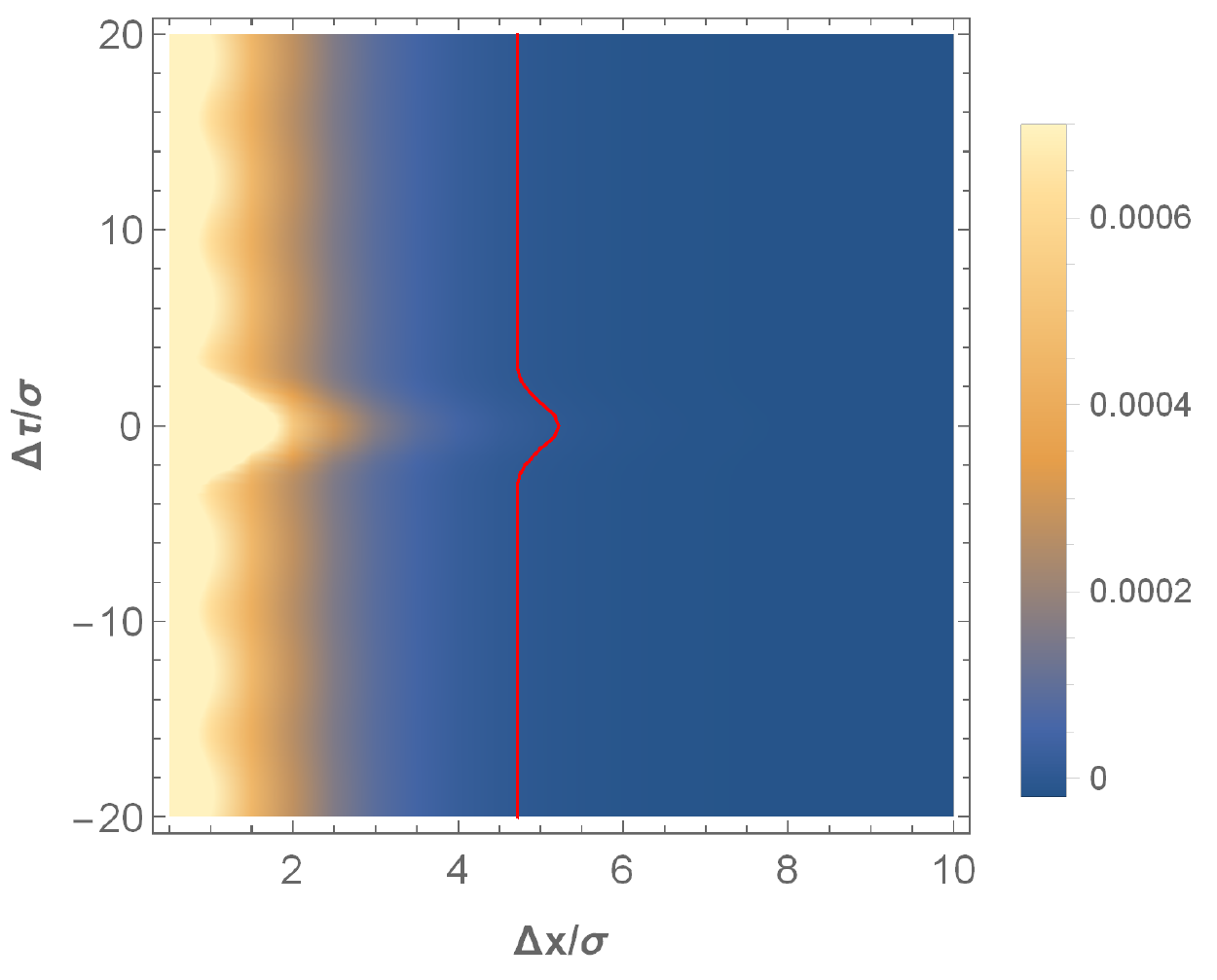}
        \label{n2_rt_L1E2_transparent_graph}
    \end{subfigure}
    \qquad
    \begin{subfigure}[hbtp]{0.3\textwidth}
        \caption{Neumann boundary condition}
        \includegraphics[width=\textwidth]{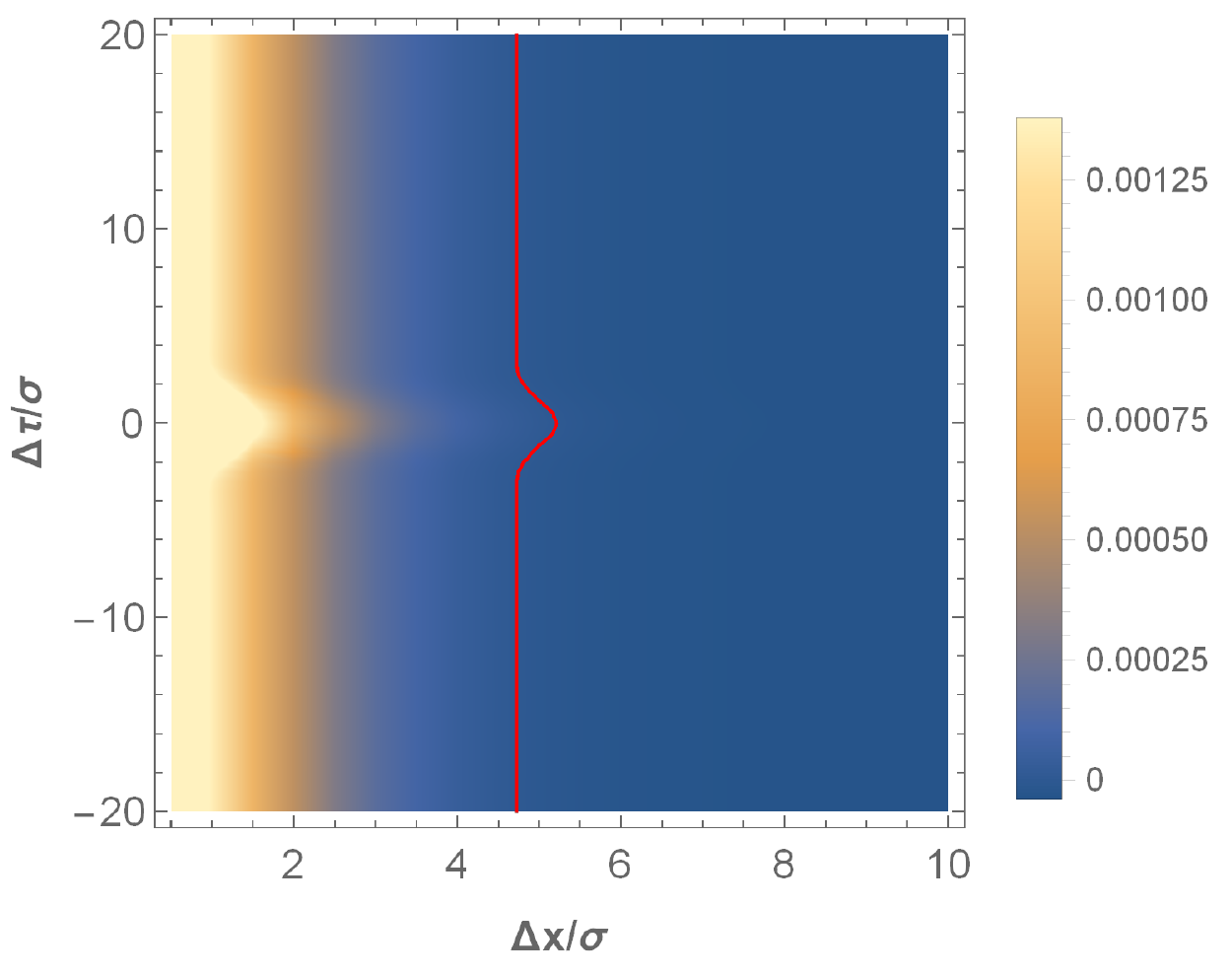}
        \label{n2_rt_L1E2_neumann_graph}
    \end{subfigure}
    
    \caption{Negativity for geodesic detectors, one at the origin, as a function of separation in space and time. $L=\sigma, \Omega=2/\sigma.$ }
    \label{n2_rt_L1E2_graph}
\end{figure*}

However, the apparent periodicity of $\mathcal N^{(2)}$ is not perfect: the maximum at $\Delta t = 0$ is different from the others. 
%\tcb{\bf[There's a way to support this algebraically, but it's a little muddy. It essentially hinges around how the Wightman function is periodic, and changes sign at each reflection, but `has its sign flipped' at t=0. Not sure if I should include that argument here...]} \tcr{\bf [If it can be done in a paragraph, please insert it.]}
This is more apparent if we pick a smaller curvature scale. Careful analysis shows that in this case, while the other peaks blur into a continuum, the central peak soars above them: Figure \ref{n2_rt_L1E2_graph} shows this in dramatic form. (Of course, by symmetry, $L_{II}$ is equal and constant for detectors A and B.) That said, in both cases, the response is almost periodic in a weaker sense: there is a component of the negativity that does \textit{not} vanish as time displacement approaches infinity.  This is contrary to the situation with cavities and is evidently  a symptom of the global recurrence time phenomenon, wherein all field operators of AdS are periodic. This remains true even if the coupling is minimal. The low-$L$ case also highlights the importance of the boundary condition for extreme AdS: while the overall structure of Figs. is broadly similar, the values of entanglement attained differ wildly, due to differences in how the various reflections of the commutator interact.
We also calculated the mutual information $I(\rho_{\textsc{ab}})$ for $L=\sigma, \Omega = 2/\sigma$. As one might have expected, without the `peakiness' from the commutator part, the result has no variation with respect to $\Delta t$; we have therefore chosen to omit these graphs.

\subsection{Static Entanglement Harvesting}

\begin{figure}[hbtp]
\centering
\begin{subfigure}[hbtp]{0.9\columnwidth}
\caption{$\mathcal{N}^{(2)}$}
\includegraphics[width=\textwidth]{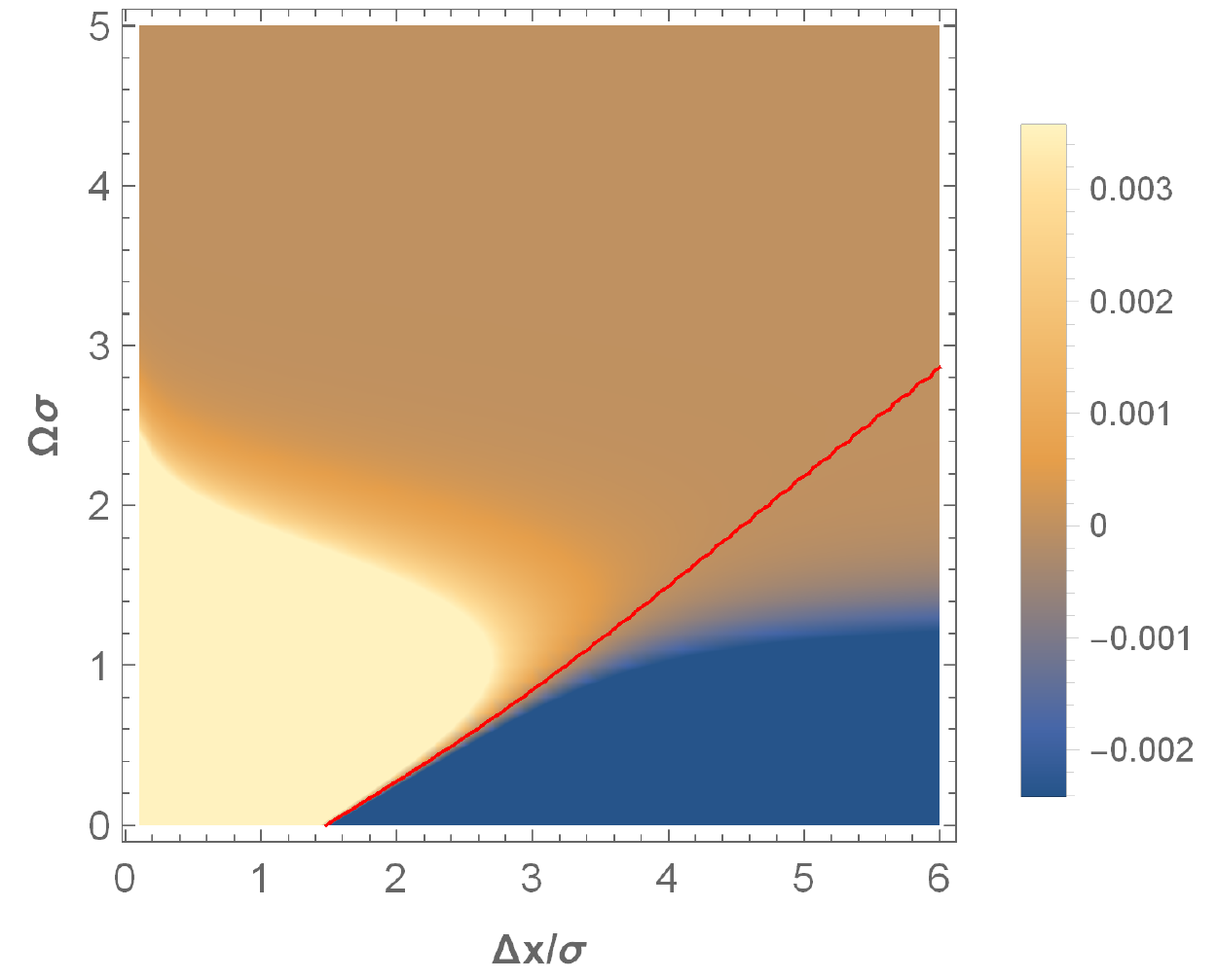}
\end{subfigure}\\
\begin{subfigure}[hbtp]{0.9\columnwidth}
\caption{$\mathcal{N}^{(2)}-\mathcal C_{\textsc{ab}}$}
\label{n2_minus_Cab_L5_static}
\includegraphics[width=\textwidth]{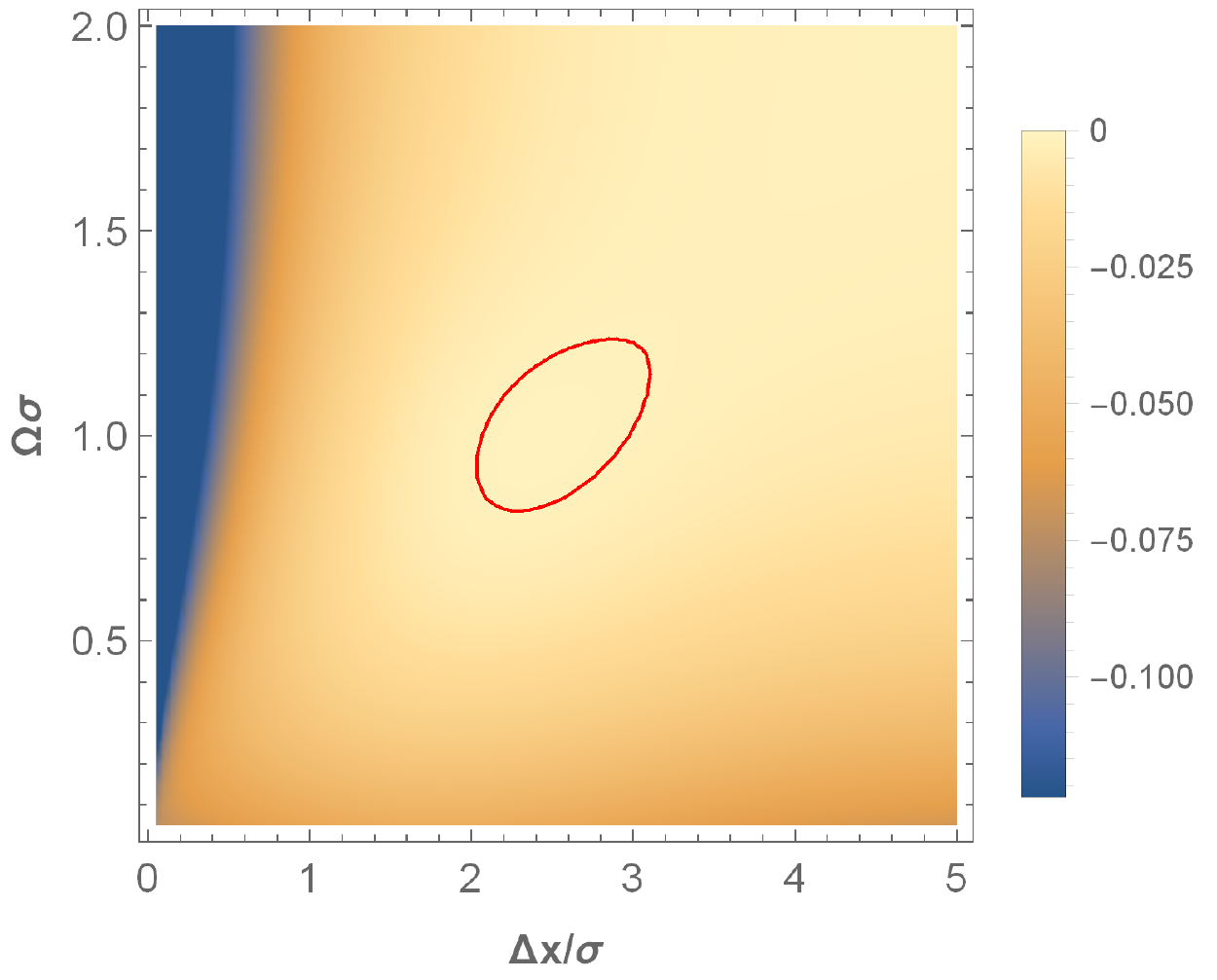}
\end{subfigure}
\caption{The negativity for a pair of static detectors separated by $\Delta x$ and having proper gap $\Omega,$ with AdS length $L=5\sigma.$ Note that coarseness of $\mathcal{N}^{(2)}$ appears to be an artifact of the plot grid. The zero contour is in red; we see it  bounds an island of spacelike entanglement.}
\label{n2_rgap_L5_static_graph}
\end{figure}

The static detector case has more interesting features. First, we plot the dependence of negativity on the proper gap of the detectors versus the proper separation, for curvature scales $L=5\sigma$ in
Fig. \ref{n2_rgap_L5_static_graph}, with qualitatively similar results (not illustrated) holding for
 $L=\sigma$. 
 We again find an island of spacelike entanglement,
 bounded by the red line in
 the plot of $\mathcal N^{(2)}-\mathcal C_{AB}$ shown in
 Fig. \ref{n2_minus_Cab_L5_static}.
 According to this estimator, ``spacelike" entanglement is only possible in a small parameter space, at a spatial separation of approximately $2.5\sigma;$ as before, even for very large $L$, this island does not significantly grow in size. Once again, whether this is the correct characterization of spacelike entanglement will need further study, especially in flat space.

\begin{figure*}[hbtp]
    \centering
    \begin{subfigure}[hbtp]{0.3\textwidth}
        \caption{Dirichlet boundary condition}
        \includegraphics[width=\textwidth]{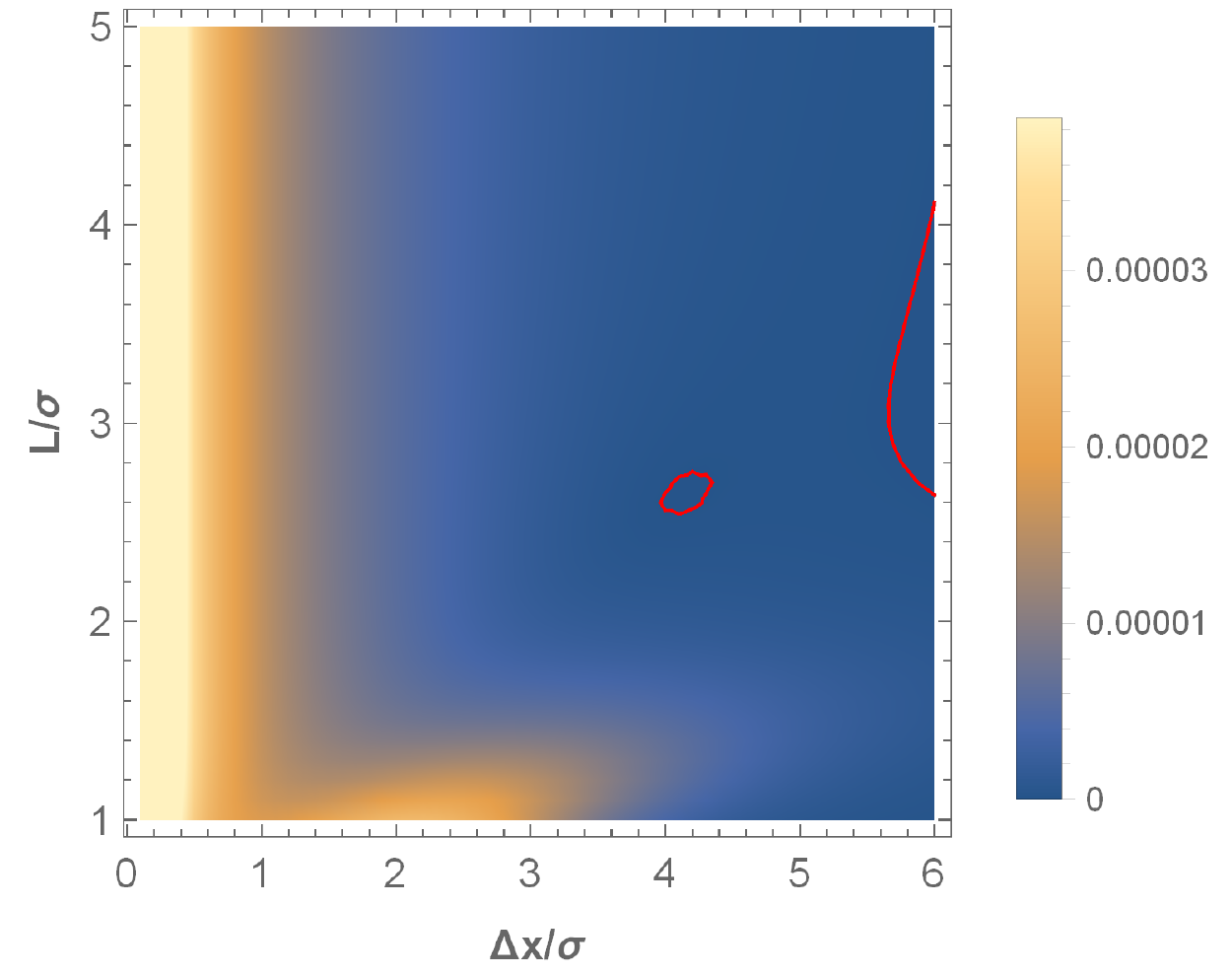}
    \end{subfigure}
    \qquad
    \begin{subfigure}[hbtp]{0.3\textwidth}
        \caption{Transparent boundary condition}
        \includegraphics[width=\textwidth]{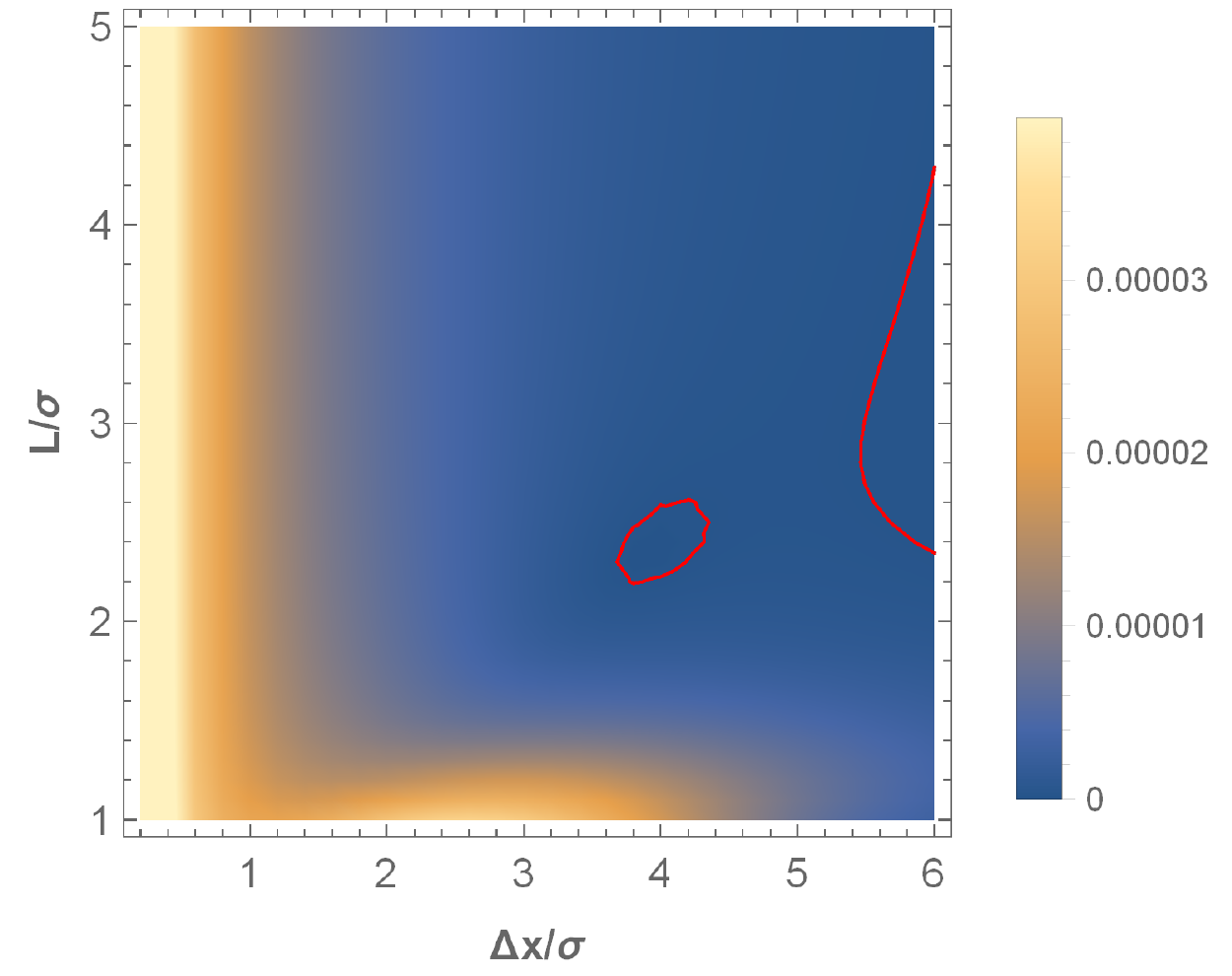}
    \end{subfigure}
    \qquad
    \begin{subfigure}[hbtp]{0.3\textwidth}
        \caption{Neumann boundary condition}
        \includegraphics[width=\textwidth]{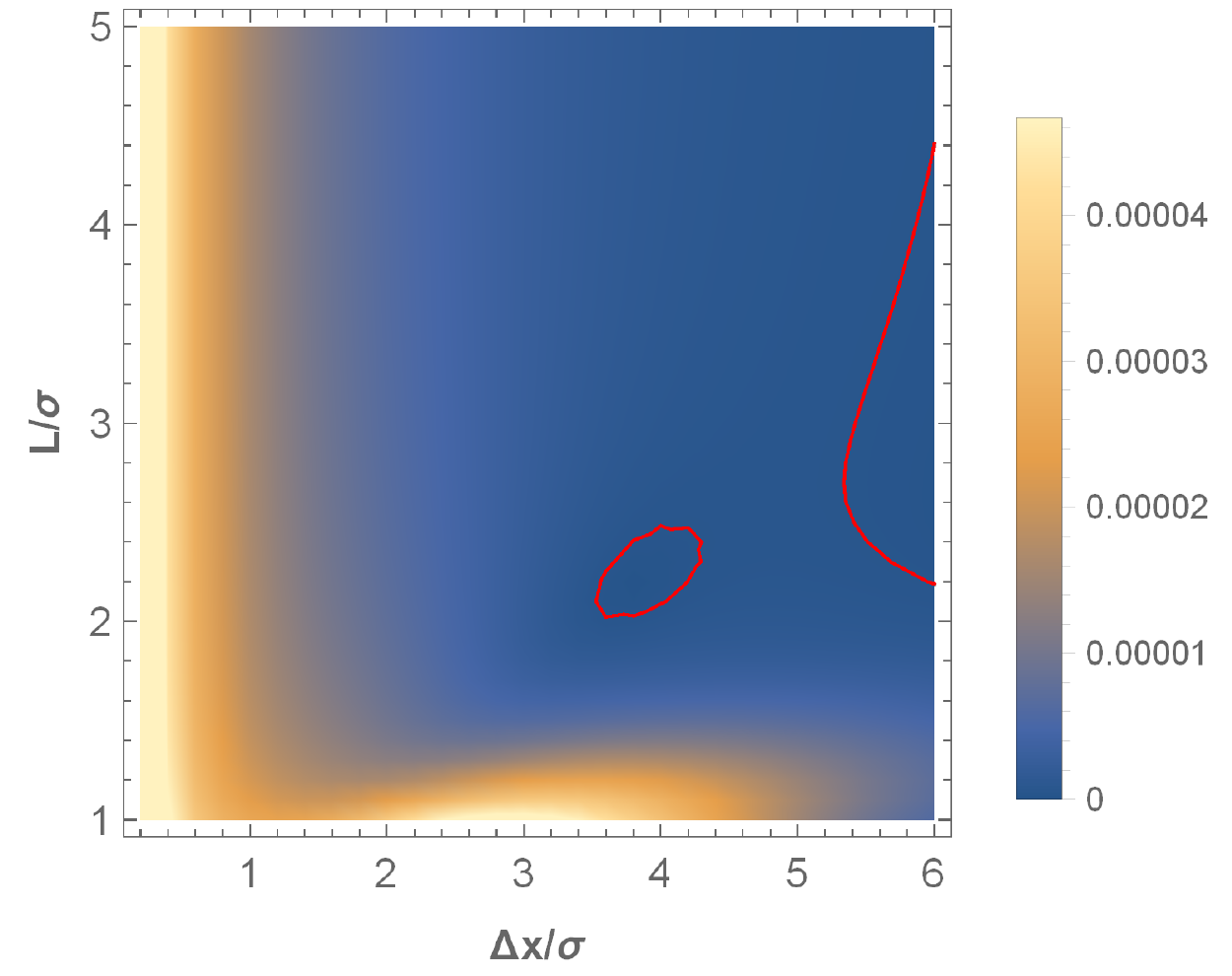}
    \end{subfigure}
    \caption{The negativity $\mathcal N^{(2)}$ for static detectors, one at the origin, as a function of curvature length $L$ and proper separation $\Delta x$.  Zero contours in red. Note the island of separability, in which there is no
    entanglement.}
    \label{n2_rL_static_graph}
\end{figure*}

Plotting next the dependence of negativity on the detector proper separation and the curvature scale, we note that in order to calculate $\mathcal L_{\textsc{bb}},$ larger separations require exponentially more resources in the static case.  While our results in Fig. \ref{n2_rL_static_graph}
are mostly as in the geodesic case, a truly bizarre feature appears: an island of separability, bounded by the red line in the figure.  All boundary conditions show this feature, albeit with slightly different sizes and positions of the island. (Notably, the Dirichlet condition has the smallest island.) However, there does not appear to be a resonance or any such numerical correspondence here.  This phenomenon is also not particular to AdS$_4$; a similar island has been 
shown to appear in AdS$_3$ in the same general region of parameter space \cite{HHMSZ}. The origin of this feature is a matter for further research. As well, the behaviour as $L$ becomes small is radically different; for sufficiently small $L$, there is a local maximum in negativity as a function of separation. Still, for even smaller $L$, we have found that the negativity does still vanish.

\begin{figure*}[hbtp]
    \centering
    \begin{subfigure}[hbtp]{0.3\textwidth}
        \caption{Dirichlet $\mathcal N^{(2)}$}
        \includegraphics[width=\textwidth]{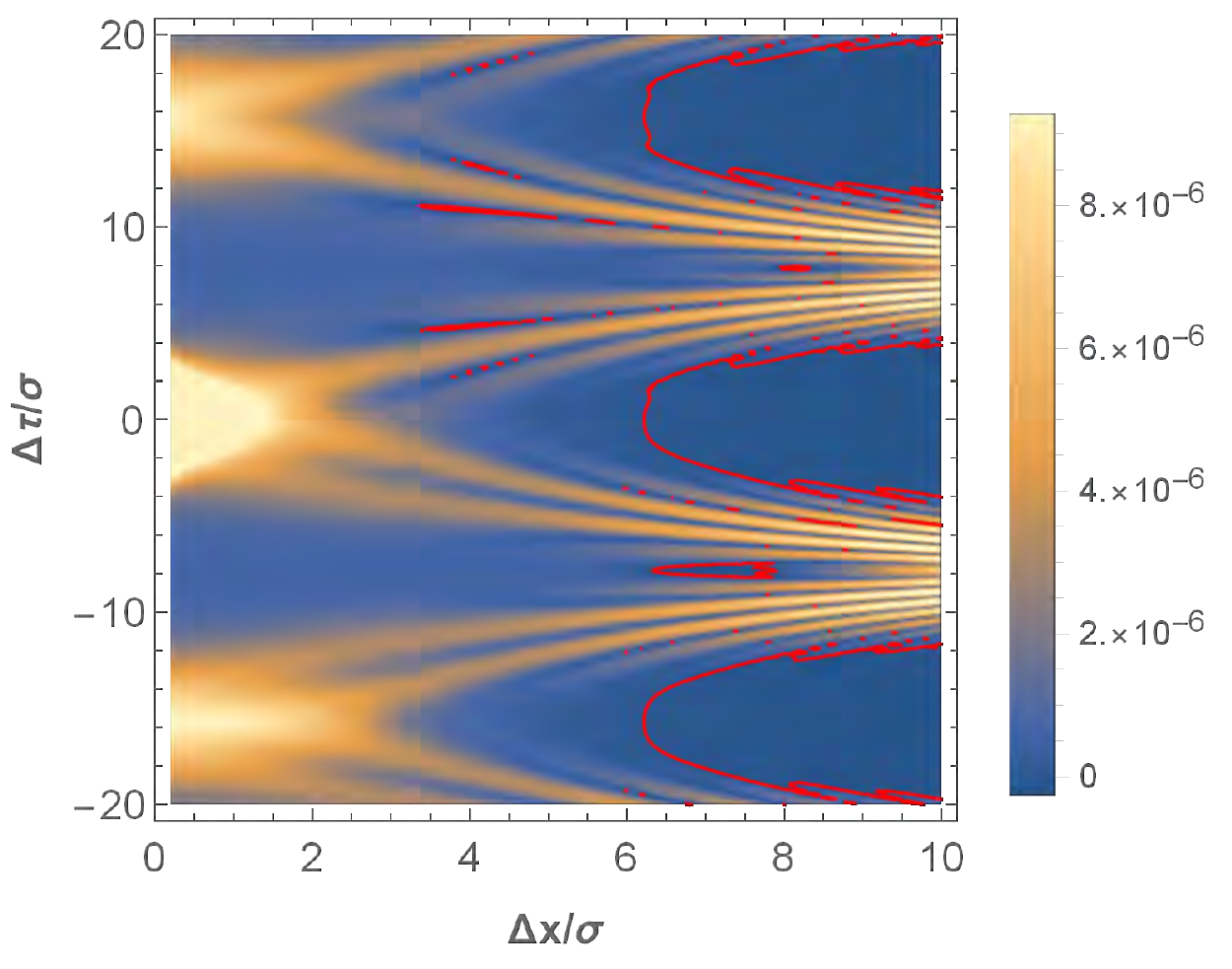}
        \label{n2_rt_L5E3_static_dirichlet_graph}
    \end{subfigure}
    \qquad
    \begin{subfigure}[hbtp]{0.3\textwidth}
        \caption{Transparent $\mathcal N^{(2)}$}
        \includegraphics[width=\textwidth]{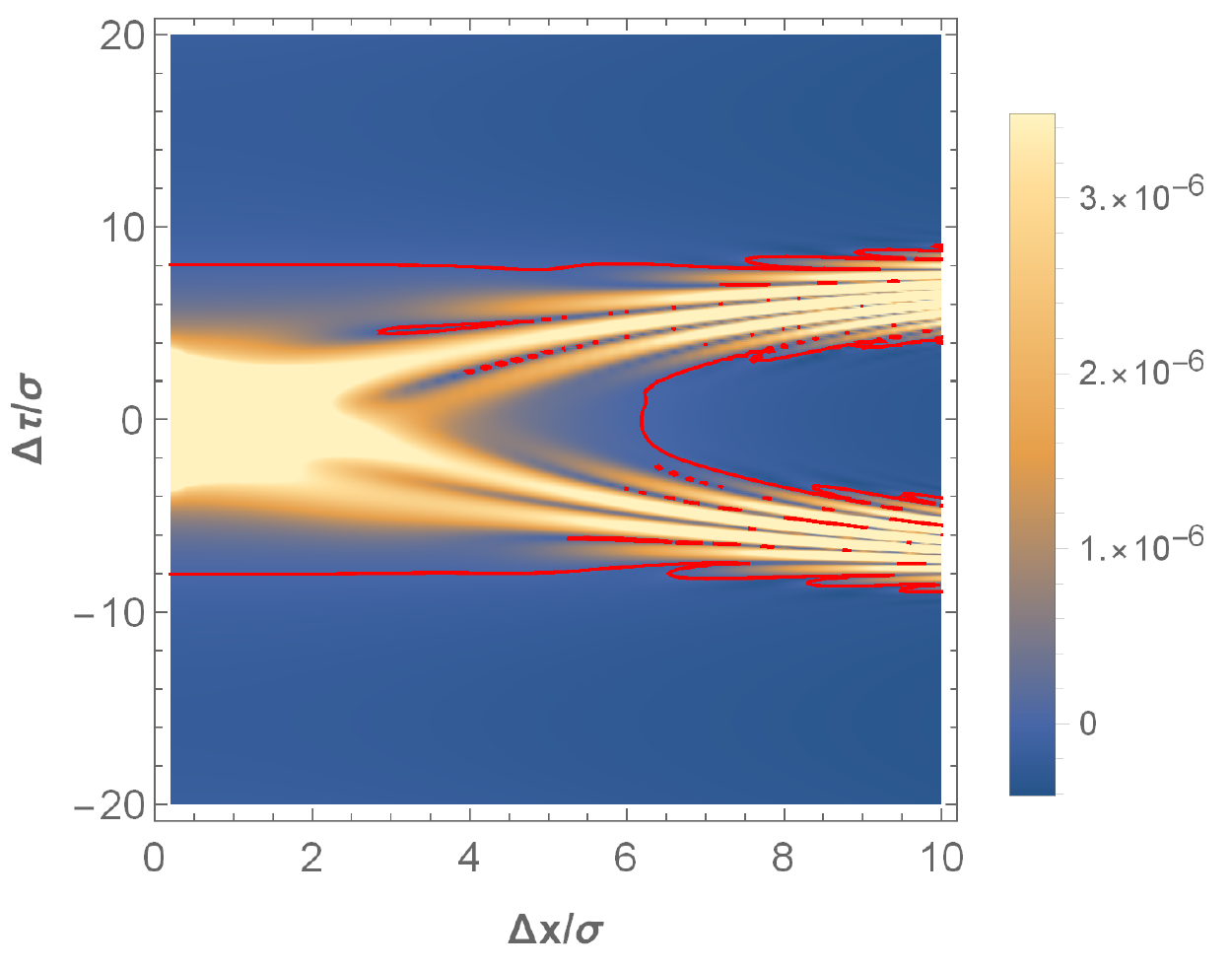}
        \label{n2_rt_L5E3_static_transparent_graph}
    \end{subfigure}
    \qquad
    \begin{subfigure}[hbtp]{0.3\textwidth}
        \caption{Neumann  $\mathcal N^{(2)}$}
        \includegraphics[width=\textwidth]{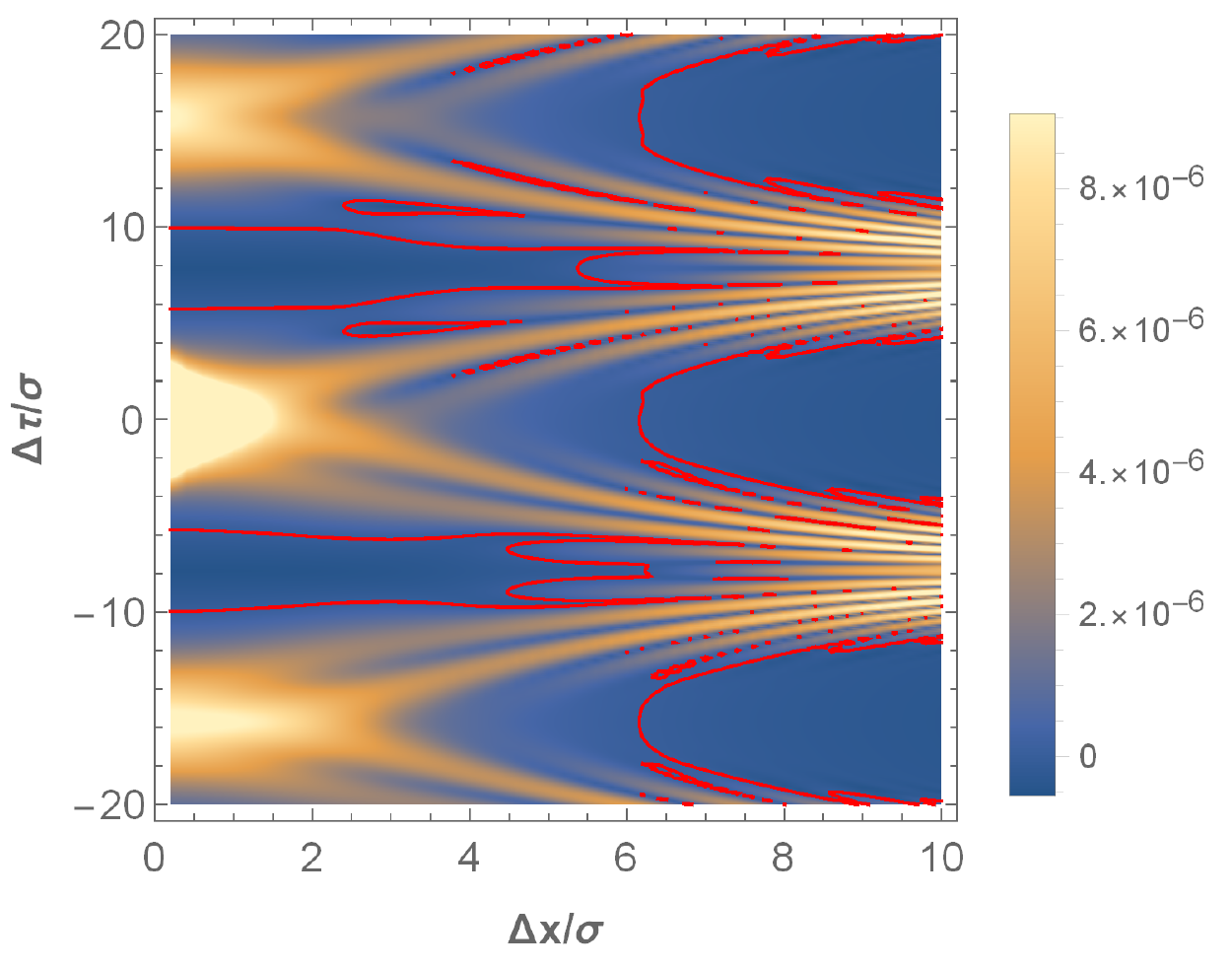}
        \label{n2_rt_L5E3_static_neumann_graph}
    \end{subfigure}
    \begin{subfigure}[hbtp]{0.3\textwidth}
        \caption{Dirichlet $I(\rho_{\textsc{ab}})$}
        \includegraphics[width=\textwidth]{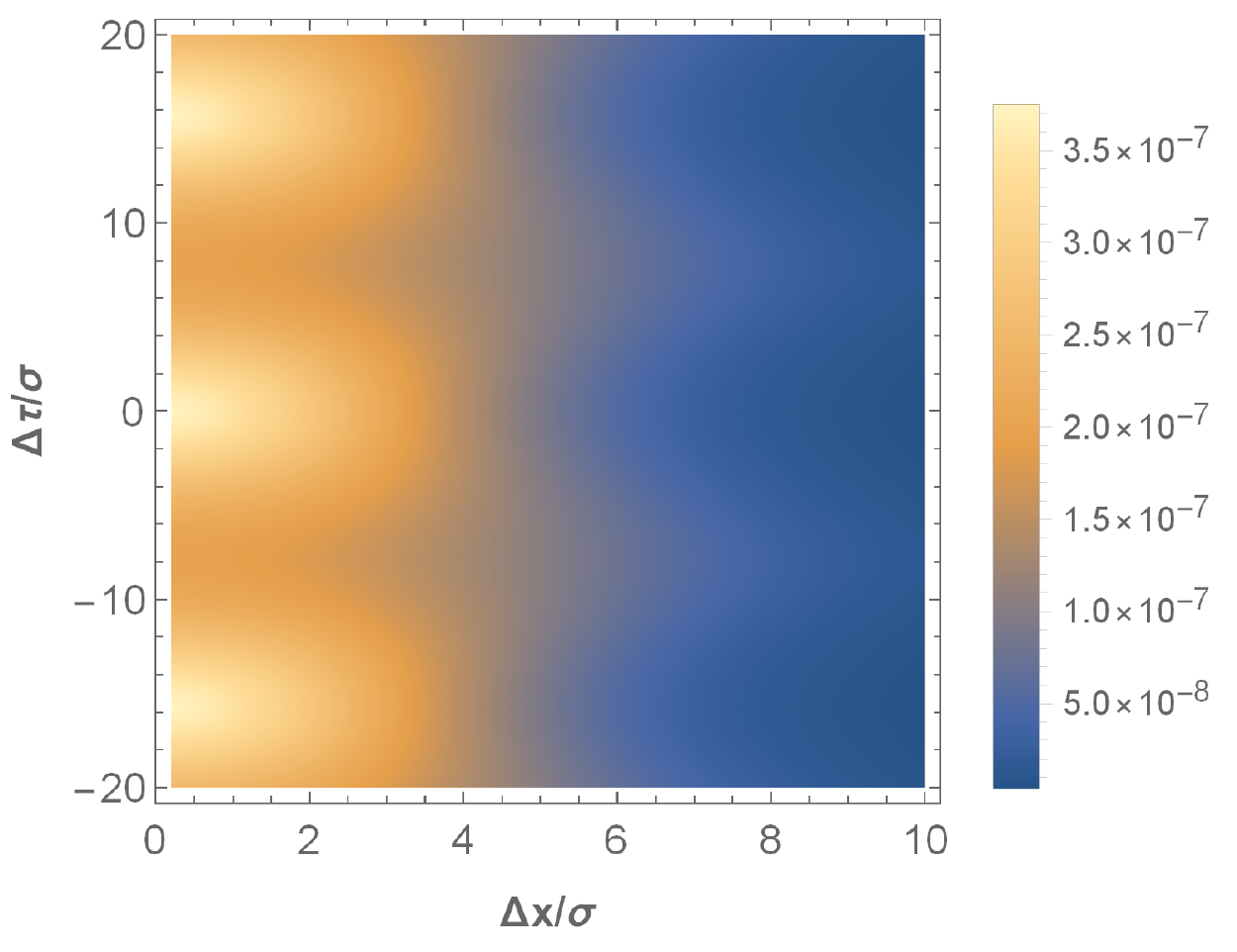}
        \label{Iab_rt_L5E3_static_dirichlet_graph}
    \end{subfigure}
    \qquad
    \begin{subfigure}[hbtp]{0.3\textwidth}
        \caption{Transparent $I(\rho_{\textsc{ab}})$}
        \includegraphics[width=\textwidth]{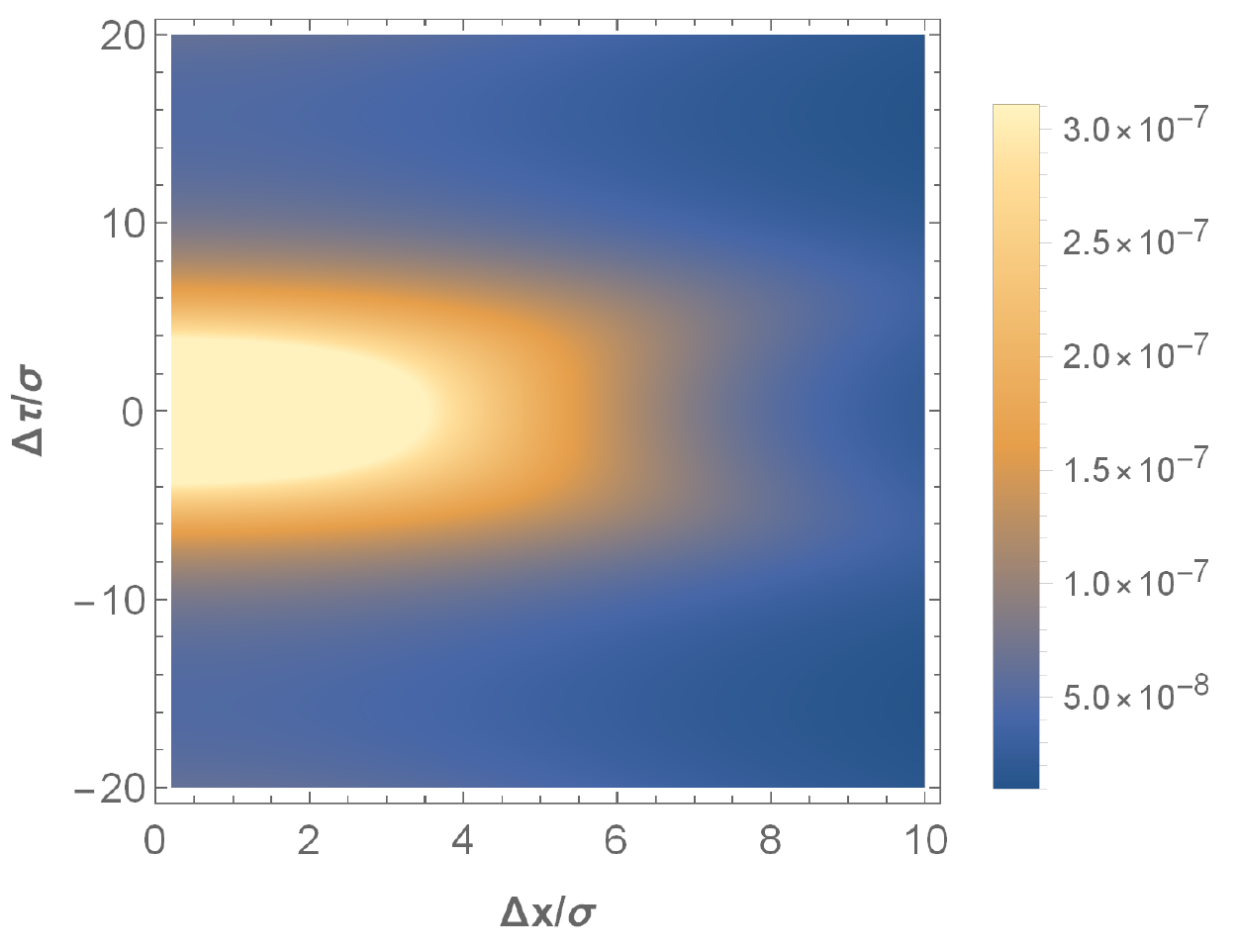}
        \label{Iab_rt_L5E3_static_transparent_graph}
    \end{subfigure}
    \qquad
    \begin{subfigure}[hbtp]{0.3\textwidth}
        \caption{Neumann $I(\rho_{\textsc{ab}})$}
        \includegraphics[width=\textwidth]{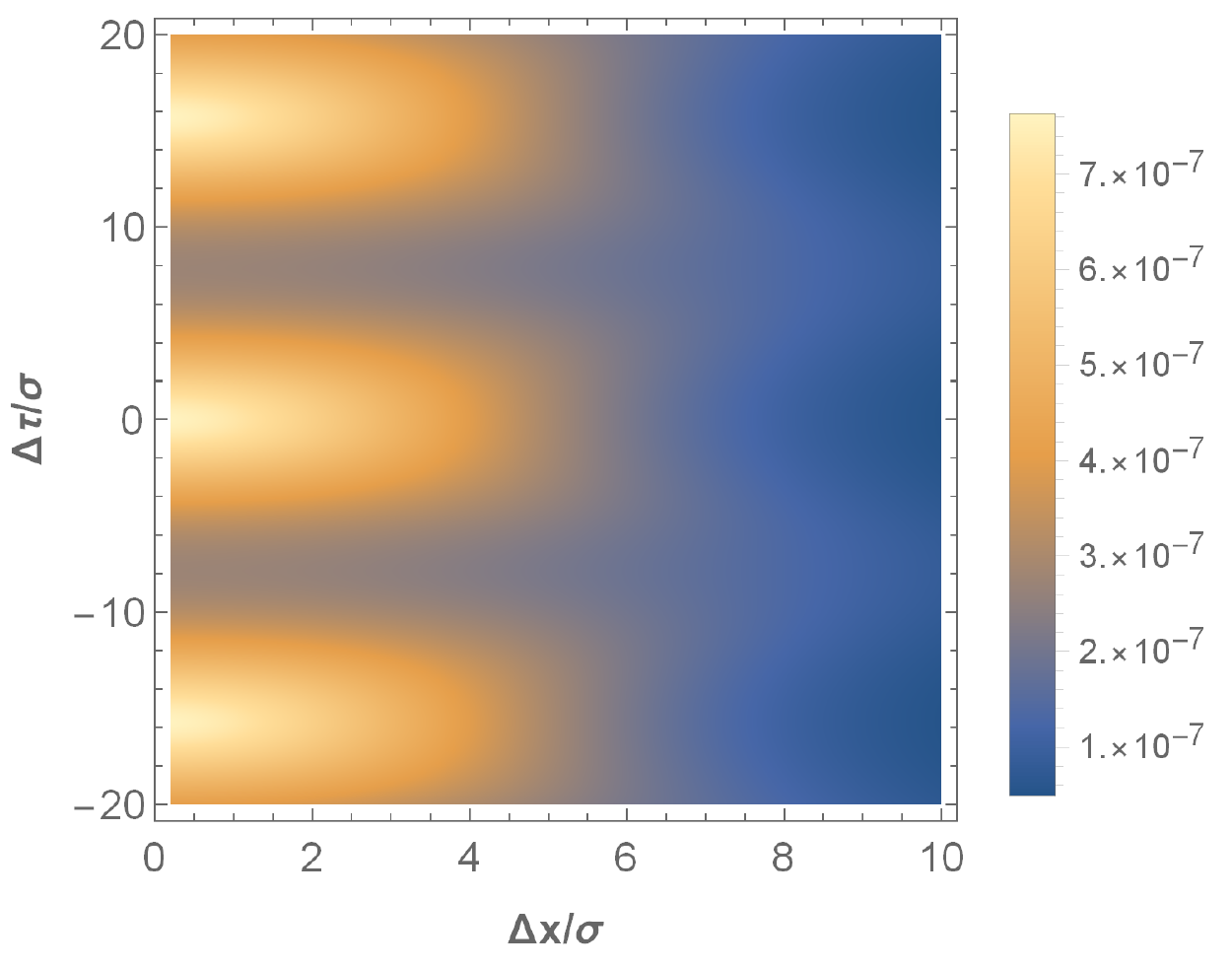}
        \label{Iab_rt_L5E3_static_neumann_graph}
    \end{subfigure}
    \caption{The negativity and mutual information for static detectors, one at the origin, as a function of separation in space and time. $L=5\sigma, \Omega=3/\sigma$.}
    \label{n2_rt_L5E3_static_Graph}
\end{figure*}

Next we plot in Fig. \ref{n2_rt_L5E3_static_Graph} the negativity  and the mutual information against the spatial and temporal separations of the detectors, for parameters $L=5\sigma, \Omega=3/\sigma.$ The picture here is rather more complicated. The far region where entanglement vanishes is reduced to a tiny dot compared to the geodesic case. However, a grid of peaks and troughs also appears in the negativity. In fact, careful analysis shows that there are tiny pockets where entanglement extinction occurs. These features appear to be the result of interference between the $\mathcal{M}^+$ and $\mathcal{M}^-$ terms, especially since no such feature is inherent to either individually,  nor in the mutual information.; however, there does not appear to be a simple way to predict their relative phase.

\begin{figure*}[hbtp]
    \centering
    \begin{subfigure}[hbtp]{0.3\textwidth}
        \caption{Dirichlet $\mathcal N^{(2)}$}
        \includegraphics[width=\textwidth]{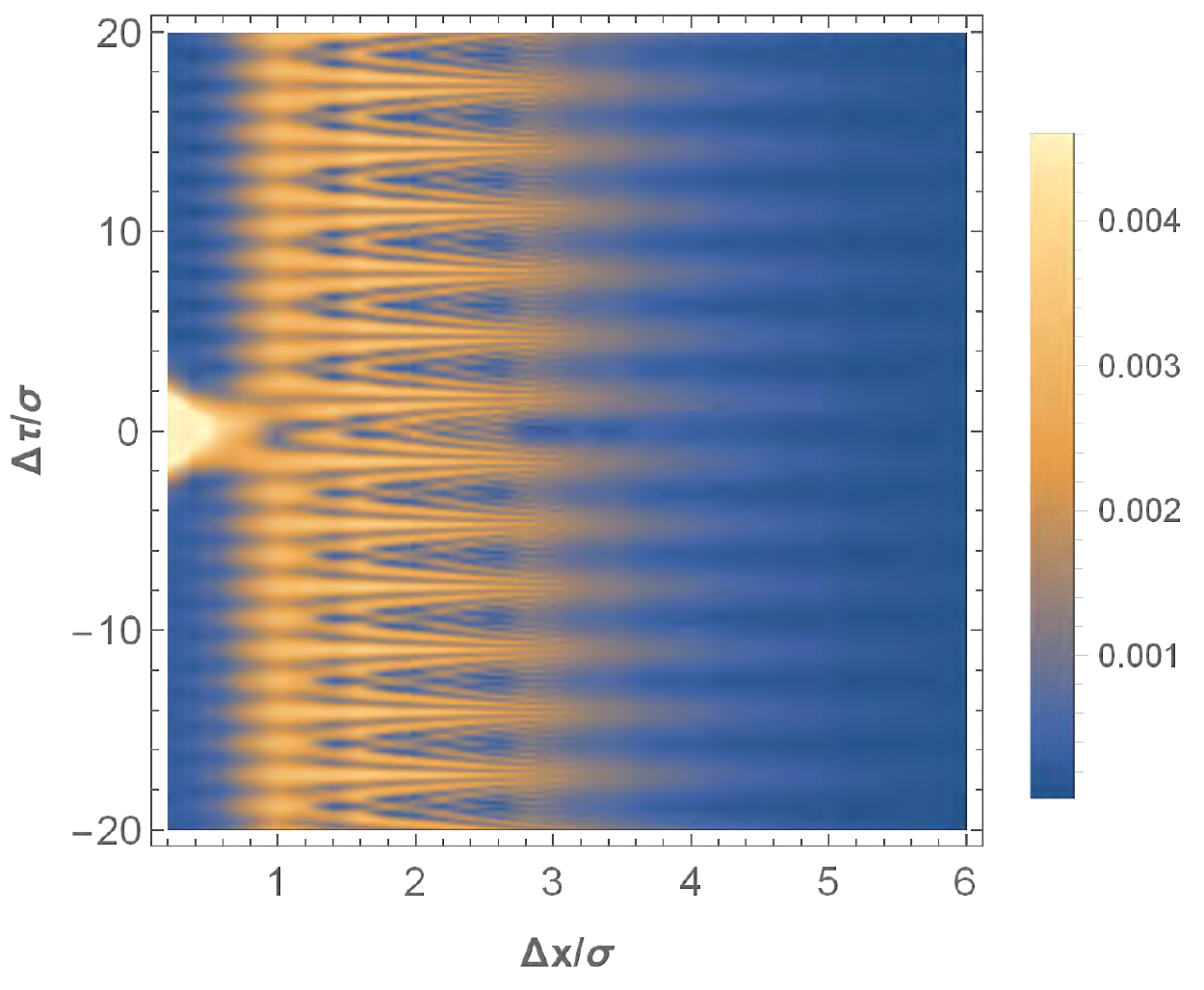}
        \label{n2_rt_L1E2_static_dirichlet_graph}
    \end{subfigure}
    \qquad
    \begin{subfigure}[hbtp]{0.3\textwidth}
        \caption{Transparent $\mathcal N^{(2)}$}
        \includegraphics[width=\textwidth]{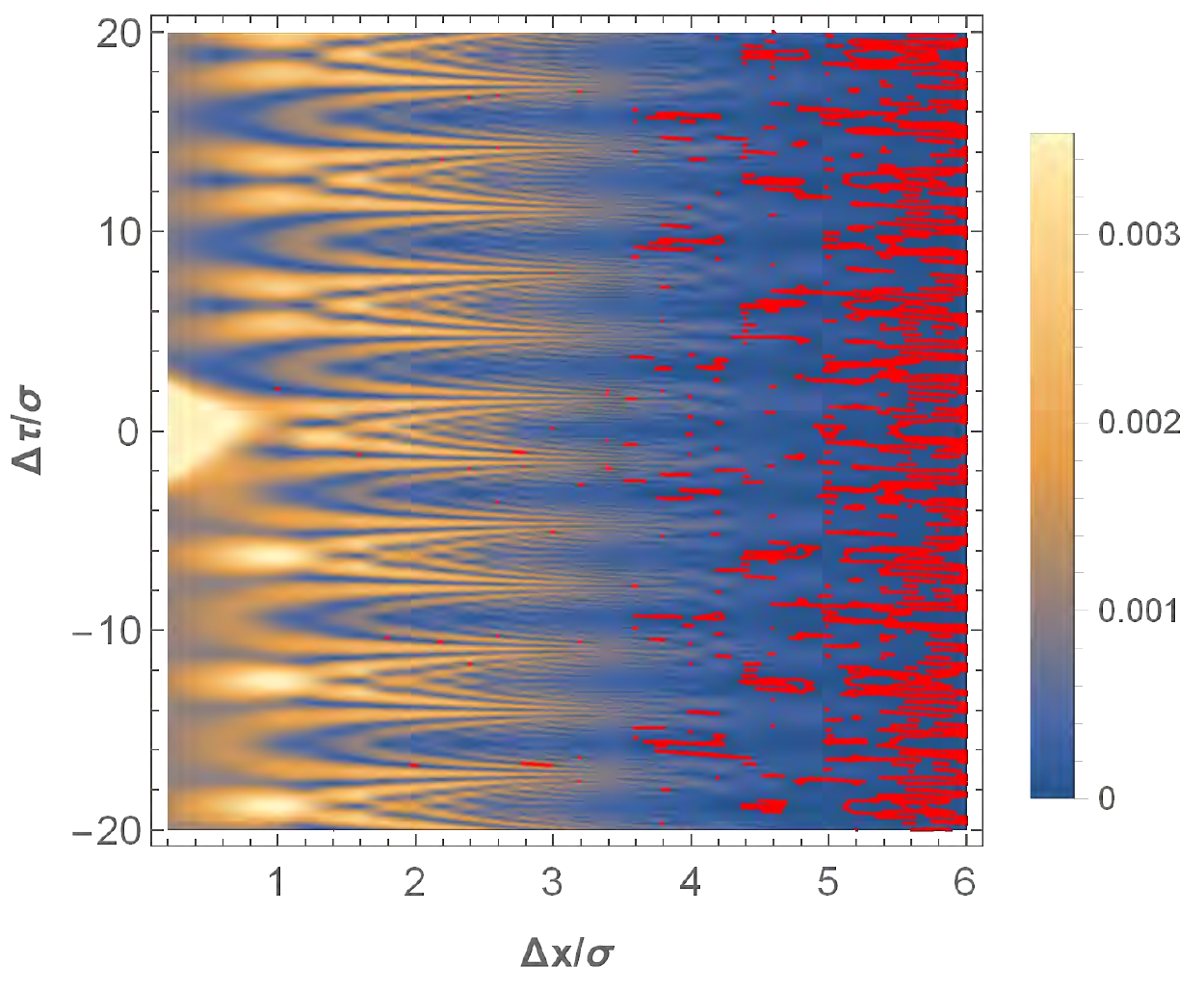}
        \label{n2_rt_L1E2_static_transparent_graph}
    \end{subfigure}
    \qquad
    \begin{subfigure}[hbtp]{0.3\textwidth}
        \caption{Neumann $\mathcal N^{(2)}$}
        \includegraphics[width=\textwidth]{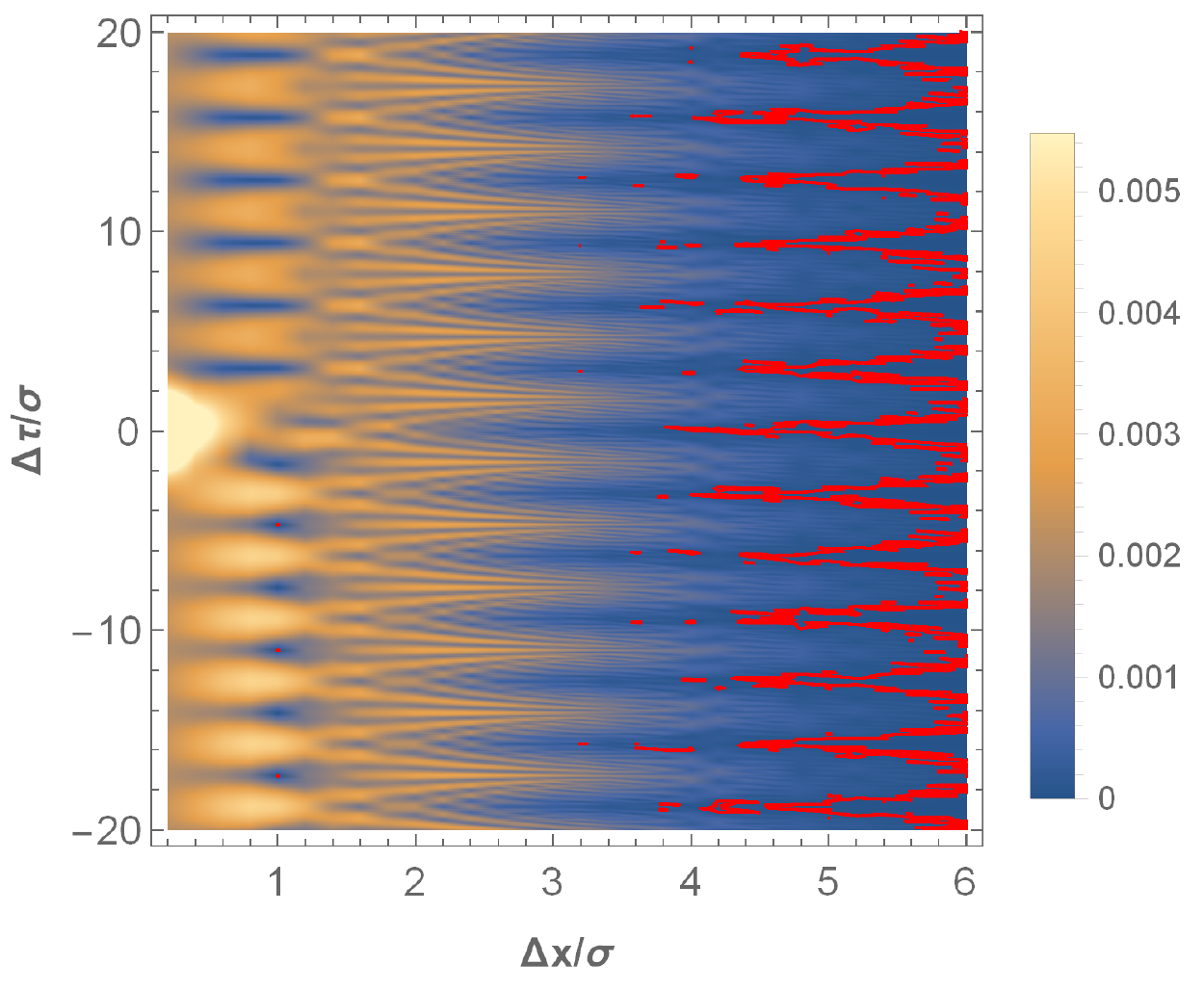}
        \label{n2_rt_L1E2_static_neumann_graph}
    \end{subfigure}
    \begin{subfigure}[hbtp]{0.3\textwidth}
        \caption{Dirichlet $I(\rho_{\textsc{ab}})$}
        \includegraphics[width=\textwidth]{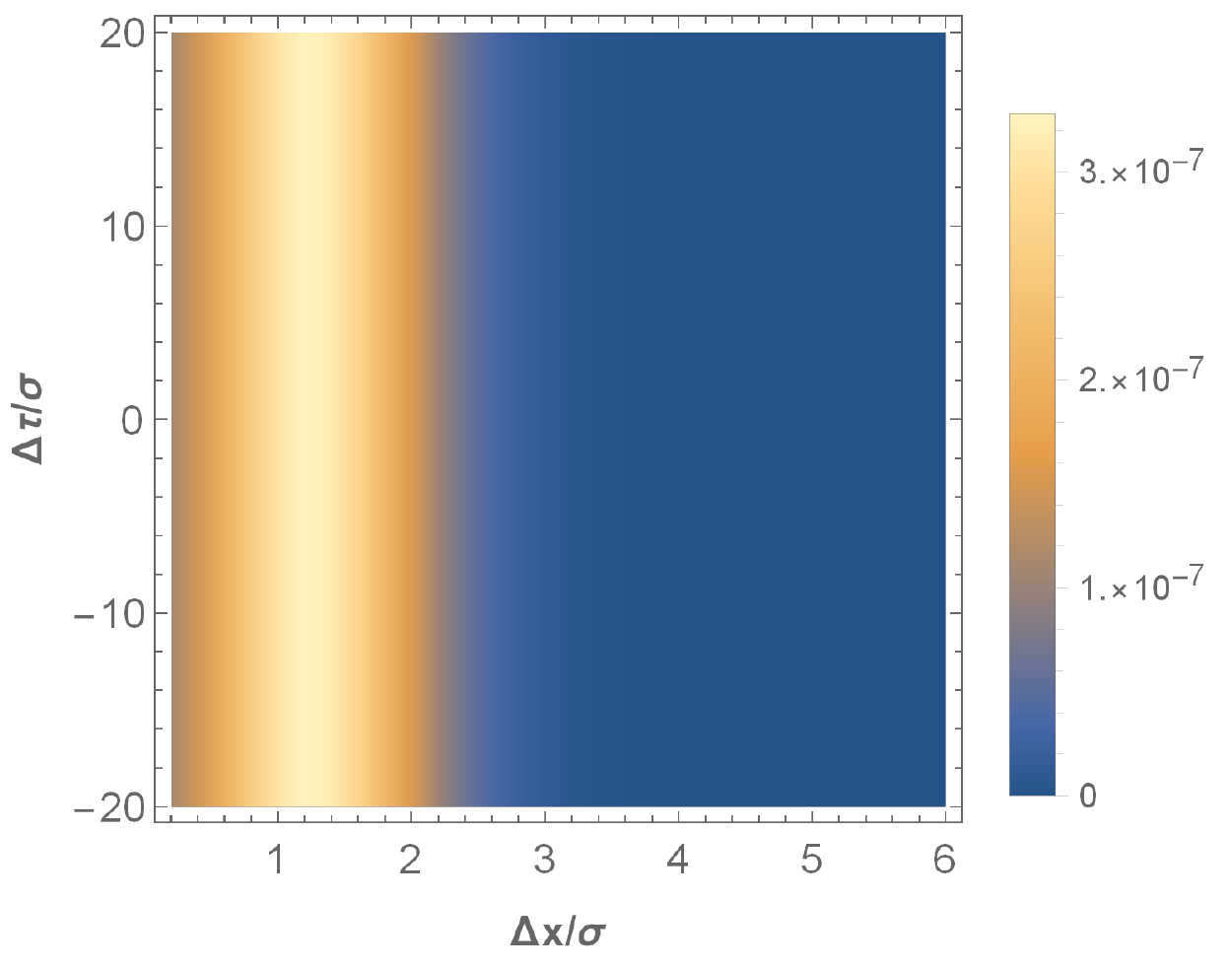}
        \label{Iab_rt_L1E2_static_dirichlet_graph}
    \end{subfigure}
    \qquad
    \begin{subfigure}[hbtp]{0.3\textwidth}
        \caption{Transparent $I(\rho_{\textsc{ab}})$}
        \includegraphics[width=\textwidth]{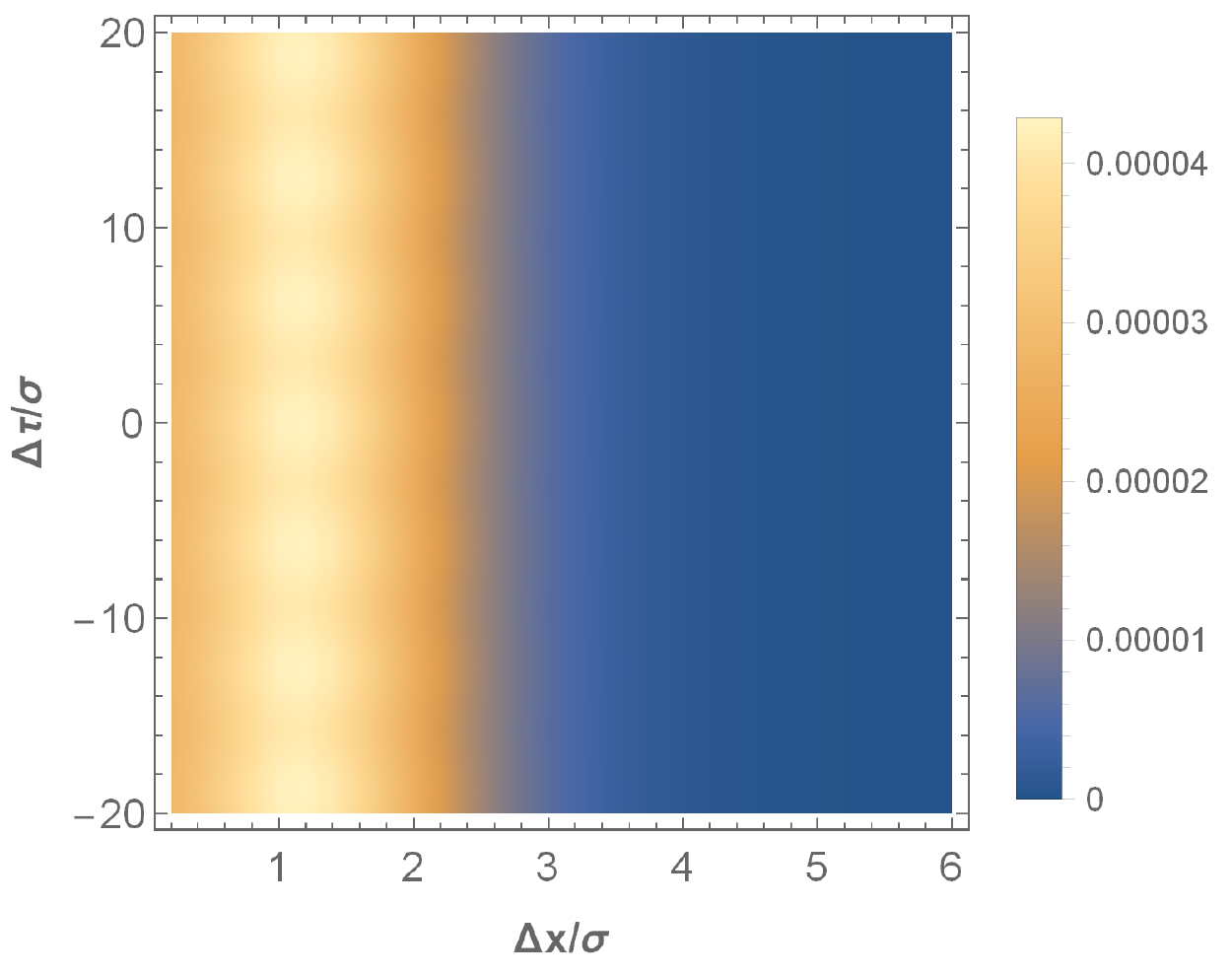}
        \label{Iab_rt_L1E2_static_transparent_graph}
    \end{subfigure}
    \qquad
    \begin{subfigure}[hbtp]{0.3\textwidth}
        \caption{Neumann $I(\rho_{\textsc{ab}})$}
        \includegraphics[width=\textwidth]{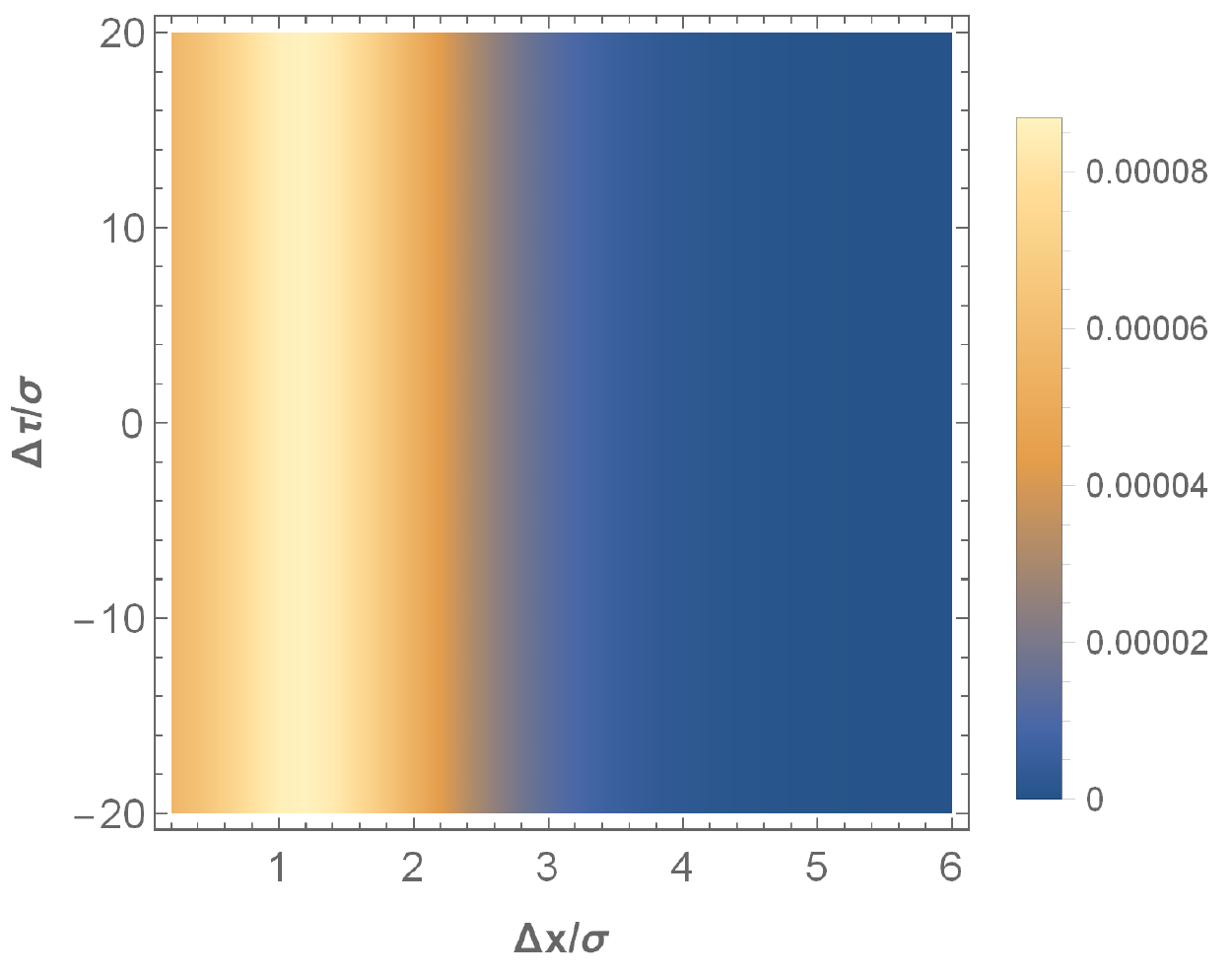}
        \label{Iab_rt_L1E2_static_neumann_graph}
    \end{subfigure}
    \caption{The negativity and mutual information for static detectors, one at the origin, as a function of separation in space and time. $L=\sigma, \Omega=2/\sigma$. Zero contours in red. }
    \label{n2_rt_L1E2_static_graph}
\end{figure*}

The negativity for $L=\sigma, \Omega=2/\sigma$ in Figs. \ref{n2_rt_L1E2_static_dirichlet_graph}-\ref{n2_rt_L1E2_static_neumann_graph} is broadly similar: entanglement reaches further than in the geodesic case, but a puzzling series of waves also appears. This particular choice of parameters  has much smaller regions of zero entanglement in the waves, however. Even more interesting is the fact that as a result of the entanglement, there is a range of separations for which entanglement is stronger than in the limit $\Delta x \rightarrow 0$. While the existence of this latter fact appears to be related to the smaller curvature scale, and the difference in redshifts, our understanding of this feature is once again limited.

We also note a degree of time-asymmetry in the negativity for all three boundary conditions, slightly more exaggerated for Dirichlet. This effect, while small, is indeed present.  The same effect appears in AdS$_3$ \cite{HHMSZ}, where it is considerably more pronounced, with much larger concurrence for $\Delta t > 0$, when detector $A$ switches first.  In both dimensions the effect vanishes for large AdS length.  We suspect that the diminished asymmetry in AdS$_4$ is due to the effect of the Huygens principle.

Surprisingly, the mutual information, shown in Figs. \ref{Iab_rt_L5E3_static_dirichlet_graph}-\ref{Iab_rt_L5E3_static_neumann_graph} and \ref{Iab_rt_L1E2_static_dirichlet_graph}-\ref{Iab_rt_L1E2_static_neumann_graph}, is quite similar to the geodesic case. It appears that the entanglement contains far more information than the mutual information does. Notably, there is still an apparent violation of what one might call ``time reversal symmetry": the quantity of mutual information depends on which detector switches on first. Of course, in this context, that may not be surprising, since the two detector configurations are quite distinct. However, one mystery is still present: evidently, in the static case, there is a finite non-zero distance at which mutual information is maximized. We believe this is another consequence of the different redshifts; however, there is no corresponding analysis in Minkowski space to which we can compare.

To summarize this section, the static detector case is much richer in features than the geodesic detector case. We hypothesize  that the ``waves" are a consequence of the detector gaps being unequal with respect to coordinate time; this means that the relative phase of the components $\mathcal M^\pm$ evolves with coordinate time. However, more study is required to come to this conclusion. In particular, even in the flat case, information about the unequal detector gap case is limited. As well, the other known cases of static detector entanglement in curved space are not immediately comparable (e.g. being in (2+1) dimensions, as in \cite{HHMSZ,Henderson:2017yuv}). Many questions yet remain.

\section{Conclusions}

Although an analytic expression for the Wightman function exists in AdS$_4,$ using it to compute the entanglement of detectors within that spacetime is not trivial. We chose to use a mode sum formalism, which can be generalized to any Killing spacetime, and then characterized the response of single detectors, as well as investigated the entanglement structure of pairs of detectors, within AdS$_4$. Because AdS$_4$ is Huygens, we were able to use a novel expression of $\mathcal{M}$ to evaluate the entanglement structure, but we believe that the expression should still be helpful in more general cases, especially if the two detectors remain spacelike separated. We then used our expressions to analyze two scenarios: one where one detector is in geodesic motion at fixed proper separation from the other  at the origin,  and another where one detector is static at a finite distance from the center.

 Still, the importance of the Huygens property of AdS$_4$ should not be understated. It is not a generic property of AdS, nor of four-dimensional spacetime individually: picking different dimensions, or picking another spacetime, will usually cause the commutator to become nonzero inside the light cone, making $\mathcal{M}^-$ more difficult to calculate. This, in a nutshell, is the largest fundamental difference between our AdS$_4$ and the previously analyzed BTZ \cite{Henderson:2017yuv} and AdS$_3$ cases \cite{HHMSZ}.  We believe this may also be the cause of some of the more qualitative differences between our results and those of AdS$_3,$  most notably related to the dependence of the negativity on which detector switches first: while the commutator contribution persists inside the lightcone in AdS$_3,$ this cannot happen in AdS$_4$, leading to some genuinely different results. 
 
A brief summary of the entanglement structure of AdS$_4$ is ``like Minkowski space locally, but more like a cavity globally.'' Similar to our analysis of the gravitating cavity, the detector is unable to distinguish AdS$_4$ from flat space unless switched for times comparable to the scale of the curvature. However, once that threshold is reached, the detectors begin to behave as though they were in a cavity: we find that the single detector exhibits mode resonances, while the pair of detectors experience periodicity in time. However, the degree of complexity introduced by forcing the detectors to remain static (and, therefore, at different redshifts and proper accelerations) spoils this simple picture. The complex structure that results  (even in Minkowski space) merits further investigation.

While the geodesic case is most symmetric, we believe that the static case is more generic, and thus more likely to be useful in the general case. Our methods should allow for calculation of the two-detector statistics of any static spacetime, even when the analytic form of the Wightman function is not known. As such, we will soon investigate the Schwarzschild solution, and hope to learn much about its entanglement structure. We hope that these methods will be employed to study many other spacetimes in the future.

\section*{Acknowledgements}
This work was supported in part by the Natural Science and Engineering Research Council of Canada. E. M-M. acknowledges funding of the Ontario Early Researcher Award.  We are grateful for discussions with Laura Henderson, Robie Hennigar, Alexander Smith, and Jialin Zhang.

\appendix
\section{Static detector causality estimator}

%\tcb{\bf[All new. Completely unverified! And not how I did it numerically. Although this might be a faster way than how I'm doing it now. Assuming it's correct.]}
Here we will perform substitute our Gaussian switching functions into \eqref{Cab_static} and integrate over time.
First, we make our expression more symmetrical by translating the time variable for each term. This also will allow us to use the previously found expressions for the products of the switching functions.
\begin{widetext}
\begin{align}
    \mathcal{C}_{AB}&=\frac{\ii\lambda_A \lambda_B}{8\pi L^2 \tan \varrho_B}\sum_{N=0}^\infty \int_{-\infty}^{\infty}dt\nonumber\\
    &\left[(e^{\ii((\tilde\Omega_A+\tilde\Omega_B) t)+(\tilde\Omega_A-\tilde\Omega_B)(\varrho_B+2N\pi))}+e^{\ii((\tilde\Omega_A-\tilde\Omega_B) t)+(\tilde\Omega_A+\tilde\Omega_B)(\varrho_B+2N\pi))})\right.\nonumber\\
    &\qquad\times\tilde\chi_A\left(t+\frac{\varrho_B+2N\pi}{2}\right)\tilde\chi_B\left(t-\frac{\varrho_B+2N\pi}{2}\right)\nonumber\\
    &+\varepsilon (e^{\ii((\tilde\Omega_A+\tilde\Omega_B) t)+(\tilde\Omega_A-\tilde\Omega_B)(-\varrho_B+(2N+1)\pi))}+e^{\ii((\tilde\Omega_A-\tilde\Omega_B) t)+(\tilde\Omega_A+\tilde\Omega_B)(-\varrho_B+(2N+1)\pi))})\nonumber\\
    &\qquad\times\tilde\chi_A\left(t+\frac{-\varrho_B+(2N+1)\pi}{2}\right)\tilde\chi_B\left(t-\frac{-\varrho_B+(2N+1)\pi}{2}\right)\nonumber\\
    &-\varepsilon (e^{\ii((\tilde\Omega_A+\tilde\Omega_B) t)+(\tilde\Omega_A-\tilde\Omega_B)(\varrho_B+(2N+1)\pi))}+e^{\ii((\tilde\Omega_A-\tilde\Omega_B) t)+(\tilde\Omega_A+\tilde\Omega_B)(\varrho_B+(2N+1)\pi))})\nonumber\\
    &\qquad\times\tilde\chi_A\left(t+\frac{-\varrho_B+(2N+1)\pi}{2}\right)\tilde\chi_B\left(t-\frac{-\varrho_B+(2N+1)\pi}{2}\right)\nonumber\\
    &-(e^{\ii((\tilde\Omega_A+\tilde\Omega_B) t)+(\tilde\Omega_A-\tilde\Omega_B)(-\varrho_B+(2N+2)\pi))}+e^{\ii((\tilde\Omega_A-\tilde\Omega_B) t)+(\tilde\Omega_A+\tilde\Omega_B)(-\varrho_B+(2N+2)\pi))})\nonumber\\
    &\qquad\times\left.\tilde\chi_A\left(t+\frac{-\varrho_B+(2N+2)\pi}{2}\right)\tilde\chi_B\left(t-\frac{-\varrho_B+(2N+2)\pi}{2}\right)\right]
\end{align}
\end{widetext}

Next, substitute in the switching function found in \eqref{chiachib}, and do the integration over $t$. The switching functions are all Gaussian, so this can be done analytically. The result should be:  
\begin{widetext}
\begin{align}
    \mathcal{C}_{AB}&=\frac{\ii\lambda_A \lambda_B}{8\pi \sin \varrho_B}\tilde\sigma_{AB}\sqrt{2\pi}\sum_{N=0}^\infty\nonumber\\
    &\left[e^{-\frac{(\varrho_B+2N\pi+t_0)^2}{2(\tilde\sigma_A^2+\tilde\sigma_B^2)}}\left(e^{-(\tilde\Omega_A+\tilde\Omega_B)^2/2\tilde\sigma_{AB}^2+\ii(\tilde\Omega_A+\tilde\Omega_B)(\varrho_B+2N\pi+t_0)\frac{\tilde\sigma_A^2-\tilde\sigma_B^2}{\tilde\sigma_A^2+\tilde\sigma_B^2}+\ii(\tilde\Omega_A-\tilde\Omega_B)(\varrho_B+2N\pi)}\right.\right.\nonumber\\
    &\qquad\left.+e^{-(\tilde\Omega_A-\tilde\Omega_B)^2/2\tilde\sigma_{AB}^2+\ii(\tilde\Omega_A-\tilde\Omega_B)(\varrho_B+2N\pi+t_0)\frac{\tilde\sigma_A^2-\tilde\sigma_B^2}{\tilde\sigma_A^2+\tilde\sigma_B^2}+\ii(\tilde\Omega_A+\tilde\Omega_B)(\varrho_B+2N\pi)}\right)\nonumber\\
    &+\varepsilon e^{-\frac{(-\varrho_B+(2N+1)\pi+t_0)^2}{2(\tilde\sigma_A^2+\tilde\sigma_B^2)}}\left(e^{-(\tilde\Omega_A+\tilde\Omega_B)^2/2\tilde\sigma_{AB}^2+\ii(\tilde\Omega_A+\tilde\Omega_B)(-\varrho_B+(2N+1)\pi+t_0)\frac{\tilde\sigma_A^2-\tilde\sigma_B^2}{\tilde\sigma_A^2+\tilde\sigma_B^2}+\ii(\tilde\Omega_A-\tilde\Omega_B)(-\varrho_B+(2N+1)\pi)}\right.\nonumber\\
    &\qquad\left.+e^{-(\tilde\Omega_A-\tilde\Omega_B)^2/2\tilde\sigma_{AB}^2+\ii(\tilde\Omega_A-\tilde\Omega_B)(-\varrho_B+(2N+1)\pi+t_0)\frac{\tilde\sigma_A^2-\tilde\sigma_B^2}{\tilde\sigma_A^2+\tilde\sigma_B^2}+\ii(\tilde\Omega_A+\tilde\Omega_B)(-\varrho_B+(2N+1)\pi)}\right)\nonumber\\
    &-\varepsilon e^{-\frac{(\varrho_B+(2N+1)\pi+t_0)^2}{2(\tilde\sigma_A^2+\tilde\sigma_B^2)}}\left(e^{-(\tilde\Omega_A+\tilde\Omega_B)^2/2\tilde\sigma_{AB}^2+\ii(\tilde\Omega_A+\tilde\Omega_B)(\varrho_B+(2N+1)\pi+t_0)\frac{\tilde\sigma_A^2-\tilde\sigma_B^2}{\tilde\sigma_A^2+\tilde\sigma_B^2}+\ii(\tilde\Omega_A-\tilde\Omega_B)(\varrho_B+(2N+1)\pi)}\right.\nonumber\\
    &\qquad\left.+e^{-(\tilde\Omega_A-\tilde\Omega_B)^2/2\tilde\sigma_{AB}^2+\ii(\tilde\Omega_A-\tilde\Omega_B)(\varrho_B+(2N+1)\pi+t_0)\frac{\tilde\sigma_A^2-\tilde\sigma_B^2}{\tilde\sigma_A^2+\tilde\sigma_B^2}+\ii(\tilde\Omega_A+\tilde\Omega_B)(\varrho_B+(2N+1)\pi)}\right)\nonumber\\
    &-e^{-\frac{(-\varrho_B+(2N+2)\pi+t_0)^2}{2(\tilde\sigma_A^2+\tilde\sigma_B^2)}}\left(e^{-(\tilde\Omega_A+\tilde\Omega_B)^2/2\tilde\sigma_{AB}^2+\ii(\tilde\Omega_A+\tilde\Omega_B)(-\varrho_B+(2N+2)\pi+t_0)\frac{\tilde\sigma_A^2-\tilde\sigma_B^2}{\tilde\sigma_A^2+\tilde\sigma_B^2}+\ii(\tilde\Omega_A-\tilde\Omega_B)(-\varrho_B+(2N+2)\pi)}\right.\nonumber\\
    &\qquad\left.\left.+e^{-(\tilde\Omega_A-\tilde\Omega_B)^2/2\tilde\sigma_{AB}^2+\ii(\tilde\Omega_A-\tilde\Omega_B)(-\varrho_B+(2N+2)\pi+t_0)\frac{\tilde\sigma_A^2-\tilde\sigma_B^2}{\tilde\sigma_A^2+\tilde\sigma_B^2}+\ii(\tilde\Omega_A+\tilde\Omega_B)(-\varrho_B+(2N+2)\pi)}\right)\right]
    \label{Cab_static_full}
\end{align}
\end{widetext}
For $\mathcal C_{BA},$ substitute $t_0$ with $-t_0,$ and exchange $\tilde\Omega_A$ with $\tilde\Omega_B,$ and $\tilde\sigma_A$ with $\tilde\sigma_B$. We leave fully expanding the tilded terms with respect to $\varrho_B$, $L$, $\Omega$ and $\sigma$ as an exercise for the reader.

\bibliography{ads_entanglement}

\end{document}